\documentclass[oneside,british,english]{book}

\usepackage[T1]{fontenc}
\usepackage[latin9]{inputenc}
\usepackage[paperwidth=17cm,paperheight=24cm]{geometry}
\usepackage{babel}
\usepackage{float}
\usepackage{amsthm}
\usepackage{amsmath}
\usepackage{amssymb}
\usepackage{graphicx}
\usepackage{esint}
\usepackage[unicode=true,pdfusetitle,
 bookmarks=true,bookmarksnumbered=false,bookmarksopen=false,
 breaklinks=false,pdfborder={0 0 1},backref=false,colorlinks=false]
 {hyperref}
\hypersetup{
 hidelinks}

\makeatletter

\let\SF@@footnote\footnote
\def\footnote{\ifx\protect\@typeset@protect
    \expandafter\SF@@footnote
  \else
    \expandafter\SF@gobble@opt
  \fi
}
\expandafter\def\csname SF@gobble@opt \endcsname{\@ifnextchar[%]
  \SF@gobble@twobracket
  \@gobble
}
\edef\SF@gobble@opt{\noexpand\protect
  \expandafter\noexpand\csname SF@gobble@opt \endcsname}
\def\SF@gobble@twobracket[#1]#2{}
\providecommand{\tabularnewline}{\\}

\numberwithin{equation}{section}

\usepackage{amsfonts,amssymb}
\usepackage{amsmath}
\usepackage{epsfig,epstopdf,graphicx}
\usepackage[hypcap]{caption}
\usepackage[T1]{fontenc}
\usepackage{ae,aecompl}
\usepackage{titlesec}

\makeatletter %only needed in preamble
\renewcommand\Huge{\@setfontsize\Huge{21pt}{26}}
\makeatother

\makeatother

\begin{document}
\begin{center}
{\LARGE Gravitational collisions and the quark-gluon plasma}
\par\end{center}{\large \par}

\vspace{1.2cm}

\begin{center}
{\large Wilke van der Schee,}
\par\end{center}{\large \par}

\begin{center}
Institute for Theoretical Physics and Institute for Subatomic Physics,\\
Utrecht University, Leuvenlaan 4, 3584 CE Utrecht, The Netherlands
\par\end{center}

\noindent \thispagestyle{empty}
\vspace{1.2cm}

\noindent \begin{center}
\textbf{\large Abstract}
\par\end{center}{\large \par}

\vspace{0.5cm}

\noindent This thesis addresses the thermalisation of heavy-ion collisions
within the context of the AdS/CFT duality. The first part clarifies
the numerical set-up and studies the relaxation of far-from-equilibrium
modes in homogeneous systems. Less trivially we then study colliding
shock waves and uncover a transparent regime where the strongly coupled
shocks initially pass right through each other. Furthermore, in this
regime the later plasma relaxation is insensitive to the longitudinal
profile of the shock, implying in particular a universal rapidity
shape at strong coupling and high collision energies. Lastly, we study
radial expansion in a boost-invariant set-up, allowing us to find
good agreement with head-on collisions performed at the LHC accelerator.\newline

\noindent As a secondary goal of this thesis, a special effort is
made to clearly expose numerical computations by providing commented
\emph{Mathematica }notebooks for most calculations presented%
\footnote{\emph{Mathematica} notebooks and sample simulations can be found at:

\href{http://sites.google.com/site/wilkevanderschee/phd-thesis}{sites.google.com/site/wilkevanderschee/phd-thesis}%
}. Furthermore, we provide interpolating functions of the geometries
computed, which can be of use in other projects.

\vspace{0.4cm}

\noindent \begin{flushleft}
Promotors: Gleb Arutyunov and Thomas Peitzmann
\par\end{flushleft}

\vspace{2cm}

\begingroup \renewcommand{\vspace}[2]{}% Gobble 2 arguments after \vspace

\tableofcontents{}\endgroup

\chapter*{Publications}

This thesis is based on the following publications:
\begin{enumerate}
\item Jorge Casalderrey-Solana, Michal Heller, David Mateos and Wilke van
der Schee, Longitudinal Coherence in a Holographic Model of Asymmetric
Collisions, \\ Physical Review Letters \textbf{112}, 221602 (2014)
or arxiv:1312.2956
\item Wilke van der Schee, Paul Romatschke and Scott Pratt, \\ A fully
dynamical simulation of central nuclear collisions, \\ Physical Review
Letters \textbf{111}, 222302 (2013) or arxiv:1307.2539 
\item Jorge Casalderrey-Solana, Michal Heller, David Mateos and Wilke van
der Schee, From full stopping to transparency in a holographic model
of heavy ion collisions, Physical Review Letters \textbf{111}, 181601
(2013) or arxiv:1305.4919 
\item Michal Heller, David Mateos, Wilke van der Schee and Miquel Triana,
\\ Holographic isotropization linearized, JHEP \textbf{09} (2013)
026 or arxiv:1304.5172 
\item Wilke van der Schee, Quarks, gluonen en zwarte gaten, \\ Nederlands
Tijdschrift voor Natuurkunde \textbf{79} 112-114 (mei 2013) 
\item Wilke van der Schee, Holographic thermalization with radial flow,
\\ Physical Review D \textbf{87}, 061901 (R) (2013) or arxiv:1211.2218
\item Michal Heller, David Mateos, Wilke van der Schee and Diego Trancanelli,
\\ Strong coupling isotropization simplified, \\ Physical Review
Letters \textbf{108}, 191601 (2012) or arxiv:1202.0981
\end{enumerate}

\chapter{Introduction}

The theory of quarks and gluons, quantum chromodynamics (QCD), has
been well established for decades now. But while the basic Lagrangian
of the theory is well known, the non-perturbative nature of this strong
force makes it hard to make practical use of the theory, especially
in situations that are out-of-equilibrium. In particular, it is still
poorly understood how a quark-gluon plasma forms in collisions of
relativistic heavy ions, such as performed at the RHIC and LHC accelerators. 

In this thesis we try to address this problem using the AdS/CFT duality.
Although this duality is only understood for theories related to QCD,
it is especially well suited to treat strong coupling and may as such
teach us about similar phenomena in QCD. In the future, the hope is
to get a better understanding of non-perturbative quantum theories,
such as QCD.

\section{Relativistic heavy ion collisions}

Colliding highly relativistic nuclei can create a very dense and hot
plasma of quarks and gluons, the so-called quark-gluon plasma. The
temperature of this plasma can reach over $10^{12}K$, which is as
hot as the universe a millisecond after the big bang. Of course, the
scale is very small: a typical collision lasts only $10\,\text{fm}/c$
and takes place within a sphere of radius $15\,\text{fm}$. Nevertheless,
at the Large Hadron Collider (LHC) these collisions can create about
26.000 particles, the analysis of which teaches us about conditions
shortly after the big bang, and more importantly about QCD in general.

Colliders such as RHIC (Relativistic Heavy Ion Collider) and LHC collide
gold nuclei (79 protons and 118 neutrons) or lead nuclei (82 protons
and 126 neutrons) respectively. At the highest energy RHIC can achieve,
each proton and neutron has an energy of 100 GeV, so they are Lorentz
contracted by a factor of one hundred. LHC achieves an even higher
energy of 1.38 TeV, giving a Lorentz factor of more than a thousand.
In both these colliders the ions move with equal energies in opposite
directions in the beam line. At the location of the detectors both
beam lines cross, such that some nuclei will hit each other, thereby
creating the quark-gluon plasma.

For our purposes we can approximate a nucleus as a smooth distribution
of energy, shaped as a Lorentz boosted sphere with radius $R\simeq6.5\,\text{fm}$.
In this simplification, the two colliding nuclei will hit randomly,
where the chance $p$ of the distance between both centres $b$ (impact
parameter) being less than $r$ is given by 
\begin{equation}
p(b<r)=r^{2}/4R^{2}.\label{eq:centrality}
\end{equation}

\begin{figure}[H]
\begin{centering}
\includegraphics[width=8cm]{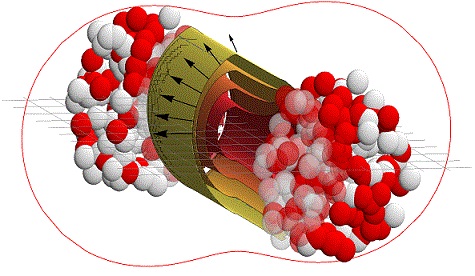}
\par\end{centering}

\caption{\label{fig:A-cartoon-of-HIC}A cartoon of a typical heavy-ion collision
of impact parameter $b=6\,\text{fm}$. The two Lorentz contracted
ions move ultrarelativistically along the z-axis; the nucleons in
the non-overlapping are called spectators and just fly on. The nucleons
in the collision region (opaque) collide and most of their energy
ends up in a quark-gluon plasma. The elliptical shape of this region
has larger pressure gradient in the short axis, causing particles
to be pushed in this direction, as indicated with the red line (elliptic
flow). The succes of this hydrodynamic picture was crucial evidence
that the dynamics is strongly coupled.}
\end{figure}

The protons and neutrons of the nucleus which do not hit the other
nucleus are called spectators since they have little effect on the
collision, as illustrated in \ref{fig:A-cartoon-of-HIC}. This means
that events with a small impact parameter (lower centrality) will
produce many more particles, making a reliable measurement of centrality
relatively straightforward.

\begin{figure}
\begin{centering}
\includegraphics[width=6.5cm]{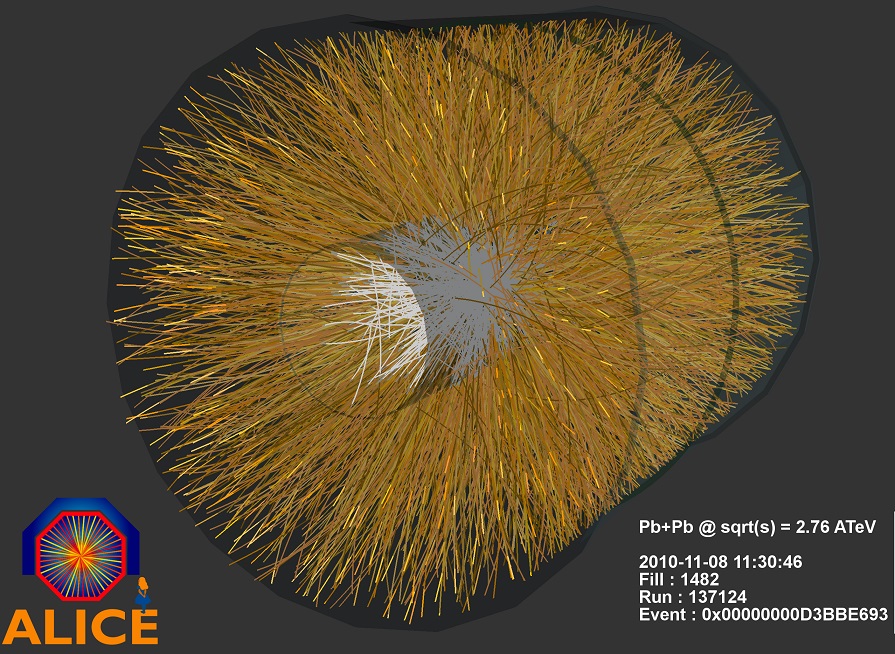}
\par\end{centering}

\caption{\label{fig:A-typical-heavy-ion}A heavy-ion collision, as registered
by the ALICE detector. It is a challenge to gain information about
the formed quark-gluon plasma from these thousands of tracks, but
using correlations and the billions of registered collisions a lot
has been learnt. Note that this detector is several meters in size,
about $10^{15}$ times bigger than the actual events studied.}
\end{figure}

The interesting challenge is to model such collisions theoretically
and predict the spectra of the resulting particles spray (fig. \ref{fig:A-typical-heavy-ion}).
Of particular interest is the averaged momentum anisotropy in the
transverse plane, usually expanded in spherical harmonics \cite{Heinz:2013th}:
\begin{equation}
\frac{d\bar{N}}{d\varphi}=\frac{\bar{N}}{2\pi}\left(1+2\sum_{n=1}^{\infty}\bar{v}_{n}\cos(n(\varphi-\bar{\Psi}_{n}))\right),
\end{equation}
with $\varphi$ the angle in the transverse plane, $\bar{\Psi}_{n}$
defined such that there are no sine terms, $\bar{N}$ the average
number of particles of interest per event and $\bar{v}_{n}$ the anisotropic
flow coefficients. The most studied is called the elliptic flow coefficient
$\bar{v}_{2}$, which is relatively large for non-central collisions
due to the approximately ellipsoidal shape of the interaction region,
as illustrated in figure \ref{fig:A-cartoon-of-HIC}. Crucially, a
(hydrodynamic) expansion will convert this elliptical shape in real
space into a similar shape in momentum space, which is experimentally
accessible and can thus provide important insights in the details
of the expansion.

Although it is currently difficult to make definite statements about
the first stages of the quark-gluon plasma by analysing the data,
there is good reason to be optimistic for significant future improvements.
The large number of events measured (many billions) makes a constraining
data set, which has a large potential for distinguishing both the
initial stage and the subsequent evolution. For this, one should not
only look at for instance $\bar{v}_{2}$ averaged over all particles
and events, but one can look at $v_{n}$ depending on transverse momentum,
rapidity and particle species, or one can look at four and higher
order particle correlations, fluctuations from event to event, or
even correlations between two different $v_{n}$. Furthermore, one
can vary the energy of the colliding nuclei or change the nuclei themselves,
thereby changing the collision geometry. At RHIC this has recently
been started, which resulted in a large amount of data, which interestingly
does not seem to be fully captured by current hydrodynamic models
\cite{Adamczyk:2012ku}.

One of the main uncertainties in current models of heavy-ion collisions
concerns the initial stage directly after the collision, before the
quark-gluon plasma is formed. This initial stage is problematic, since
it concerns the real-time evolution of many quarks and gluons, which
should in principle be described by a fully non-perturbative calculation
in QCD. Currently this cannot be achieved, and typically one resorts
to weakly coupled calculations, such as used by the colour glass condensate
\cite{McLerran:1993ni,Gelis:2012ri}. In this thesis we take a different
approach, by using holography, which allows doing a full strongly
coupled calculation, albeit not in QCD itself. The hope is that a
combination of both weakly and strongly coupled methods will then
lead to a better understanding of the initial stage of a heavy-ion
collision.

\section{Holography}

The concept of holography goes back to two very old ideas. The first
idea is from 't Hooft, in 1974 \cite{'tHooft:1973jz}. There he noticed,
with a very general argument, that strongly coupled $SU(N_{c})$ gauge
theories with coupling constant $g$ may simplify when $N_{c}$ is
large. When examining Feynman diagrams, a simple counting argument
shows that all diagrams of leading order in $N_{c}$, while keeping
the 't Hooft coupling $\lambda=g^{2}N_{c}$ fixed%
\footnote{It is worth mentioning that in the large $N_{c}$ limit with fixed
$\lambda=g^{2}N_{c}$ the coupling $g$ goes to zero. This, nevertheless,
describes a strongly coupled theory, as the expansion in the coupling
constant has an effective expansion parameter $\lambda$.%
}, are planar (they can be drawn on a plane). If one combines this
result with the idea that the path integral may be rewritten as a
string theory, one sees that this string theory will not contain loops
if $N_{c}$ is large, thereby making computations much easier.

A relatively independent argument comes from black hole thermodynamics,
where it was noticed early on that the entropy of a black hole scales
with the area of the black hole horizon \cite{Bekenstein:1973ur}.
The implications of this scaling go much beyond the study of just
black holes. In a thought experiment one can imagine collapsing any
region of spacetime to a black hole, whereby entropy necessarily needs
to increase. The only possible conclusion is that any region of spacetime
has an entropy bounded by its area \cite{'tHooft:1989kz,'tHooft:1993gx,Susskind:1994vu}.
This is of little practical concern, due to the large prefactor, but
it naturally leads to the idea that any theory with gravity is fundamentally
holographic: it can be described by a theory with one dimension less.

More recently in 1997, Maldacena made both ideas precise in an extraordinary
paper \cite{Maldacena:1997zz} (see also \cite{Witten:1998qj,Aharony:1999ti}).
Here he conjectured an exact duality, AdS/CFT, between type IIB string
theory on a five dimensional anti-de-Sitter (AdS) background%
\footnote{Type IIB string theory in its full form lives in ten dimensions, of
which five form $AdS_{5}$ and five others a 5-sphere. Dynamics on
this sphere, however, can in many cases be consistently decoupled
and will therefore not be considered in this thesis.%
} with conformal super-Yang-Mills gauge theory with four supersymmetries,
a $SU(N_{c})$ gauge group and living in four dimensions (the CFT:
$\mathcal{N}=4$ SYM). So indeed he found a string theory dual of
a $SU(N_{c})$ gauge theory, which simplifies to a free string theory
when $N_{c}$ is large. Moreover, the string theory reduces to gravity
in the classical, non-string, limit and this duality provides the
precise holographic dictionary to a non-gravitational theory living
in one dimension less \cite{Susskind:1998dq}.

Although the above arguments are very general it has so far only been
possible to find a precise dual for specific gauge theories, with
usually quite some supersymmetry. In particular, a realistic dual
to (large $N_{c}$) QCD is still far away, and the weak coupling in
the ultra-violet (UV) of QCD implies that a full solution will require
solving the planar string theory, at least in the UV.

On the other hand, the AdS/CFT correspondence can be generalised and
allows studying a wide variety of field theories. Importantly, these
field theories do not need to be conformal and can for instance have
confinement \cite{Witten:1998zw} or a running coupling resembling
QCD quite closely, such as in improved holographic QCD \cite{Gursoy:2007cb}.
The field theory can have (some of) the supersymmetry broken \cite{Klebanov:1999tb}.
Including D7-branes in the bulk can include flavoured quarks \cite{Karch:2002sh}.

\section{Relativistic hydrodynamics and fluid/gravity\label{sec:Relativistic-hydrodynamics-and}}

This thesis deals with the collision of heavy ions, particularly with
the initial far-from-equilibrium evolution to a quark-gluon plasma
describable using relativistic hydrodynamics. One would be tempted
to call this transition `thermalisation', but strictly speaking this
is not correct, since we consistently find states described by hydrodynamics
where pressures in different directions are different. A completely
thermalised fluid cell would have equal pressures, which will be achieved
much later than the moment hydrodynamics becomes applicable.

Formally, hydrodynamics can be viewed as a gradient expansion around
thermal equilibrium. One starts with an exact (boosted) thermal solution
with constant energy density and fluid velocity and then promotes
these two to a field, both assumed to vary slowly compared to other
scales, which gives the following constitutive relations, up to first
order in gradients \cite{Baier:2007ix}: 
\begin{eqnarray}
T_{\mu\nu} & = & e\, u_{\mu}u_{\nu}+p[e]\Delta_{\mu\nu}+\pi_{\mu\nu},\text{ where,}\label{eq:hydro-constituive}\\
\Delta_{\mu\nu} & = & g_{\mu\nu}+u_{\mu}u_{\nu}\text{ and }\\
\pi_{\mu\nu} & = & -\eta[e]\,\sigma_{\mu\nu}+\mathcal{O}(\partial^{2}),\text{ with}\\
\sigma_{\mu\nu} & = & \Delta_{\mu\alpha}\Delta_{\nu\beta}(\nabla^{\mu}u^{\nu}+\nabla^{\nu}u^{\mu})-\frac{2}{d-1}\Delta_{\mu\nu}\Delta_{\alpha\beta}\nabla^{\alpha}u^{\beta},\label{eq:sigma}
\end{eqnarray}
where $e$ is the local energy density, $p[e]$ is the equation of
state, $u_{\mu}$the local fluid velocity, $\pi_{\mu\nu}$ the shear
tensor and $\eta$ the shear viscosity, which is the only non-vanishing
transport coefficient in first order conformal hydrodynamics. Note
that the fluid velocity and energy density are defined as the timelike
eigenvector and associated eigenvalue of the stress tensor ($T_{\mu\nu}u^{\nu}=e\, u_{\mu})$
and that $\sigma$ is transverse and traceless: $u^{\mu}\sigma_{\mu\nu}=\sigma_{\,\mu}^{\mu}=0$.
Alternatively one can say that the fluid velocity is defined such
that when boosting $T_{\mu\nu}$ with velocity $u_{\mu}$ there is
no momentum flow, i.e. $T'_{0i}=0$, which is called the Landau frame.
Having written down the hydrodynamic constitutive relations it will
be essential in this thesis to check if \ref{eq:hydro-constituive}
holds for resulting stress tensors. Also, one can use the conservation
equation $\nabla^{\mu}T_{\mu\nu}=0$ to evolve an initial energy density
and fluid velocity forward in time.

Problematically, relativistic first-order hydrodynamics contains modes
propagating faster than light\cite{Hiscock:1985zz}, as can be seen
by looking at the dispersion relation at high momenta. These modes
contain large gradients and are hence outside the regime of the applicability
of hydrodynamics. The acausal modes are therefore not a fundamental
problem, but they nevertheless cause instabilities when solving the
equations numerically. For this purpose second-order hydrodynamics
has been extensively studied \cite{Baier:2006gy}, which is a causal
and numerically stable theory for suitable transport coefficients
\cite{Baier:2007ix}. Interestingly, in all microscopic theories where
these transport coefficients could be computed the second-order hydrodynamics
is causal, but it is still an open question if this is always true.

In a recent paper \cite{Heller:2013fn} it is shown that the hydrodynamic
gradient expansion is not necessarily convergent. This is particularly
clear in the example of section \ref{sec.linearized}, where we can
explicitly find degrees of freedom not described by hydrodynamics,
the so-called quasi-normal modes. In \cite{Heller:2013fn} this was
made precise within AdS/CFT by computing the hydrodynamic expansion
up to order 240 in derivatives and identifying in the re-summed series
the lowest quasi-normal mode.

While the above describes hydrodynamics on itself, the idea of a link
between hydrodynamics and gravity dates back to the eighties. First,
the membrane paradigm \cite{Thorne:1986iy} proposed that an outside
observer may view the horizon of a black hole as a membrane. This
membrane would behave very much like a fluid, with temperature, heat
flow, electrical conductivities and so on. Later on this could be
made much more precise using AdS/CFT, where a CFT in the hydrodynamic
regime can be precisely identified with the corresponding gravitational
system.

Importantly, this fluid/gravity correspondence \cite{Bhattacharyya:2008jc,Hubeny:2011hd}
is not a duality between hydrodynamics and gravity, since the gravitational
side also contains non-hydrodynamic degrees of freedom. In this thesis
we will mostly be interested in this far-from-equilibrium regime,
where there is no hydrodynamic description. We will see, however,
that relatively quickly hydrodynamics does become applicable, which
can in some sense be rephrased as that black hole horizons equilibrate
fast.

\section{Holography and heavy-ion physics}

As is now clear it is possible to use AdS/CFT to study strongly coupled
theories in the thermodynamic limit, but it can not be used for any
gauge theory, in particular not for QCD itself. Nevertheless, AdS/CFT
is one of the only tools to study strongly coupled theories, especially
in real-time dynamics where the sign problem makes lattice simulations
almost impossible.

There are excellent reviews \cite{Mateos:2007ay,Gubser:2009md,DeWolfe:2013cua,Janik:2013qua}
and a recent book \cite{Casalderrey:2014} on AdS/CFT applied to heavy-ion
collisions. These applications typically focus on three topics. The
first and oldest application studies the transport coefficients during
the hydrodynamic phase, most famously the shear viscosity \cite{Policastro:2001yc},
$\eta=s/4\pi$ with $s$ the entropy density. While it is already
a major achievement of AdS/CFT that one can compute the shear viscosity
from a microscopic theory, it is also offers a natural explanation
in terms of dissipation near a black hole horizon. The latter suggests
that this (small) shear viscosity may be far more universal than just
$\mathcal{N}=4$ SYM theory, and indeed heavy-ion experiments suggest
a value close to the prediction by AdS/CFT \cite{Heinz:2013th}.

A second topic often studied is jet quenching. At the very first moment
of the collision it is possible to form a pair of ultra-energetic
quarks, with energies as large as 100 GeV. These quarks then have
to pass through (part of) the quark-gluon plasma, whereby they can
lose energy. As the energy of the quark jets can be compared between
themselves and also with similar results from (simpler) proton-proton
collisions the energy loss can be well estimated experimentally. In
QCD itself it is hard to study such an energy loss, but at strong
coupling several interesting estimates have been made using AdS/CFT
around 2006 \cite{Gubser:2006bz,Herzog:2006gh,Liu:2006ug} and also
more recently \cite{Chesler:2013cqa,Ficnar:2013wba,Chesler:2014jva}.
This may be of special interest as these quarks have energies much
above the quark-gluon plasma temperature and can therefore be used
to study QCD at higher energies, whereby the coupling constant is
weaker.

Lastly, the far-from-equilibrium initial stage of the collision is
an excellent example of real-time dynamics and AdS/CFT is the only
available tool to study this if the coupling is strong. In this case
the formation of a thermal quark-gluon plasma, dual to a black hole
in AdS, really corresponds to black hole formation. There has been
previous works on this \cite{Horowitz:1999jd,Chesler:2008hg,Chesler:2010bi,Heller:2011ju},
suggesting that this black hole forms `as fast as possible', within
a time shorter than a thermal wavelength. This thesis will focus on
this avenue and push these earlier studies to more realistic settings,
aiming at a comparison with experimental data.

\section{Outline}

In this thesis we try to address the initial stage of a heavy-ion
collision before hydrodynamics becomes applicable within the framework
of AdS/CFT. Since computations within general relativity can still
be challenging, the problem is studied from three different viewpoints.
Chapter \ref{chap:General-relativity-in} studies the transition from
far-from-equilibrium to hydrodynamics in a completely homogeneous
setting. While far from realistic, the two interesting results are
a universal `fast' thermalisation, and a simplification in terms of
linearised equations. Furthermore, this chapter allows us to introduce
the so-called characteristic formulation to solve Einstein equations
numerically.

Chapter \ref{chap:Colliding-planar-shock} assumes homogeneity in
the transverse plane, allowing us to study the longitudinal dynamics
of the collision. Several profiles were studied, resulting in fully
stopped nuclei, transparent collisions and asymmetric collisions,
which may model qualitative features of RHIC, LHC and asymmetric proton-lead
collisions. One of the main results is the profile of the local energy
density as a function of rapidity. Against expectations this profile
turns out not to be boost-invariant, but has a universal shape, even
at asymptotically high energies. We comment on experimental consequences.

Chapter \ref{chap:Thermalisation-with-radial} assumes boost-invariance
along the collision axis and rotational symmetry in the transverse
plane, allowing us to study the radial dynamics of the collision.
This radial expansion is crucial for the transverse particle spectra,
and this will be used to present a fully dynamical simulation, all
the way from far-from-equilibrium to viscous hydrodynamics, to a hadronic
gas cascade, to the final (measured) particle spectra. The model fits
the data surprisingly well, especially considering that the simulation
is much more constrained than previous attempts.

In the end, the hope is expressed that a combination of these methods
may provide a full picture of a heavy-ion collision at strong coupling,
noting especially that the longitudinal dynamics is (initially) much
faster than the transverse dynamics.

\chapter{General relativity in the characteristic formulation\label{chap:General-relativity-in}}

\chaptermark{The characteristic formulation}

Solving Einstein's equations numerically can be a very difficult task.
It was for instance only in 2005 that it became possible to fully
simulate the merger of two black holes \cite{Pretorius:2005gq}. However,
within the context of AdS/CFT Einstein's equations can be naturally
rewritten in a much simpler numerical scheme. This so-called `characteristic'
formulation was first discovered by Bondi \cite{Bondi,Bondi1960Gravitational}
and Sachs \cite{Sachs} in the 1960s while studying gravitational
waves in flat space, after which Chesler and Yaffe \cite{Chesler:2008hg,Chesler:2013lia}
pioneered this formulation within AdS.

The key simplification is to write the coupled partial differential
equations into a nested set of linear ordinary differential equations
(ODE). For this three steps are essential:
\begin{enumerate}
\item Fix (part of) the diffeomorphism invariance by employing generalised
ingoing Eddington-Finkelstein coordinates, where paths of varying
radial coordinate $r$ (with other coordinates fixed) are null geodesics.
\item The determinant of the spatial part of the metric needs to be a single
function.
\item Instead of writing Einstein's equations directly in terms of time
derivatives, one should use derivatives along outgoing null rays.
\end{enumerate}
In the community of numerical general relativity the characteristic
formulation is not very popular. This is firstly due to the required
null slices, which in particular should not form caustics. In typical
problems in numerical general relativity, such as the collision of
black holes, gravitational lensing does quite generally form caustics.
In typical problems studied in AdS/CFT on the other hand, caustics
are unlikely to arise. Furthermore, more generally caustics are only
expected in the far infrared, and can presumably be neglected for
most purposes.

Secondly, in flat space a constant time slice is usually a natural
starting point, which leads to evolution using the ADM formalism (developed
by Richard Arnowitt, Stanley Deser and Charles Misner \cite{Arnowitt:1962hi}).
From the point of view of the boundary of AdS a null slice is perhaps
more natural, and these light rays are indeed used as a mapping from
boundary to horizon, in the fluid/gravity correspondence \cite{Hubeny:2011hd}.
In the context of holographic thermalisation there is one study using
the ADM formulation in a boost-invariant setting with interesting
results \cite{Heller:2011ju}, but the numerics in this study is somewhat
complicated.

\section{The metric ansatz and AdS/CFT\label{sec:The-metric-ansatz}}

We use coordinates $r$, $t$ and $x_{i}$%
\footnote{We use $x_{i}$, $x_{\mu}$ and $x_{M}$ to denote boundary space,
boundary space-time and AdS spacetime coordinates respectively. Furthermore,
we use units where the size of AdS $L_{AdS}=1$. Though many methods
are applicable to other dimensions, we restrict ourselves to 3+1 dimensions
in the CFT, which gives 4+1 dimensions in AdS.%
}, where $r=\infty$ corresponds to the boundary of AdS, $x_{i}$ are
the spatial coordinates of the boundary and $t$ is the time coordinate
on the boundary, which is null inside AdS. This fixes the metric to
be of the form
\begin{equation}
ds^{2}=dt\left[-Adt+\beta dr+2F_{i}dx^{i}\right]+S^{2}\, h_{ij}dx^{i}dx^{j},\label{eq:metricEF}
\end{equation}
where $A$, $\beta$, $F_{i}$, $S$ and $h_{ij}$ are functions of
all coordinates, and $\det(h_{ij})=1$. This metric is still invariant
under arbitrary reparameterisations of $r$, which need to be fixed
for a well-posed initial-value problem. Bondi \cite{Bondi,Bondi1960Gravitational}
and Sachs \cite{Sachs} did this by choosing $S=r$, appropriate for
their spherical coordinates, and a similar choice was also used more
recently in \cite{Balasubramanian:2013yqa}. Here, we follow Chesler
and Yaffe \cite{Chesler:2013lia} and choose to fix $\beta=2$. 

While it is possible to do a fully covariant analysis of the Einstein
equation in this gauge \cite{Chesler:2013lia}, we choose to illustrate
the solutions by the examples presented in this thesis. However, a
few general remarks are in order:
\begin{itemize}
\item As an initial condition, encoding the full quantum state of the CFT,
it is sufficient to provide $h_{ij}(t=0,\, r,\, x^{i}$) and boundary
conditions (at $r=\infty$) for $A$ and $F_{i}$, where the latter
can be thought of as energy density and momentum flow. Conveniently,
Einstein's equations fix the other metric components, which is easier
than in the ADM-evolution, where providing consistent initial conditions
is a non-trivial problem.
\item In normal Cauchy evolution one would always initially specify the
metric and its first time derivative. In the null form \ref{eq:metricEF}
this is not necessary, but one has to provide extra boundary conditions
at the boundary. Importantly, these boundary conditions can causally
influence the whole domain instantaneously in $t$, which is a major
difference with Cauchy evolution.
\item For the simplification of solving nested linear ODEs it is essential
to first compute derivatives of all functions $h_{ij}$ in the direction
of outgoing null rays, $(\partial_{t}+\frac{1}{2}A\partial_{r})h_{ij}\equiv\dot{h}_{ij}$,
and subsequently compute $A$. Computing $\dot{h}$ is done using
the $ij$ components of Einstein equations, which are invariant under
the residual gauge transformation $r\rightarrow r+\xi(x^{\mu})$ (presented
below), just like $\dot{h}$ is. As $A$ is not invariant it follows
that these equations do not contain $A$.
\end{itemize}

\subsection{Holographic renormalisation and near-boundary expansions\label{sub:Holographic-renormalisation-and}}

Clearly, we need some dictionary to translate observables in AdS to
observables in the CFT. It is important that only observable and hence
gauge independent quantities can be matched. The AdS/CFT dictionary
is simply that the partition sums of the AdS and the CFT theories
should be equal (see \cite{Aharony:1999ti}, page 63): 
\begin{equation}
\langle e^{\int d^{4}x\phi_{0}(x^{\mu}){\cal O}(x^{\mu})}\rangle_{CFT}={\cal Z}_{string}\Bigg[\phi(x^{\mu},\, z)\Big|_{z=0}=\phi_{0}(x^{\mu})\Bigg],\label{eq:AdS-CFT}
\end{equation}
where in this case $\phi(x^{\mu},\, z)$ is a scalar field (dilaton)
in AdS, which has $\phi_{0}(x^{\mu})$ as its boundary condition,
which in turn is a source in the CFT for a scalar operator dual to
the dilaton ${\cal O}(x^{\mu})$. While written out here for a scalar,
the same logic applies to other fields, in particular the stress tensor
$T_{\mu\nu}$, which is dual to the metric field.

One of the problems is that both partition sums diverge, which reflects
the UV divergence of quantum field theories, and the IR divergence,
or infinite volume, of theories in AdS. The renormalisation of these
divergences and the matching of observables thereafter goes under
the name of holographic renormalisation \cite{deHaro:2000xn}.

This holographic renormalisation is typically most conveniently done
by writing the metric in the Fefferman-Graham form:
\begin{equation}
ds^{2}=\frac{dz^{2}+g_{\mu\nu}dx^{\mu}dx^{\nu}}{z^{2}},\label{eq:metricFG}
\end{equation}
where $g_{\mu\nu}$ depends on all coordinates and now the boundary
is located at $z=0$. As is clear from \ref{eq:AdS-CFT} (local) CFT
observables are determined near the boundary, so that it is natural
to expand AdS fields near the boundary:
\begin{equation}
g_{\mu\nu}(z,\, x^{\mu})=\sum_{n=0}^{\infty}g_{\mu\nu}^{(n)}(x^{\mu})\, z^{n}+\tilde{g}_{\mu\nu}^{(n)}(x^{\mu})\, z^{n}\log[z].\label{eq:NB-expansion}
\end{equation}
While solving Einstein equations numerically can be technically involved,
performing a (high-order) analytic near-boundary expansion is much
easier. Order by order one plugs \ref{eq:NB-expansion} into the Einstein
equations, which are linear algebraic equations at each order:%
\footnote{For an efficient implementation one may consult the \emph{Mathematica}
notebook accompanying chapter \ref{chap:Colliding-planar-shock},
where this expansion is performed for a metric and gauge field with
planar homogeneity.%
}
\begin{equation}
R_{MN}-\frac{1}{2}g_{MN}R=8\pi G_{N}T_{MN}=6g_{MN},\label{eq:Einstein}
\end{equation}
where we used our units where $L_{AdS}=1$ and $G_{N}=1/8\pi$. As
Einstein equations are second order, one will find two undetermined
terms: $g_{\mu\nu}^{(0)}(x^{\mu})$ and $g_{\mu\nu}^{(4)}(x^{\mu})$.
The first is non-normalisable and should be thought of as a source
term in the CFT corresponding to the (non-dynamical) metric the CFT
lives on. Indeed, one can study a CFT on a curved spacetime, thereby
sourcing energy into the spacetime \cite{Gibbons:1977mu}, which was
for instance explored in \cite{Chesler:2008hg}. The normalisable
mode $g_{\mu\nu}^{(4)}(x^{\mu})$ can be thought of as the CFT stress
tensor, which is traceless and conserved with respect to the CFT metric
$g_{\mu\nu}^{(0)}$. All logarithmic terms $\tilde{g}_{\mu\nu}^{(n)}(x^{\mu})$
are completely fixed in terms of $g_{\mu\nu}^{(0)}$, and they vanish
if $g_{\mu\nu}^{(0)}$ is flat.

In reference \cite{deHaro:2000xn} it was shown how to carefully subtract
all counterterms, leading to a renormalised CFT stress tensor in terms
of AdS observables:
\begin{equation}
<T_{\mu\nu}>=\frac{N_{c}^{2}}{2\pi^{2}}\left(g_{\mu\nu}^{(4)}-\frac{1}{8}g_{\mu\nu}^{(0)}\left[(\text{Tr}\, g^{(2)})^{2}-\text{Tr}\, g^{(2)}{}^{2}\right]-\frac{1}{2}(g^{(2)\,2})_{\mu\nu}+\frac{1}{4}g^{(2)}{}_{\mu\nu}\text{Tr}\, g^{(2)}\right),
\end{equation}
where we reinstated $G_{N}=\frac{\pi}{2N_{c}^{2}}$, which is valid
for a $\mathcal{N}=4$ SYM SU($N_{c}$) dual. Fortunately, if the
CFT metric is flat then $g^{(2)}=\tilde{g}^{(n)}=0$, making the numerics
significantly simpler.

For numerical evolutions the form \ref{eq:metricFG} is not convenient,
as there is a coordinate singularity at the horizon. Much better are
the Eddington-Finkelstein coordinates (eqn. \ref{eq:metricEF}), which
will be used throughout this thesis. To obtain the stress-tensor one
therefore needs to compute the transformation between both frames
near the boundary, which can again be easily computed order-by-order
by solving linear algebraic equations. More details are given in subsection
\ref{subsec:FG-to-EF}, where also the complete transformation will
be computed.

One subtlety arises when computing the near-boundary expansion of
\ref{eq:metricEF}, where Einstein equations leave 3 terms of the
expansion of $A(r,\, x_{\mu})$ undetermined. This reflects a residual
gauge symmetry of the metric under $r\rightarrow r+\xi(x^{\mu})$,
leaving the form of the metric intact (albeit transforming non-trivially
$A(r,\, x_{\mu})\rightarrow A(r+\xi(x_{\mu}),\, x_{\mu})-2\partial_{t}\xi(x^{\mu})$
and $F_{i}(r,\, x_{\mu})\rightarrow F_{i}(r+\xi(x_{\mu}),\, x_{\mu})+\partial_{i}\xi(x^{\mu})$
). In practice this gauge symmetry will be essential to get a rectangular
computational domain, thereby simplifying numerics significantly.

\section{Numerics and a homogeneous background\label{sec:Numerics-and-a}}

This section will study the simplest non-trivial example of thermalisation
using the characteristic formulation, which will illustrate both the
numerical method and the (CFT) physics involved. The simplest set-up
assumes complete homogeneity in the three boundary coordinates $x^{i}=(x_{L},\,\mathrm{\mathbf{x}}_{\text{T}})$,
but allows for a time dependent anisotropy: $T_{x_{L}x_{L}}(t)\neq T_{x_{T}x_{T}}(t)$,
thereby assuming rotational symmetry in the two transverse directions.
This symmetry allows the metric \ref{eq:metricEF} to be further simplified
into
\begin{equation}
ds^{2}=2dtdr-Adt^{2}+S^{2}e^{-2B}dx_{\text{L}}^{2}+S^{2}e^{B}d\mathrm{\mathbf{x}}_{\text{T}}^{2}\,,\label{metricEF-homogeneous}
\end{equation}
where $A$, $S$ and $B$ are functions of time $t$ and the radial
coordinate $r$. The link between the form of the field theory stress
tensor and the dual metric ansatz becomes clear after solving Einstein's
equations in the near-boundary (large $r$) expansion, where we include
the extra gauge freedom $\xi(t)$ described above:\begin{subequations}
\begin{eqnarray}
A & = & (r+\xi(t))^{2}-2\partial_{t}\xi(t)+\frac{a_{4}}{r^{2}}+\cdots\,,\label{nearbdryexpansions.a}\\[2mm]
B & = & \frac{b_{4}(t)}{r^{4}}+\frac{\partial_{t}b_{4}(t)-4b_{4}(t)\xi(t)}{r^{5}}+\cdots\,,\label{nearbdryexpansions.b}\\[2mm]
S & = & r+\xi(t)-\frac{b_{4}(t)^{2}}{7r^{7}}+\cdots\,.\,\,\,\,\,\,\,\label{nearbdryexpansions.c}
\end{eqnarray}
\label{eq:nearbdry}\end{subequations}We identify $a_{4}$ and $b_{4}(t)$
as the normalisable modes which are related to the components of the
stress tensor through holographic renormalisation (see section \ref{subsec.numerics}
and \cite{deHaro:2000xn}):
\begin{equation}
\langle T_{\mu\nu}\rangle=\frac{N_{\text{c}}^{2}}{2\pi^{2}}\mathrm{diag}\Big[\mathcal{E},\,\mathcal{P}_{\text{L}}(t),\,\mathcal{P}_{\text{T}}(t),\,\mathcal{P}_{\text{T}}(t)\Big]\,,\,\text{with}
\end{equation}
\begin{equation}
\mathcal{E}=\mathcal{P}_{\text{L}}(t)+2\mathcal{P}_{T}(t)=-3a_{4}/4\quad\mathrm{and}\quad\Delta\mathcal{P}(t)=\mathcal{P}_{\text{L}}(t)-\mathcal{P}_{T}(t)=3b_{4}(t)\,.\label{related-1}
\end{equation}
Note that the Einstein equations, as well as energy conservation,
imply that the field theory energy density $\mathcal{E}$ is constant
in our homogeneous setting. As the only possible static state with
finite energy density is the isotropic and homogeneous plasma \cite{Janik:2008tc},
the final state is known already from the start. This seems to be
a rather non-generic feature of our setup, which we discuss in the
last section of this chapter.

In (\ref{eq:nearbdry}) we suppressed the near-boundary expansion
at relatively low order, but it is important to stress that the expansion
has infinitely many terms carrying arbitrarily high derivatives of
the pressure anisotropy. This inevitably leads to a general conclusion
that a state given by the form of the geometry on a constant time
slice is (partly) specified by infinitely many derivatives of the
dual stress tensor, in our case the pressure anisotropy.

\subsection{Solving Einstein's equations\label{sub:Solving-Einstein's-equations}}

As anticipated at the beginning of this chapter and originally noted
in \cite{Chesler:2008hg}, the Einstein equations \ref{eq:Einstein}
are particularly simple:\begin{subequations}
\begin{eqnarray}
0 & = & S''+{\textstyle \frac{1}{2}}B'^{2}\, S\,,\label{Ct}\\
0 & = & S\,(\dot{S})'+2S'\,\dot{S}-2S^{2}\,,\label{eq:Seq}\\
0 & = & S\,(\dot{B})'+{\textstyle \frac{3}{2}}\big(S'\dot{B}+B'\,\dot{S}\big)\,,\label{Beq}\\
0 & = & A''+3B'\dot{B}-12S'\,\dot{S}/S^{2}+4\,,\label{Aeq}\\
0 & = & \ddot{S}+{\textstyle \frac{1}{2}}\big(\dot{B}^{2}\, S-A'\,\dot{S}\big)\,,\label{Cr}
\end{eqnarray}
\label{Eeqns} \end{subequations} where 
\begin{equation}
h'\equiv\partial_{r}h\quad\mathrm{and}\quad\dot{h}\equiv\partial_{t}h+{\textstyle \frac{1}{2}}A\,\partial_{r}h\label{eq.defdirderivs}
\end{equation}
denote respectively derivatives along the ingoing and outgoing radial
null geodesics. We will be interested in solving the initial-value
problem, i.e.~given the geometry on the initial-time slice we want
to obtain the evolution of the dual stress tensor by computing the
bulk spacetime outside the event horizon.

Not all equations among (\ref{Eeqns}) are evolution equations, i.e.~specify
the form of the metric on a neighboring time slice. Equations (\ref{Cr})
and (\ref{Ct}) are constraints in the sense that the remaining components
of the Einstein's equations can be shown to guarantee that they are
obeyed provided that (\ref{Cr}) holds at the boundary and (\ref{Ct})
holds on the initial-time slice \cite{Chesler:2008hg}.

The characteristic formulation leads to a nested algorithm for solving
the initial-value problem in which one uses as evolution equations
(\ref{Ct})-(\ref{Aeq}) and at each time step one only needs to solve
linear ordinary differential equations in $r$. The precise scheme
that we will follow is a slight modification of the one originally
introduced in \cite{Chesler:2008hg}, and consists of the following
steps: 
\begin{enumerate}
\item we start with $B$ as a function of $r$ and the energy density $\mathcal{E}$
(constant in our setup); 
\item the constraint equation (\ref{Ct}) allows us to solve for $S$ as
a function of $r$; 
\item we then solve (\ref{eq:Seq}) for $\dot{S}$, with $\mathcal{E}$
being the integration constant; 
\item having $B$, $S$ and $\dot{S}$, we solve (\ref{Beq}) for $\dot{B}$; 
\item with $B$, $S$, $\dot{B}$ and $\dot{S}$ at hand we can integrate
(\ref{Aeq}) for $A$; 
\item knowing $\dot{B}$ and $A$ and using (\ref{eq.defdirderivs}) we
get $\partial_{t}B$; 
\item we proceed to the next time step using a finite difference scheme
(for details see section \ref{subsec.numerics}). 
\end{enumerate}
In our set-up the constraint (\ref{Cr}) is implemented when solving
the Einstein equations as a near-boundary expansion. Equivalently,
it encodes the conservation of the stress tensor in the dual gauge
theory, which can in some sense be seen as a check of the AdS/CFT
duality. In the homogeneous case this translates into the rather trivial
$\partial_{t}a_{4}=0$, which is indeed implied by constraint (\ref{Cr}).
In the next chapters this constraint is more non-trivial (eqn. \ref{eq:SEconservation-shocks}
and eqn. \ref{eq:SEconservation-radial}), but it is important that
it is still only imposed at the boundary. In order to monitor the
accuracy of the numerical code we check the value of this constraint
in the full bulk when evaluated on the numerical solution (see also
subsection \ref{subsec.numerics}).

The algorithm outlined above needs to be supplemented with the initial
conditions $B(r)$ and $\mathcal{E}$, the choice of which we discuss
in the next subsection.

\subsection{Specifying initial states \label{sec.specinistates}}

Gravity encodes dual initial states in the form of the geometry on
a bulk initial-time slice. The conditions on the initial data arise
from three sources: the constraint (\ref{Ct}), the near-boundary
expansion (\ref{eq:nearbdry}) and bulk regularity. By the latter
we mean that all possible singularities in the initial data must be
hidden inside the event horizon.

One way to obtain a non-equilibrium state while automatically satisfying
the conditions above is to start with vacuum AdS and perturb it by
turning on a non-normalisable mode of the bulk metric or some other
bulk field for a finite period of time \cite{Chesler:2008hg}. The
alternative approach, that we adopt here and which was used also in
\cite{Beuf:2009cx,Heller:2011ju,Heller:2012je}, is to start with
non-equilibrium states defined as solutions of the constraints on
the initial-time slice without invoking the way in which a particular
state was created.

Equation (\ref{Ct}) imposes a constraint between the forms of $B$
and $S$ on the initial-time slice. Since $B$ enters (\ref{Ct})
quadratically, we choose to specify the initial state through $B$
and then use (\ref{Ct}) to solve for $S$. Note that this equation,
together with the asymptotic behaviour linear in $r$ (\ref{nearbdryexpansions.c}),
implies that $S$ must be a convex function and hence that it must
vanish for some $r\geq0$, with the inequality being saturated only
for vacuum AdS and the Schwarzschild-AdS black brane. Alternatively
one can say that since $S\sim r$ asymptotically and since $S''\leq0$
we find that $S\leq r$, implying that $S=0$ for $r\geq0$. As our
coordinate frame is spanned by the ingoing radial null geodesics and
$S$ measures the transverse area of the congruence, $S=0$ implies
reaching a caustic and hence the breakdown of our coordinate frame.

For the \emph{successful} evolution of the initial data specified
by some $B$ we thus need to make sure that the locus where $S$ vanishes
is hidden behind the event horizon on the initial-time slice. As the
event horizon is a teleological object, this cannot be verified a
priori - we need to try to run a simulation and when it is successful
we know that the initial state we started with was legitimate.

The contrary is not necessarily the case, as a caustic, a priori,
is just a breakdown of a coordinate system. However, we verified numerically
that in the neighbourhood of a point where $S$ vanishes we obtain
very large curvatures. This suggests that this point \emph{must} be
hidden inside the event horizon.

We thus see there is an interesting interplay between the choice of
$B$ and the choice of the (initial) energy density $\mathcal{E}$.
Both quantities, a priori, seem to be very much independent when it
comes to specifying the initial state. If, however, the point where
$S$ vanishes corresponds to a genuine curvature singularity, which
is the case for the Schwarzschild-AdS black brane and which our numerical
studies also indicate, then there must be a minimal energy density
$\mathcal{E}$ for which this singularity is still covered by the
event horizon on the initial-time slice. 

When interpreting $B$ as relating to the field theory anisotropy
and time derivatives thereof, then this suggests that for a given
energy density the field theory can only sustain a limited amount
of anisotropy. Note, however, that the anisotropy itself, $3b_{4}(t)$,
is practically unbounded, but that the full function $B(r)$ has to
be small enough such that there is no curvature singularity outside
the event horizon. This discussion suggests that it is possible to
find states \emph{maximally far from equilibrium}, for which the initial
position of the event horizon is close to the point where $S=0$.

In our set-up, we have a freedom of preparing arbitrary initial conditions,
i.e.~we can specify $B$ as a function of $r$ on the initial-time
slice and $\mathcal{E}>0$, as long as $B$ obeys the near-boundary
expansion of the form (\ref{nearbdryexpansions.b}) and there are
no naked singularities. We use this freedom to prepare and follow
the evolution of states in which $B$ has support very close to the
boundary, very close to the horizon or spreads over a large range
of the radial direction. In order to generate a large number of non-equilibrium
initial states we followed the following procedure: 
\begin{enumerate}
\item without loss of generality we choose units such that $a_{4}=-1$,
or equivalently $\mathcal{E}=\frac{3}{4}$; 
\item we generate the initial $B$ as a ratio of two 10th order polynomials
in $1/r$ with random coefficients in the range $(0,1)$; 
\item we subtract from it a cubic expression so that the near-boundary expansion
for $B$ of the form (\ref{nearbdryexpansions.b}) is obeyed; 
\item the whole expression is then normalised so that the maximal value
of the $B$ between the boundary and the position of the final event
horizon is $\frac{1}{2}$; 
\item we then run a binary search algorithm to find the factor that $B$
needs to be multiplied by such that the code is just stable, while
storing successful runs. Typically, we repeat this step about 6 times
per seed function generated in step 2. 
\end{enumerate}
In this way we can generate states which are as far from equilibrium
as our numerical code allows. In the end this means there is some
sensitivity to the number of grid points, since increasing the number
of grid points would improve the stability.

Finally, it is interesting to note that a constraint of exactly the
form (\ref{Ct}) also holds for metric ansätze corresponding to a
dual plasma expanding in one dimension \cite{Chesler:2009cy,Chesler:2010bi}.
This implies that our discussion about the specification of the initial
states, including the maximally far from equilibrium ones, also applies
in these other setups. However, if we relax the assumption of a homogeneity
in the transverse plane, then $S$ is no longer forced to be convex
(chapter \ref{chap:Thermalisation-with-radial}) and there might be
bulk states which do not lead to caustics/apparent singularities in
the way described above.

\subsection{Numerical implementation - pseudo-spectral methods%
\footnote{The examples of this subsection are fully worked out in the \emph{Mathematica}
notebook `spectral\_example.nb'.%
}\label{subsec.numerics}}

In this subsection pseudo-spectral methods \cite{boyd2001chebyshev}
(see also \cite{Wu:2011ab,lrr-2009-1,Chesler:2013lia}) are introduced
using two examples: a linear and a non-linear ordinary differential
equation. The first example will be the basis for almost all computations
performed, while the second example is used to find the apparent horizon
in the geometries of chapters \ref{chap:Colliding-planar-shock} and
\ref{chap:Thermalisation-with-radial}.

The linear differential equation considered is
\begin{equation}
y''(x)+y'(x)-20x\, y(x)=0\label{eq:spectralexample1}
\end{equation}
with boundary conditions $y(-1)=5$ and $y(1)=-1$. In spectral methods
the idea is to expand the function $y(x)$ in therms of $n$ Chebyshev
basis functions:
\begin{equation}
y(x)\approx\sum_{i=0}^{n-1}c_{i}T_{i}(x)\label{eq:spectral}
\end{equation}
where $T_{i}(\cos(x))\equiv\cos(ix)$. The $i$-th Chebyshev polynomial
can be written as an $i$-th order polynomial in $x$, and it is therefore
also said that a spectral approximation provides an `all-order' interpolation
of the function on the grid. This automatically implies that the numerical
error made will scale as $\delta x^{-n}$, with $\delta x$ the largest
grid distance, which is sometimes called `exponential convergence'.

Pseudo-spectral make use of \ref{eq:spectral} only indirectly, by
specifying $y(x)$ by its grid point values $y_{i}\equiv y(x_{i})$,
instead of specifying the $c_{i}$-s. Here one has to use the pseudo-spectral
grid points $x_{i}=\cos\left(\frac{\pi i}{n-1}\right)$, which are
denser near the boundaries, thereby avoiding Runge's phenomenon of
interpolation. For solving \ref{eq:spectralexample1} we need $y'(x)$,
which is a linear operation on $y(x)$:
\begin{equation}
y'(x_{i})=D_{ij}y(x_{j}),
\end{equation}
where $D_{ij}$ is determined through \ref{eq:spectral}, and can
be found in \cite{boyd2001chebyshev} or the \emph{Mathematica }notebook
`spectral\_example.nb' accompanying this subsection. Equation \ref{eq:spectralexample1}
can therefore be written as a matrix equation:
\begin{equation}
(D_{ij}^{2}+D_{ij}-20\mathbb{I}_{ij}x_{j})y_{j}=0.\label{eq:spectralexample1-matrix}
\end{equation}

This matrix equation is linearly dependent, as the problem is underdetermined
without providing boundary conditions%
\footnote{Sometimes the matrix is directly invertible without explicitly specifying
boundary conditions. This happens for instance if a homogeneous solution
diverges on the domain, for instance $y(x)=C/x$. These diverging
solutions cannot be expanded using \ref{eq:spectral} and are therefore
absent in the solution, which means that there is an implicit boundary
condition $C=0$. So in this case one directly finds the correct solution
without providing boundary conditions explicitly. In this thesis all
our equations are written into this form, except where a boundary
condition is necessary on physical grounds.%
}. These can be provided by modifying the first and last row of \ref{eq:spectralexample1-matrix}
with the conditions $y_{0}=5$ and $y_{n-1}=-1$. Due to the exponential
convergence, already with $n$ as low as 30 one can achieve the analytic
solution with 19 digits accuracy%
\footnote{This is shown in 'spectral\_example.nb', where 50 digit precise numbers
are used. When using only standard double precision (15 digits) one
can naturally only achieve around 15 digits of precision. Note also
that eqn. \ref{eq:spectralexample1} was chosen to be analytically
solvable, thereby facilitating a comparison with the analytical solution.%
}. Note, however, that this convergence does rely on $y(x)$ being
sufficiently smooth. Especially when $y(x)$ includes $\log(x)$'s
or fractional powers $x^{1/k}$, the diverging derivatives will reduce
the convergence, which means one either has to increase the number
of grid points or treat the non-analytic terms analytically.

The following equation is chosen as a non-linear problem:
\begin{equation}
y''(x)-y(x)^{2}-1=0\label{eq:spectralexample2}
\end{equation}
with boundary conditions $y(-1)=1$ and $y(1)=2$. In this case we
aim to solve the following discretised problem 
\begin{equation}
D_{ij}^{2}y_{j}-y_{i}^{2}-1=\theta_{i}=0,\label{eq:spectralexample2-matrix}
\end{equation}
which we will try with Newton's method. For this one needs an initial
guess, often inspired by the (physical) problem at hand. In this case
we take the simplest vector satisfying the boundary conditions, $y_{i}^{(0)}=(x_{i}+3)/2$.
In Newton's method one linearises the problem around the trial solution,
$\theta_{i}\approx\theta_{i}|_{y=y^{(0)}}+\frac{\partial\theta_{i}}{\partial y_{j}}(y-y^{(0)})_{j}$,
after which one can again solve $\theta_{i}=0$ as a matrix equation,
and repeat the process to obtain a better approximation:
\begin{equation}
y_{j}^{(n)}=y_{j}^{(n-1)}-\left(\frac{\partial\theta_{i}}{\partial y_{j}}\right)^{-1}\theta_{i}|_{y=y^{(n-1)}},
\end{equation}
where for eqn. \ref{eq:spectralexample2-matrix} $\frac{\partial\theta_{i}}{\partial y_{j}}=D_{ij}^{2}-2y_{j}\mathbb{I}_{ij}$.
Also here one has to provide boundary conditions, in this case $y^{(n)}=y^{(n-1)}$
at the boundaries, since the first trial already satisfies the boundary
conditions.

In this problem one obtains within 5 steps an accuracy of more than
15 digits, with only 30 grid points. This, however, depends somewhat
on a good initial trial. Fortunately in most real-time evolutions
one can just use the solution of the previous time step. Sometimes
it is also convenient to solve an easier problem first and use that
solution as a trial, progressively making the problem more complicated,
as is for instance done in subsection \ref{subsec:The-apparent-horizon}.

The examples presented work for a wide variety of problems. One can
for instance easily change the computational domain $(-1,\,1)$ by
a linear or more complicated coordinate transformation. With the latter
it would be possible to place more grid points in a region where the
function has more structure, thereby improving the accuracy (used
in chapter \ref{chap:Thermalisation-with-radial}). For periodic problems
one would use Fourier series, only slightly modifying $D_{ij}$. 

It is also easily possible to use a finite difference scheme by just
replacing $D_{ij}$. In such schemes a $l$-th order approximation
of the derivative is made by just using $l+1$ neighbouring points.
Finite difference schemes can also be very accurate if $l$ is big
enough and have the advantage that derivatives are only locally determined,
thereby reducing the susceptibility to numerical errors and possibly
improving the stability of an algorithm.

\begin{figure}
\begin{centering}
\includegraphics[width=6cm]{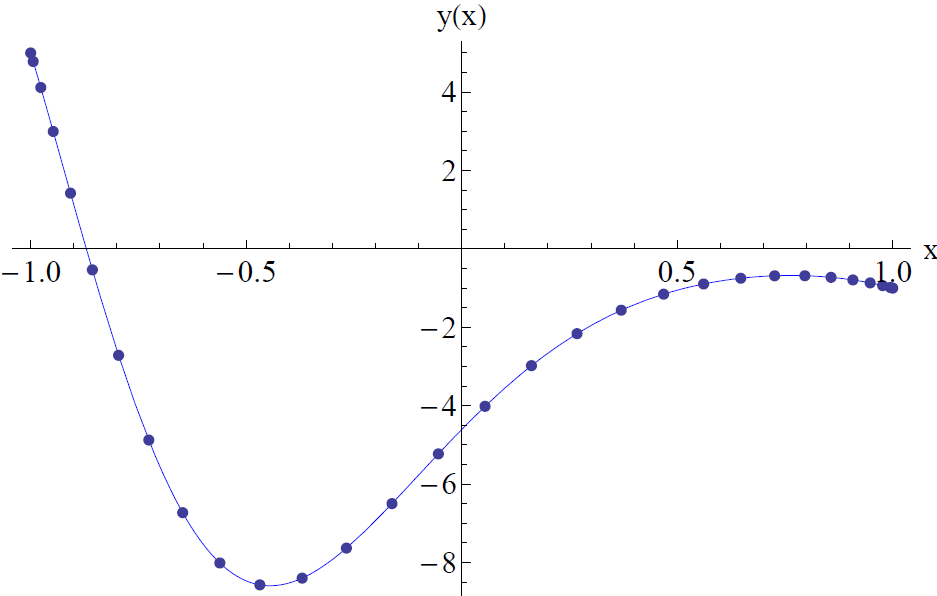}\includegraphics[width=6cm]{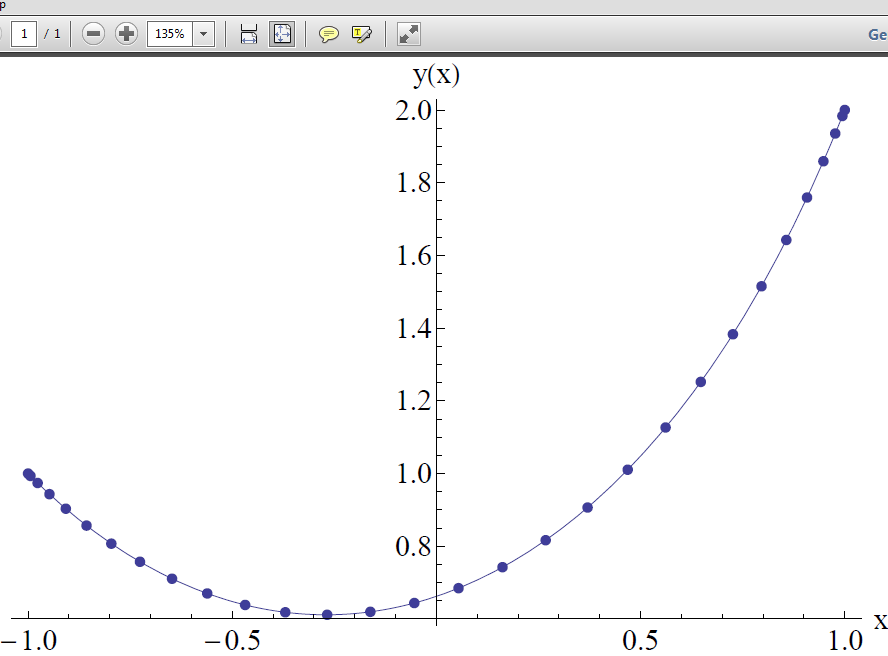}
\par\end{centering}

\caption{The analytic and pseudo-spectral solutions of eqn. \ref{eq:spectralexample1}
(left) and eqn. \ref{eq:spectralexample2} (right), using $n=30$
grid points. Both require only a few fast matrix inversions, and obtain
solutions accurate to 19 and 15 digits respectively.\label{fig:spectralexamples}}
\end{figure}

\subsection{Thermalisation criterion\label{sub:Thermalisation-criterium}}

In this homogeneous setting there is no momentum flow by symmetry,
and hence there are no hydrodynamic modes. Alternatively one can say
that all gradients vanish, and hence the hydrodynamic stress tensor
\ref{eq:hydro-constituive} equals its ideal part. This means that
in this particular case the applicability of hydrodynamics (hydrodynamisation)
is equivalent to an isotropic plasma with $\Delta\mathcal{P}(t)=0$.

Although $\mathcal{E}$ is constant in time, the physical temperature
can only be assigned to the system once the equilibrium is reached.
In this regime $\mathcal{E}=3\pi^{4}T^{4}/4$ and the metric takes
the form 
\begin{equation}
ds^{2}=2dtdr-r^{2}\left(1-\frac{(\pi T)^{4}}{r^{4}}\right)dt^{2}+r^{2}d\vec{x}^{2}\,,\label{eq.AdSSchwarzschildLineElem}
\end{equation}
or in terms of $A$, $S$ and $B$ 
\begin{equation}
A=r^{2}\left(1-\frac{(\pi T)^{4}}{r^{4}}\right),\quad S=r\quad\mathrm{and}\quad B=0,\label{eq.AdSSchwarzschildComponents}
\end{equation}
and describes the Schwarzschild-AdS solution between the boundary
and the future event horizon covering also the black brane interior.

Although equilibration of a holographic system can be sampled with
different field theory probes, including expectation values of local
operators, two point functions, entanglement entropy and Wilson loops,
in this study we will primarily focus on tracing the evolution of
the one-point function of the stress tensor. There are two reasons
for this. In the first place this is the quantity of interest if one
wants to make a phenomenological contact with the fast applicability
of hydrodynamics at RHIC and LHC. Secondly, after the stress tensor
eventually settles down to its thermal value, the geometry becomes
a patch of the Schwarzschild-AdS black brane and from this moment
on there is no need to evolve the Einstein's equations further.

We will hence define thermalisation time as the isotropisation time
$t_{\text{iso}}$, i.e.~the time after which the stress tensor anisotropy
$\Delta\mathcal{P}(t)$ remains small compared to the energy density
and eventually decays to zero. In our calculations, as in \cite{Heller:2012km},
we adopt the following criterion for $t_{\text{iso}}$: 
\begin{equation}
\Delta\mathcal{P}(t>t_{\text{iso}})\leq0.1\,\mathcal{E},\label{eq.tiso}
\end{equation}
but it is important to keep in mind that thermalisation is never a
clean-cut event and the threshold on the RHS of (\ref{eq.tiso}) can
be always slightly raised or lowered without altering much the results.
Note that in later chapters (subsection \ref{sub:Hydrodynamisation})
we will analogously define $t_{\text{hyd}}$ in non-homogeneous settings,
but then $\Delta\mathcal{P}$ is given by the difference between the
real pressure and the pressure computed within hydrodynamics.

\subsection{Numerical implementation - Einstein equations\label{sub:Numerical-implementation--EE}}

In the numerical implementation instead of the variable $r$ we used
its inverse 
\begin{equation}
z=1/r,
\end{equation}
so that the AdS boundary is at $z=0$. We furthermore chose units
$a_{4}=-1$ and $\xi(t)=0$, such that the horizon of the final black
brane will be located at $z=1$, which can be seen from eqn. \ref{eq.AdSSchwarzschildLineElem}
with $a_{4}=\pi T^{4}.$ Note however that, for definiteness, all
dimensionful quantities that we will provide will be specified in
terms of the energy density or, equivalently, the temperature of the
final black brane, which is the only scale at the moment of thermalisation.

The Einstein equations \ref{Eeqns} and the functions $A(z,\, t)$
and $S(z,\, t)$ diverge at the boundary. For numerics one could in
principle solve this problem by multiplying the Einstein equations
with the right power of $z$, and redefining $A(z,\, t)\rightarrow A(z,\, t)z^{2}$
and $S(z,\, t)\rightarrow S(z,\, t)z$. This, however, is not the
most effective way of solving the equations, as all dynamics take
place at the order of the normalisable modes $b_{4}(t)$ and higher
order. The leading order behaviour near the boundary is solely governed
by the non-normalisable modes, which are fixed and therefore known
analytically.

This leads us to propose to rewrite all equations \emph{and} functions
in \ref{Eeqns} such that they are both finite and non-trivial at
the boundary (with the possible exception of $S$). Although this
strategy may make the equations somewhat longer, it has the advantage
of directly computing the quantity we are interested in. In practice
this leads to the following redefinitions:
\begin{equation}
z^{4}\tilde{B}=B,\quad z^{4}\tilde{S}=S-1/z,\quad z^{2}\tilde{A}=A-1/\, z^{2},\quad z^{3}\tilde{\dot{B}}=\dot{B},\quad z^{2}\tilde{\dot{S}}=\dot{S}-1/2z^{2},
\end{equation}
where we then solely compute with $\tilde{B}$, $\tilde{S}$, $\tilde{A}$,
$\tilde{\dot{B}}$ and $\tilde{\dot{S}}$. As an example, eqn. \ref{Ct}
is rewritten into:
\begin{equation}
\left(2\, z^{2}\,\partial_{z}^{2}+20\, z\,\partial_{z}+\left(z^{8}\left(z\,\partial_{z}\tilde{B}+4\tilde{B}\right)^{2}+40\right)\right)\tilde{S}=-z^{3}\left(z\,\partial_{z}\tilde{B}+4\tilde{B}\right)^{2}\,.\label{eq:modified-Einstein}
\end{equation}

This equation reduces to $\tilde{S}=0$ in the limit $z\rightarrow0$,
which in this case means that the boundary condition is already included
at the grid point $z=0$ and does not need to be imposed explicitly.
This contrasts somewhat with the idea that any differential equation
needs explicit boundary conditions, but in this example all homogeneous
solutions of $\tilde{S}$ actually diverge and are therefore put to
zero already when expanding in Chebyshev polynomials (eqn. \ref{eq:spectralexample1-matrix}).

Unlike equation \ref{eq:modified-Einstein}, some equations will contain
explicit $1/z$ terms, prohibiting a direct evaluation at $z=0$.
To resolve this we expand all equations near $z=0$ and treated this
point separately. Using these rewritten Einstein equations it is then
possible to obtain $\partial_{t}\tilde{B}$ and one can proceed to
the next time step, where we use a $4{}^{\text{th}}$ order Adams-Bashforth
stepper:
\begin{equation}
\tilde{B}_{n+4}=\tilde{B}_{n+3}+\Delta t\left(\tfrac{55}{24}\partial_{t}\tilde{B}(t_{n+3})-\tfrac{59}{24}\partial_{t}\tilde{B}(t_{n+2})+\tfrac{37}{24}\partial_{t}\tilde{B}(t_{n+1})-\tfrac{3}{8}\partial_{t}\tilde{B}(t_{n})\right)
\end{equation}

The first five time steps are done using smaller and increasing time
steps, solving for the next $\tilde{B}$ using \emph{Mathematica's}
Interpolation and Integration routines. Typically, we use time steps
of size $\Delta t=0.0025\sim1/n^{2}$.

\begin{figure}
\begin{centering}
\includegraphics[width=6.75cm]{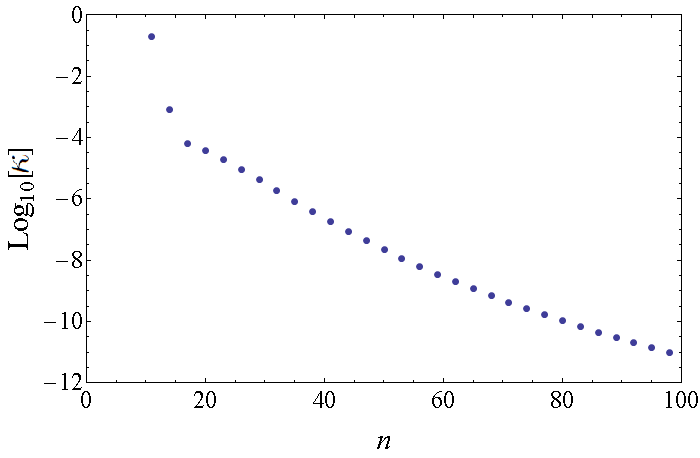}\includegraphics[width=6.75cm]{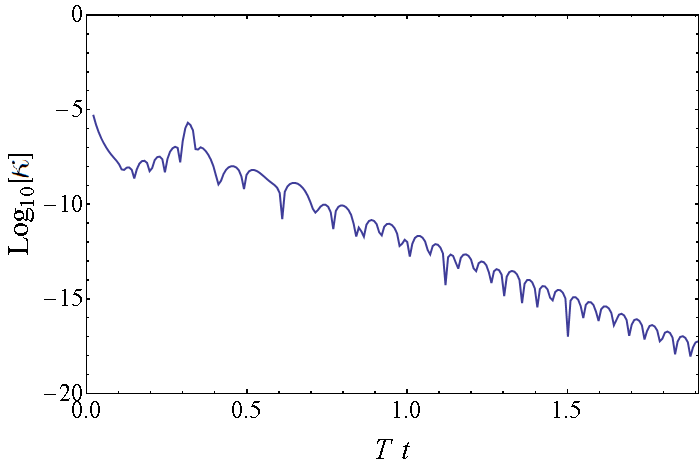} 
\par\end{centering}

\caption{The left plot shows the value of the normalised constraint $\kappa(t=0)$
as a function of the number of grid points $n$ for the evolution
of the initial profile $B(z,\, t_{\text{ini}})=\frac{8}{3}\mathcal{E}z^{4}$.
It is clear from the plot that our numerics converges exponentially
with the number of grid points. The right plot shows the evolution
of $\kappa(t)$ as a function of time for $n=26$ and one can see
there that the constraint actually decreases with time. To achieve
$\kappa(t)<10^{-9}$ one typically needs higher precision than the
standard double precision computations offer. \label{fig:constraint}}
\end{figure}

As a way of monitoring the accuracy of our code, we used the normalised
constraint (\ref{Cr}) 
\begin{equation}
\kappa(t)=\mathrm{max}_{\, r}\left(\frac{\ddot{S}+\frac{1}{2}\dot{B}^{2}S-\frac{1}{2}A'\dot{S}}{|\ddot{S}|+\frac{1}{2}\dot{B}^{2}S+\frac{1}{2}|A'\dot{S}|}\right)\Bigg|_{\mathrm{fixed}\,\, t}\label{normalizedconstraint}
\end{equation}
The convergence of our code is then illustrated by Fig.~\ref{fig:constraint},
which shows typical plots of the maximum value of the normalised constraint
$\kappa(t)$.

The last feature that needs to be discussed is the choice of the position
of the inner boundary of the computational grid. Note that the simulation
is well defined only if the grid covers the entire portion of the
spacetime outside the event horizon. Initially this is hard to predict,
since the position of the event horizon depends on the future evolution.
Therefore one typically focuses attention on the presence of the apparent
horizon because, if it can be found, it is guaranteed to lie inside
the black hole. However, quite frequently in our case there is no
apparent horizon on the initial-time slice and therefore we use the
following procedure. We first try to run simulations with the radial
cut-off put at $z=1.01$, which is right below the late-time position
of the event horizon. This often works, and when it does not we rerun
the simulation with $z=1.07$ as a cut-off. The latter point turns
out to almost always lie past the event horizon. In this way we can
successfully evolve a large number of initial states.

\begin{figure}
\begin{centering}
\includegraphics[width=6.5cm]{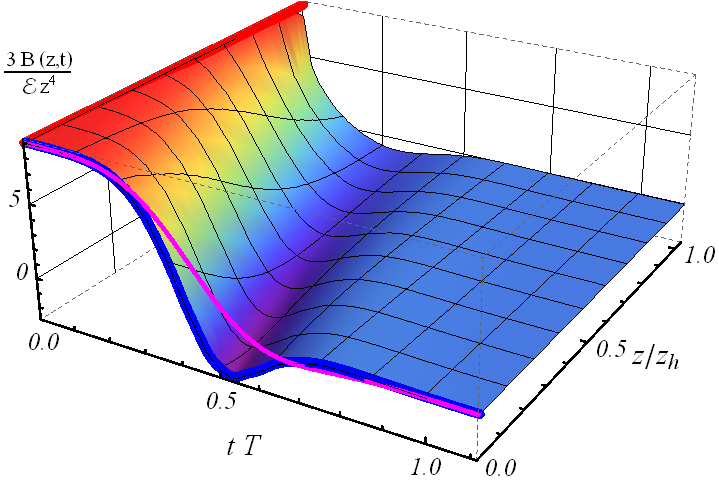} \includegraphics[width=6.5cm]{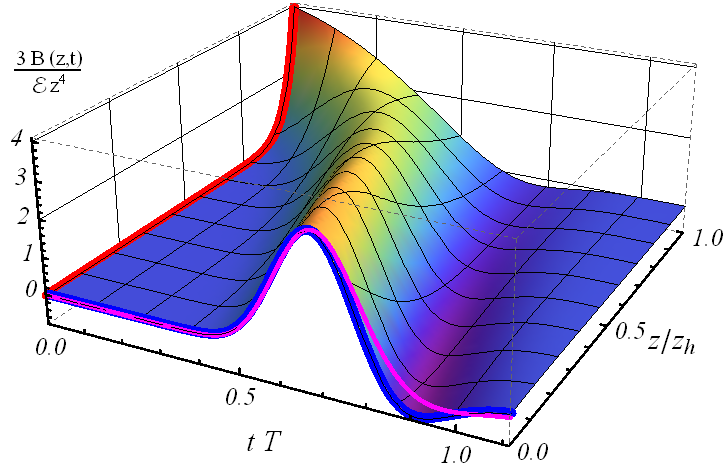}
\par\end{centering}

\caption{The left plot shows $B(z,\, t)$ for the first initial of profile
(\ref{eq:constDP}), which is shown as a thick red line at $t=0$.
The thick blue curve at $z=0$ shows the value of the gauge theory
quantity $\Delta\mathcal{P}(t)/\mathcal{E}$. The purple line shows
the linear approximation, explained in subsection \ref{sec.leadingorder}.
The right plot (second profile in (\ref{eq:constDP})) shows similar
behaviour. The initial disturbance, which is localised in the IR part
of the geometry, propagates to the boundary in a time limited by causality.
This creates the pressure anisotropy, which quickly relaxes back to
zero.\label{Bexampderivs} }
\end{figure}

As a simple example we present two specific evolutions in figure (\ref{Bexampderivs})
with initial $B(z)$ given by 
\begin{equation}
B_{\text{ini}}(z)=3\mathcal{E}\, z^{4}\:\text{and }B_{\text{ini}}(z)=(\frac{4}{3}\mathcal{E})^{6}z^{24}.\label{eq:constDP}
\end{equation}
The first has the anisotropy located at all scales, whereas the latter
is focussed in the infrared. We will come back to the surprising finding
that the linearised approximation (purple lines) performs so well,
even for these almost maximally far-from-equilibrium states (section
(\ref{sec.linearized})).

\subsection{The event and apparent horizons and their entropy}

The event horizon is defined as the causal boundary of the black hole.
It is a teleological object which can be located only after the black
hole settles down to the state of permanent equilibrium. The apparent
horizon is defined as the outermost surface where outgoing light rays
are trapped, i.e. any causal evolution of the surface decreases in
area.

We will be interested in the area of these horizons as examples of
easy-to-compute bulk observables that are directly sensitive to the
form of the geometry in the deep IR. Although no precise definition
of the entropy density exists in a truly far-from-equilibrium situation,
the change in the area density of these horizons provides a crude
measure of the total entropy produced in the thermalisation process.
For this reason we will loosely refer to the area density of these
horizons as `entropy density', but we emphasise from the start that
this terminology is rigorously applicable only near equilibrium. In
equilibrium both horizons are indeed equal, but this is not the case
during the far-from-equilibrium regime; indeed we even found many
evolutions with no apparent horizon in the initial time slice at all.

In our homogeneous setting the event horizon will be a hypersurface
defined by 
\begin{equation}
r-r_{\text{EH}}(t)=0,
\end{equation}
with the normal vector being null 
\begin{equation}
r_{\text{EH}}'(t)-\frac{1}{2}A\left(t,r_{\text{EH}}(t)\right)=0.\label{eq.horloc}
\end{equation}
The latter is the geodesic equation for the outgoing light ray and
needs to be supplemented with the following condition to be imposed
in the asymptotic future 
\begin{equation}
r_{\text{EH}}(t)\rightarrow\pi T\quad\mathrm{as}\quad t\rightarrow\infty\,.
\end{equation}
In practical terms this condition implies that when the metric eventually
approaches the form of the Schwarzschild-AdS black brane (\ref{eq.AdSSchwarzschildLineElem}),
$r_{\text{EH}}$ approaches the position of the event horizon of the
static solution. The apparent horizon in a homogeneous setting can
be found by solving the algebraic equation $\dot{S}(t,r_{\text{AH}}(t))=0$,
but see subsection (\ref{subsec:The-apparent-horizon}) for a more
non-trivial example in a non-homogeneous setting.

The area (3 dimensional) of both horizons gives rise to the following
expression for the entropy density: 
\begin{equation}
s_{\text{EH/AH}}(t)=\frac{1}{2\pi}N_{c}^{2}\, S\left(t,\, r_{\text{EH/AH}}(t)\right)^{3},\label{eq.entropy}
\end{equation}
which for the event horizon is guaranteed to be a non-decreasing function
of $t$. In (\ref{eq.entropy}) we implicitly assume that a horizon
is mapped onto the boundary along ingoing null radial geodesics, i.e.~along
lines of constant $t$ for the metric ansatz (\ref{metricEF-homogeneous}). 

\begin{figure}[h]
\begin{centering}
\includegraphics[width=7.5cm]{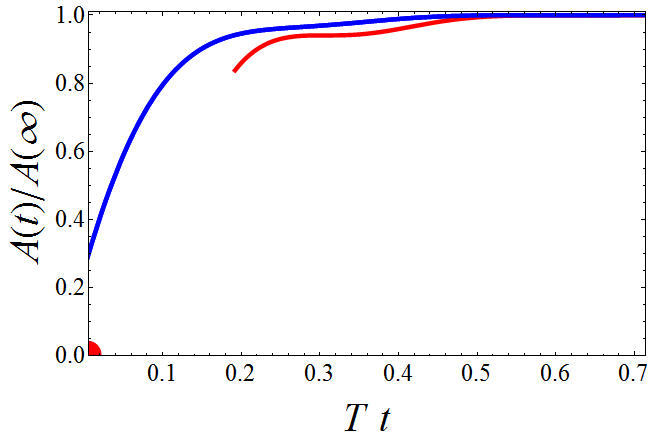}
\par\end{centering}

\caption{Time evolution of the areas of the event (blue) and apparent (red)
horizons for the initial state of figure (\ref{Bexampderivs}). The
red dot at the origin signifies that there is no apparent horizon
for this state at the initial time. From that time until the start
of the red curve there is no apparent horizon within the range of
the radial coordinate covered by our grid, but there could be one
at a deeper position. \label{fig:EHentropy}}
\end{figure}

In figure (\ref{fig:EHentropy}) the horizon areas are plotted for
the example profile of figure \ref{Bexampderivs}. Indeed there is
no initial apparent horizon, although in our AdS setting there will
always be a (small) event horizon. From this figure it is very clear
that this profile starts out far-from-equilibrium, since the black
hole area grows more than a factor of 3.

\subsection{A sample profile and expectations for thermalisation times}

To get intuition about how the dynamics proceeds on the gravity side
and to get acquainted with the features following from the choice
of a foliation by null constant-time slices, it will be instructive
to discuss in detail the dynamics of the following initial state 
\begin{equation}
B(t=0,z)=\frac{2}{15}\mathcal{E}\, z^{4}\exp\left[-\frac{150}{z_{h}^{2}}\left(z-\tfrac{1}{3}z_{h}\right)^{2}\right]\,,\label{sampleprofile}
\end{equation}
where $z_{h}=\frac{2^{1/2}}{3^{1/4}}\mathcal{E}^{1/4}$. As $B$ is
supported at intermediate values of $z$, naive intuition from the
physics of linear wave equations would suggest that the wave packet
splits into two: one propagating inwards and the other propagating
outwards. The one propagating outwards is expected to eventually reach
the boundary, bounce back and fall into the bulk. Both wave packets
will be eventually absorbed by the event horizon (which is guaranteed
to be present given that $\mathcal{E}\neq0$) leading to the increase
in its area.

\begin{figure}
\begin{centering}
\includegraphics[width=7cm]{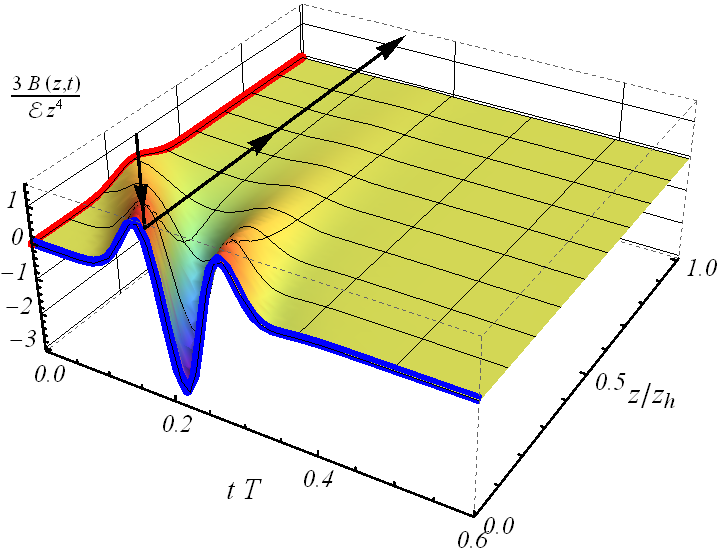}\includegraphics[width=5.5cm]{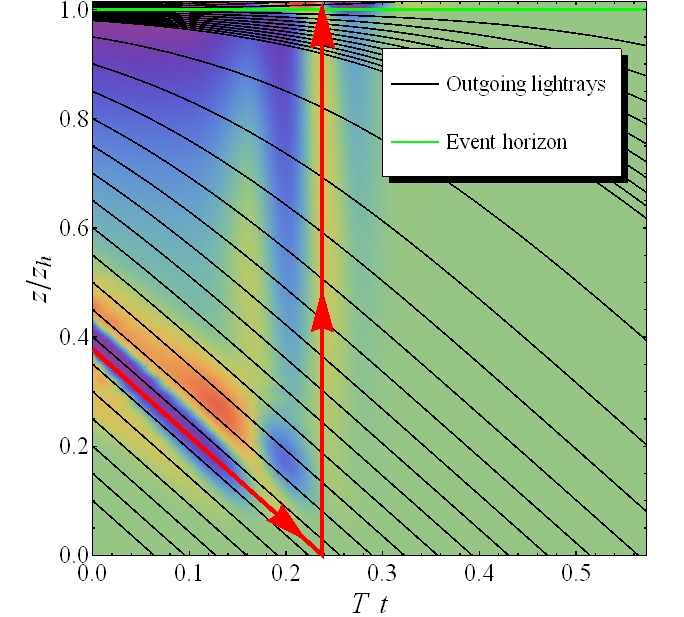} 
\par\end{centering}

\caption{The left figure shows $B$ as a function of time and radial coordinate
for the initial profile (\ref{sampleprofile}), which is shown as
a thick red curve at $t=0$ and which is initially localised near
$z=\tfrac{1}{3}z_{h}$. The blue curve at the boundary ($z=0$) depicts
the pressure anisotropy as a function of time in the gauge theory.
The right figure shows the Kretschmann scalar (with the value for
an equilibrium black brane with the same energy density subtracted)
as a function of time and radial coordinate for the same initial profile.
One clearly sees on this plot the wave bouncing off the boundary and
falling into the black brane. In the adopted generalised ingoing Eddington-Finkelstein
coordinates this happens instantaneously. \label{fig:bounce}}
\end{figure}

These expectations are confirmed by the outcome of the numerical simulation,
as illustrated by Fig.~\ref{fig:bounce}, which depicts the bulk
anisotropy (left plot) and the square of the Riemann tensor, the Kretschmann
scalar (right plot). We can clearly see the rise in the curvature
due to the outgoing wave packet as it approaches the boundary of AdS.
Closer inspection reveals also the presence of a wave packet resulting
from the bouncing off the boundary of the outgoing packet. This wave
packet, due to the null nature of our coordinate frame, propagates
towards the boundary from the horizon along lines of constant Eddington-Finkelstein
time. Note also that this signal falls through the black brane event
horizon without significant scattering. This feature persisted for
other choices of initial states and seems to be related to the high
degree of symmetry of our problem.

It is interesting to note that the initial ingoing part of the wave
packet seems to be mostly taken care of by the solution of the constraints.
Indeed, although $B$ is supported only over some small range of $z$
centred around $z_{h}/3$, the metric functions $A$ and $S$ deviate
from their vacuum values all the way from this point to the horizon,
as required by causality. In contrast, the curvature outside the outgoing
wave packet is very close to the curvature of the static black brane.

These observations suggest that the states which take the longest
time to thermalise are those that are initially localised close to
the horizon on the initial-time slice. An example is provided by $B(t=0,z)\sim z^{24}$,
whose evolution is shown in figure \ref{Bexampderivs} (right). The
reason is that the outgoing wave packet needs to escape the neighbourhood
of the horizon and travel all the way to the boundary to bounce off
and finally fall into the black brane horizon. By localising the initial
profiles close to the horizon, the longest isotropisation times that
we are able to obtain with our numerics, which uses rather moderate
grids, are about $1.1/T-1.2/T$, with $T$ the final equilibrium temperature
(see figure \ref{fig:histogram1}).

\section{A large sample of states and a linearised simplification \label{sec.linearized}}

Apart from toy-models based on the AdS-Vaidya geometry of infalling
dust (see e.g. \cite{AbajoArrastia:2010yt,Balasubramanian:2010ce,Balasubramanian:2011ur,Aparicio:2011zy},
but also \cite{Bhattacharyya:2009uu}), the only existing approximation
scheme to study holographic thermalisation processes is the amplitude
expansion, in which one linearises Einstein's equations on top of
the static black brane background. In this approximation the relaxation
towards equilibrium is described by quasinormal modes with complex
frequencies, whose imaginary parts lead to the damping of their amplitudes
with time and hence to equilibration. These modes were thought so
far to be appropriate for the description of only the late-time approach
to equilibrium, when deviations from equilibrium are sufficiently
small in amplitude \cite{Horowitz:1999jd}.

An indication that this assumption might be too restrictive comes
from black hole mergers in asymptotically flat four-dimensional spacetime.
There, in the so-called close-limit approximation, the Einstein's
equations linearised on top of the final black hole predict rather
accurately the pattern of gravitational radiation at infinity provided
the initial data have a single horizon surrounding the merging black
holes \cite{Price:1994pm,Anninos:1995vf}. This initial horizon, however,
needs not to be a small perturbation of the final black hole for the
close-limit approximation to work.

These features, together with the observation that the AdS analogue
of gravitational radiation at infinity is the expectation value of
the boundary stress tensor, motivates us to apply a linear approximation
to the simple example of far-from-equilibrium gravitational dynamics
in AdS spacetime studied above. With the algorithm to generate many
initial states (subsection (\ref{sec.specinistates})) we can then
compare the full numerical solution of the Einstein equations with
the one linearised on top of the black brane background. Quite surprisingly,
even for states which we found to be maximally far from equilibrium,
the linearised approximation always works within about 20\%. This
finding can therefore greatly simplify future computations.

\subsection{Leading order correction to the pressure anisotropy \label{sec.leadingorder}}

Linearising Einstein's equations in the setup of holographic isotropisation
can be formally phrased as an expansion in the amplitude of perturbations
on top of the Schwarzschild-AdS black brane. We thus write 
\begin{eqnarray}
A(t,z)=\frac{1}{z^{2}}(1-z^{4})+\alpha\,\delta A(t,z) & + & \mathrm{\mathcal{O}}(\alpha^{2}),\nonumber \\
S(t,z)=\frac{1}{z}+\alpha\,\delta S(t,z)+\mathcal{O}(\alpha^{2})\quad\mathrm{and}\quad B(t,z) & = & \alpha\,\delta B(t,z)+\mathcal{O}(\alpha^{2}),
\end{eqnarray}
where $\alpha$ is a formal parameter counting the order in the amplitude
expansion.

The smallness of the initial data can be physically quantified by
either measuring the total entropy production on the event horizon
or by following the amplitude of the pressure anisotropy during the
evolution process and comparing it to the energy density. It is important
to re-stress that we want to use the linearised approximation without
necessarily restricting to the initial data being small perturbations
of the Schwarzschild-AdS black brane, precisely in the spirit of the
original close-limit approximation \cite{Price:1994pm,Anninos:1995vf}
but now in the context of AdS spacetime.

The initial data for the full non-linear Einstein's equations are
given by specifying the energy density $\mathcal{E}$ and the form
of $B$ as a function of the radial coordinate on the initial-time
slice. As anticipated earlier, one of the motivations for choosing
$B$ over $S$ in specifying the initial data was that the former
appears quadratically in the constraint (\ref{Ct}). This feature
persists also with the other components of the Einstein equations
apart from the equation (\ref{Beq}), which immediately leads to 
\begin{equation}
\delta A(t,z)=0\quad\mathrm{and}\quad\delta S(t,z)=0.
\end{equation}
$\delta B(t,z)$ on the other hand remains nontrivial and is a solution
of the equation (\ref{Beq}) with $A$ and $S$ set to their form
in the Schwarzschild-AdS background given in (\ref{eq.AdSSchwarzschildComponents}).

The initial condition for this equation is \emph{the same} as the
initial condition for the full Einstein's equations, i.e. 
\begin{equation}
\delta B(t=0,z)=B(t=0,z).
\end{equation}
The energy density $\mathcal{E}$, which is constant in our setup
and is the remaining part of the initial state specification, is already
included in the background that we linearise on top of.

In full detail, the equation for $\delta B(t,z)$ reads (with the
choice of units $\mathcal{E}=\frac{3}{4}$) 
\begin{equation}
\frac{1}{2z}(3+z^{4})\,\partial_{z}\delta B-\frac{1}{2}(1-z^{4})\,\partial_{z}^{2}\delta B-\frac{3}{2z}\,\partial_{t}\delta B+\partial_{z}\partial_{t}\,\delta B=0\,.\label{eq.B1}
\end{equation}

\subsection{Connection with quasinormal modes}

\begin{figure}[h]
\begin{centering}
\includegraphics[width=6.5cm]{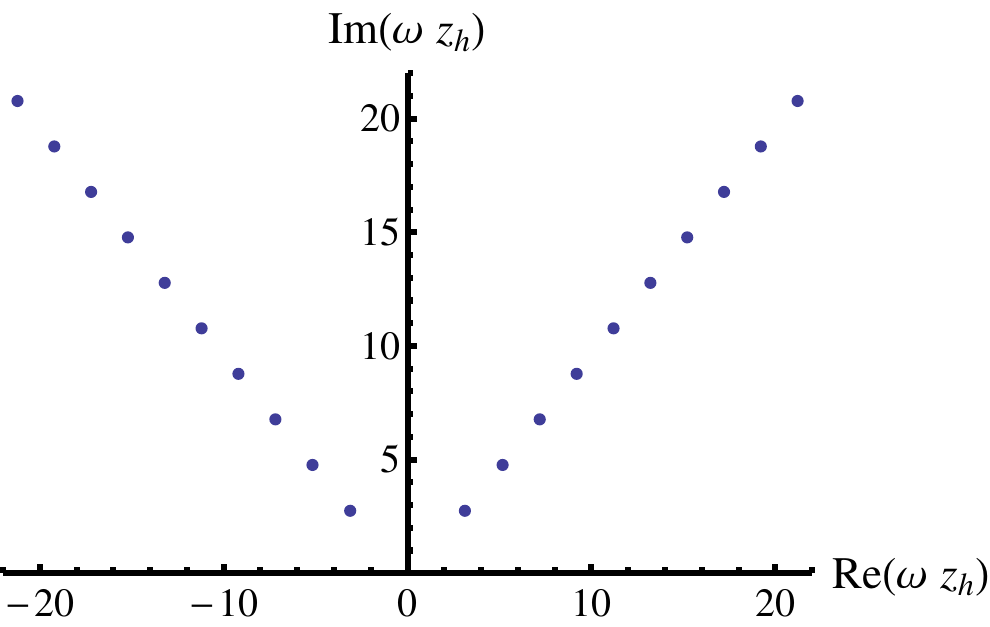}\includegraphics[width=7.5cm]{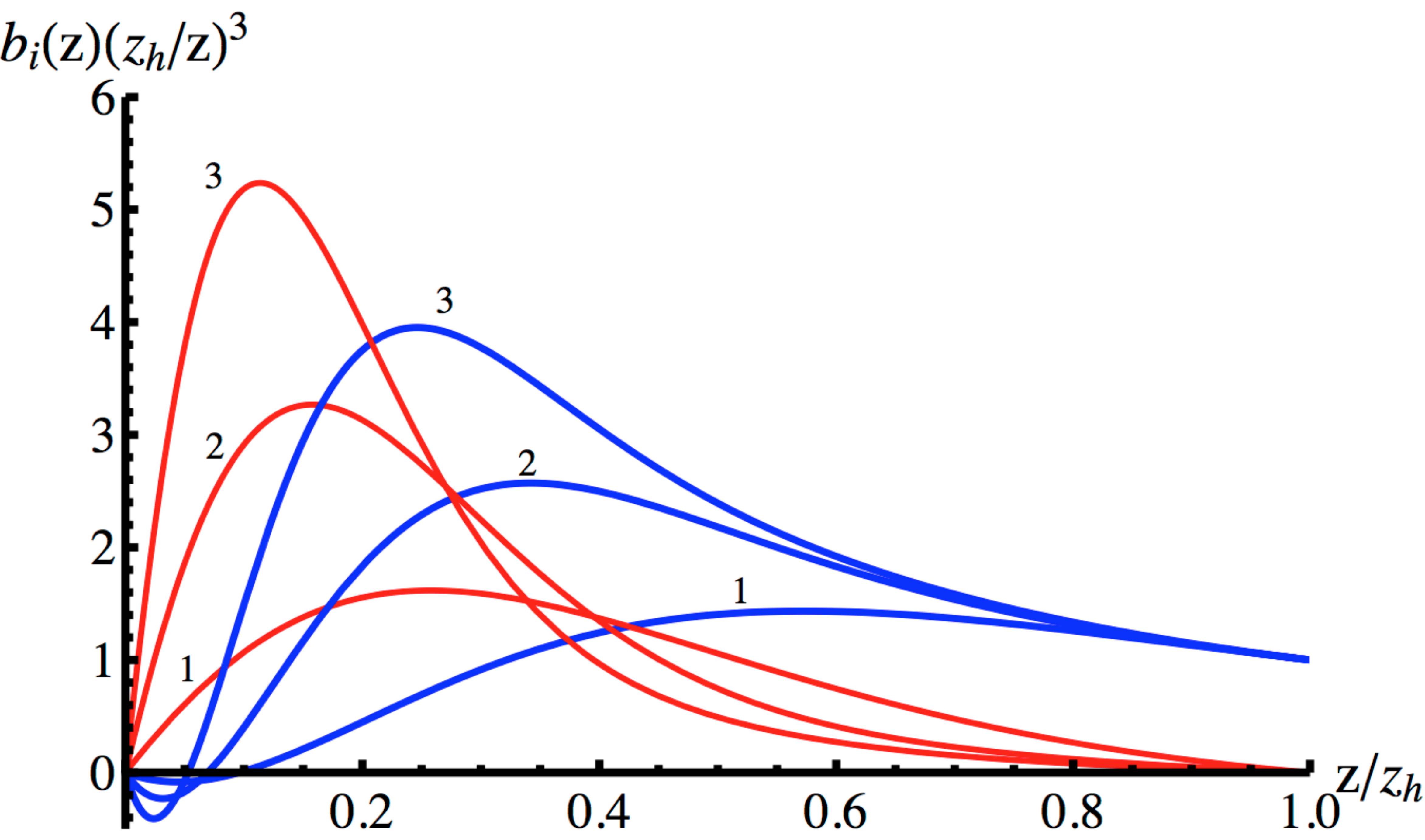} 
\par\end{centering}

\caption{The plot on the left shows the frequencies of the ten lowest quasinormal
modes including their complex conjugates. The mode with the smallest
negative imaginary part will be the dominant mode at late times. Notice
that the spacing between the modes is approximately constant (it differs
by about 0.1\%). The plot on the right displays the lowest three quasinormal
modes as a function of the radial coordinate $z$, where blue and
red denote their real and imaginary parts. The normalisation we use
is such that the real part at the horizon ($z/z_{h}=1$) is equal
to unity, whereas the imaginary part vanishes there. One clearly sees
that higher modes (which decay faster) are more dominant near the
boundary.\label{fig:lowestQNMs} }
\end{figure}

Equation (\ref{eq.B1}) can be solved either as an evolution equation
given some initial profile for $\delta B$, as discussed in the previous
section, or by decomposing $\delta B$ as a superposition of modes
with factorised time dependence: 
\begin{equation}
\delta B(t,z)\sim e^{i\omega_{j}t}\, b_{j}(z).\label{eq.qnmansatz}
\end{equation}
These modes are known as quasinormal modes, and they are characterised
by the requirements that they are normalisable near the boundary ($z=0$)
and that they obey in-going boundary conditions at the event horizon
($z=1$).%
\footnote{In the ingoing Eddington-Finkelstein coordinates the ingoing condition
at the horizon is equivalent to regularity of the solution at the
horizon \cite{Hartnoll:2009sz}.%
} The latter condition makes the frequencies $\omega_{j}$ complex
with imaginary parts responsible for the exponential decay in time.
The quasinormal modes (\ref{eq.qnmansatz}) appear in pairs, as taking
the complex conjugate of the equation (\ref{eq.B1}) for the quasinormal
mode with frequency $\omega_{j}$ leads to the equation for the quasinormal
mode with frequency $-\omega_{j}^{*}$. This feature can be seen in
figure \ref{fig:lowestQNMs}.

In the context of gravitational collapse, the \emph{lowest} quasinormal
modes are known to govern the late-time decay of black hole perturbations
(see e.g. \cite{lrr-1999-2}) and this is also expected in the current
setup. On the other hand, the results from \cite{Heller:2012km},
reviewed in the previous section, suggest that the equation (\ref{eq.B1})
predicts the full time dependence of the large-$z$ behaviour of function
$B$ rather well. Hence it is a natural question to compute the quasinormal
mode content of the perturbations that we considered.

In order to answer this question we followed the prescription of \cite{Horowitz:1999jd}
and computed the lowest 10 quasinormal modes (\ref{eq.qnmansatz})
by solving equation (\ref{eq.B1}) for the ansatz (\ref{eq.qnmansatz})
in the near-horizon expansion and evaluating the resulting expression
at the boundary to find $\omega_{j}$'s leading to normalisable modes.
The (somewhat arbitrary) normalisation of our modes is fixed by demanding
that at the horizon $(z=1)$ 
\begin{equation}
b_{j}(1)=1.\label{eq.qnmnormalization}
\end{equation}
On figure \ref{fig:lowestQNMs} we plot the obtained frequencies $\omega_{j}$
of the lowest 10 quasinormal modes, as well as bulk profiles for the
real and imaginary parts of $b_{1}(z)$, $b_{2}(z)$ and $b_{3}(z)$
normalised according to (\ref{eq.qnmnormalization}).

\begin{figure}[h]
\begin{centering}
\centerline{\includegraphics[width=4.6cm]{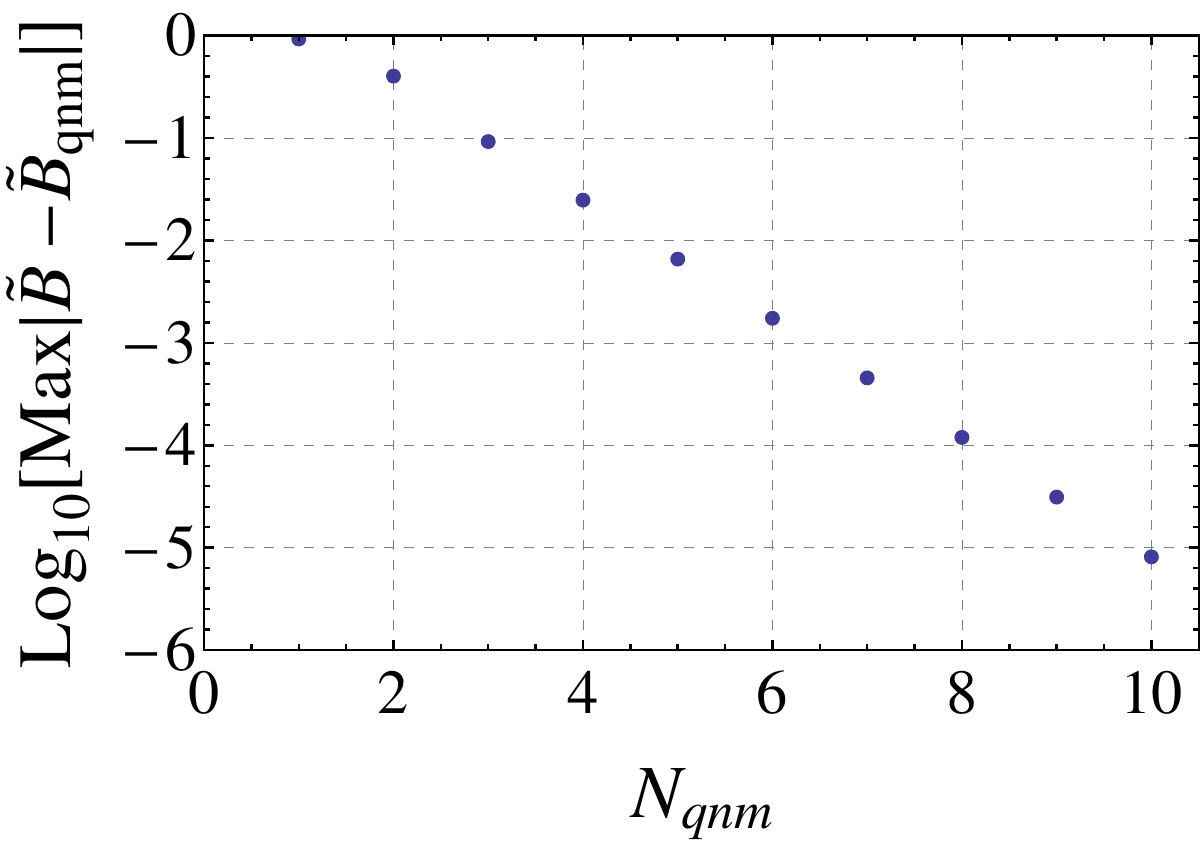}\,\includegraphics[width=5cm]{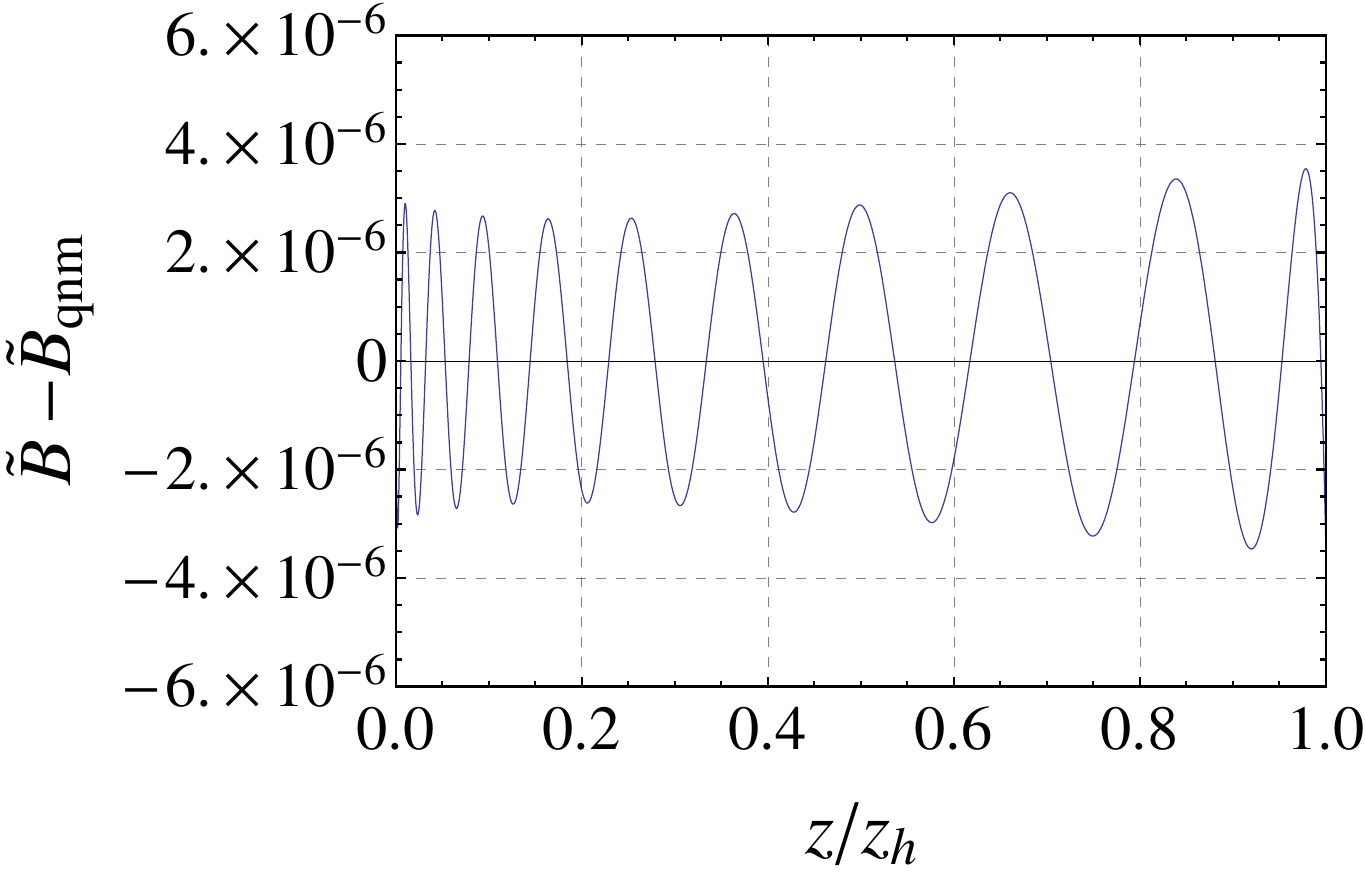}\includegraphics[width=5.1cm]{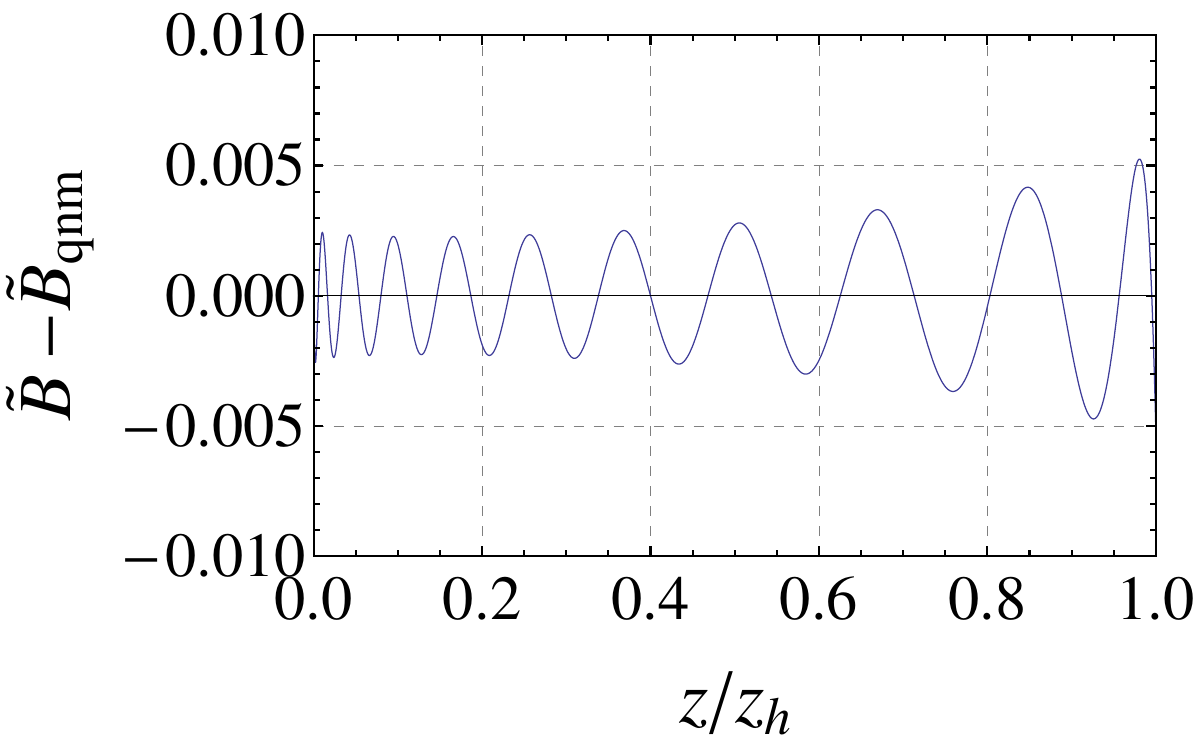}
}
\par\end{centering}

\caption{The plot on the left displays the maximum of the error when approximating
$\tilde{B}(z)=B(z)/z^{3}$ by the first $N_{\text{QNM}}$ (complex)
quasinormal modes, with $B(t=0,z)=-2a_{4}z^{4}$. The plot in the
middle shows the error for the same profile as a function of the bulk
coordinate $z$ while using the 10 lowest quasinormal modes. The right
plot displays the error for $B(t=0,z)=z^{25}$ and shows clearly that
a profile which is dominated in the IR is much harder to fit by the
quasinormal modes. This causes oscillations in the evolution, as can
be seen in figure \ref{fig:examples}. \label{fig:QNMerror}}
\end{figure}

\begin{figure}[h]
\begin{centering}
\includegraphics[width=6.5cm]{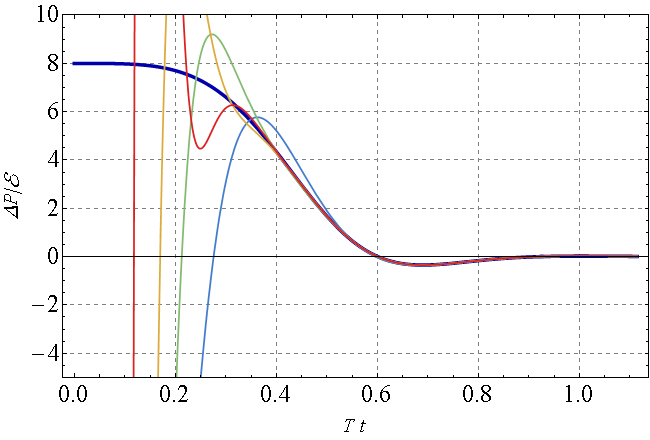}\includegraphics[width=6.5cm]{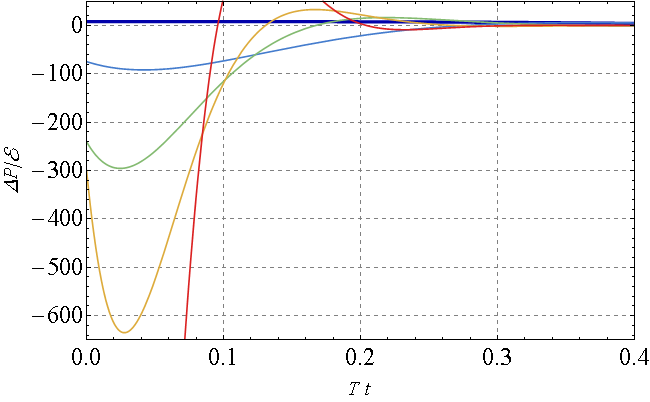} 
\par\end{centering}

\caption{On the left one sees the pressure anisotropy $\Delta\mathcal{P}/\mathcal{E}$
as predicted by the linearised evolution, or indistinguishably by
the sum of the lowest 10 quasinormal modes as a thick blue line. One
can also see there the sum of the first 1 (blue), 2 (green), 3 (orange)
or 4 (red) quasinormal modes. As becomes apparent, the late time dynamics
is well approximated already by keeping only the lowest quasinormal
mode, but if one uses more the fit starts matching earlier. Note that
the coefficients are computed such that the sum of the 10 fits the
initial state best. \protect \\
On the right we plot the individual quasinormal modes with the same
coloring. One clearly sees that each of them carries very large anisotropy,
but that their interference matches the linearised solution. \label{fig:QNMindividual}}
\end{figure}

The idea now is to use the quasinormal modes to decompose solutions
of (\ref{eq.B1}), i.e.~to write a solution of (\ref{eq.B1}) in
the form 
\begin{equation}
\delta B_{\text{QNM}}(t,z)=\text{Re}\left[\underset{i=1}{\overset{N_{\text{QNM}}}{\sum}}c_{i}\, b_{i}(z)\, e^{i\omega_{i}t}\right],\label{eq.qnmexpansion}
\end{equation}
where we truncated the expansion at some $N_{\text{QNM}}$, although
formally we could set $N_{\text{QNM}}=\infty$. In our calculations
we used $N_{\text{QNM}}=10$.

One can view (\ref{eq.qnmexpansion}) as a further simplification
as compared with solving numerically (\ref{eq.B1}), which approximates
the full Einstein's equations well. The reason for this extra simplification
is that now the solution is specified by providing a few complex numbers%
\footnote{One may construct exceptional initial profiles, which are for instance
very close to the boundary, or very rapidly oscillating. Including
more quasinormal modes (taking $N_{\text{QNM}}$ in (\ref{eq.qnmexpansion})
somewhat bigger than $10$) would allow us to treat these cases more
accurately.%
} (say 10 complex coefficients $c_{j}$'s) which due to the linearity
of the problem can be fitted on the initial-time slice to $B(t=0,\, z)$.

As a way of generating coefficients $c_{j}$'s we minimised 
\begin{equation}
\int_{0}^{1}\frac{dz}{z^{3}}\left|B(t=0,z)-\delta B_{\text{QNM}}^{(1)}(t=0,z)\right|\label{eq.leastsquaresfit}
\end{equation}
by using the least squares method on a discrete sample of the radial
position $z$. Naturally, one needs far more points than the number
of quasinormal modes included in (\ref{eq.qnmexpansion}).

The subtlety in using (\ref{eq.leastsquaresfit}) lies in the choice
of the multiplicative factor under the integral, which we set to be
$1/z^{3}$. We checked that both $1/z$ and $1/z^{4}$ do not work
well, as the first one does not take sufficiently into account and
the other overcounts the near-boundary behaviour of $B(t=0,z)$. On
the other hand, $1/z^{2}$ seems to work equally well as $1/z^{3}$,
but for definiteness we focused here on the latter.

Figure \ref{fig:QNMerror} displays the difference between $B(t=0,z)$
and $\delta B_{\text{QNM}}(t=0,z)$ as a function of the number of
quasinormal modes in two representative examples. Clearly, if a good
fit is possible, then the profile (\ref{eq.qnmexpansion}) will solve
the linearised Einstein's equations nicely since each quasinormal
mode solves them individually.

\begin{figure}[h]
\begin{centering}
\includegraphics[width=13cm]{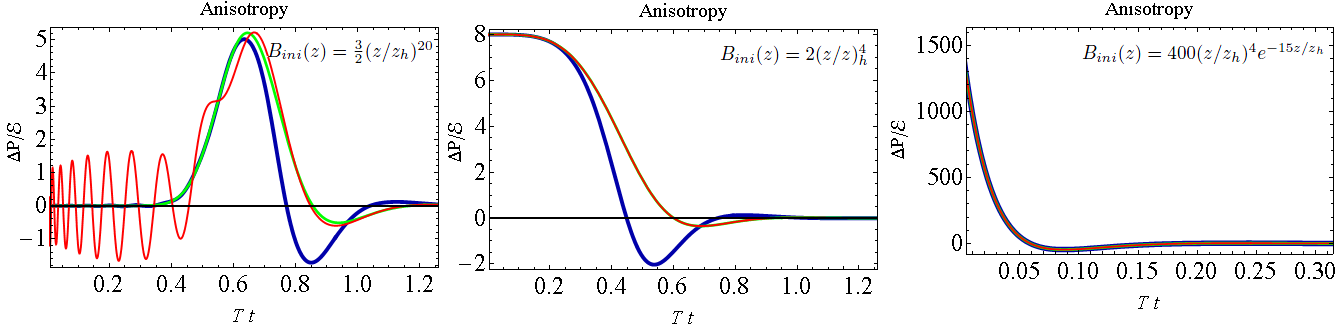}\\
 \includegraphics[width=13cm]{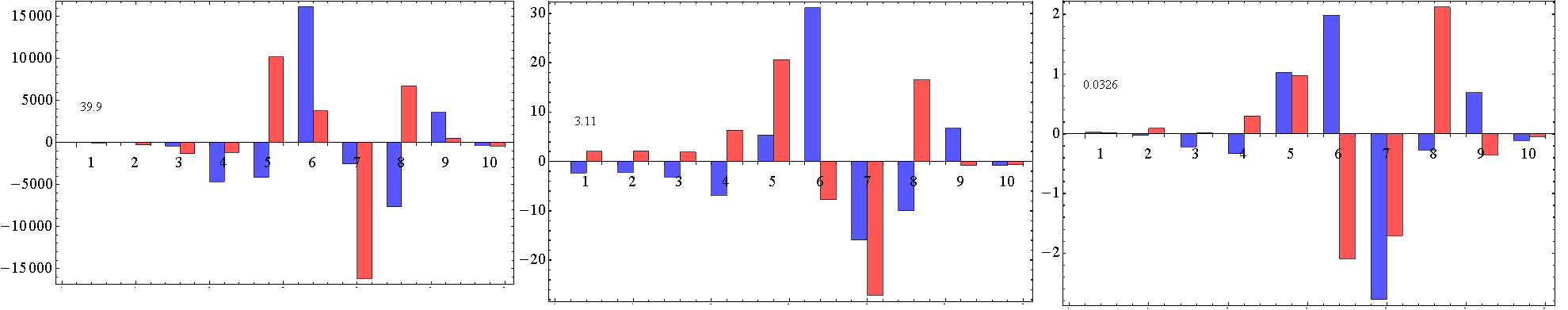} 
\par\end{centering}

\caption{In this figure we illustrate anisotropies of the full (blue), linearised
(green) and quasinormal mode (red) evolution of three representative
initial profiles, located in the IR, spread-out and located in the
UV respectively. Clearly, the initial profile located in the IR takes
some time before exciting the anisotropy at the boundary, which also
explains the late thermalisation. The UV profile can have a very large
anisotropy, but isotropises very fast. \protect \\
These features are nicely described when looking at the quasinormal
modes coefficients $c_{j}$'s (below, real (blue) and imaginary (red)
part). For the IR profile each individual contribution is very large,
but they interfere in such a way to give only moderate anisotropies.
In this way it is also possible to reach isotropisation as late as
6 times the lowest QNM e-folding time. We also see that one would
need to compute more quasinormal modes to accurately fit this profile.
\label{fig:examples}}
\end{figure}

In figure \ref{fig:QNMindividual} we compare the linearised evolution
obtained from a direct solution of (\ref{eq.B1}) and from a solution
based on a decomposition into quasinormal modes. One can see that
the contribution from each individual quasinormal mode can be large,
but that the final sum approximates the linearised evolution very
well. Finally, in figure \ref{fig:examples} we plot three representative
examples, where the profile with $B(t=0,z)$ having support mostly
in the IR displays this interference phenomenon particularly nicely.

The interference described above is important to counter a naive argument
about a bound on the thermalisation time. Naively one may argue that
a state with a maximal thermalisation time should consist fully of
the lowest quasinormal mode, as this mode decays the slowest. According
to the argument in subsection (\ref{sec.specinistates}) the amplitude
of this mode should be bounded to avoid a naked singularity, which
would then imply a bound on the thermalisation time. That this argument
fails is clear from figure \ref{fig:QNMindividual}: each individual
mode would lead to a naked singularity, but the sum is perfectly well
behaved. In fact, this leads us to believe that a profile located
as close as possible to the event horizon could have an unbounded
thermalisation time, though it is probably exponentially hard to obtain
larger and larger thermalisation times. 

\pagebreak{}This in principle unboundedness of the thermalisation
time fits well with causality in the field theory: if one starts with
a state having large correlations over a distance $\ell$, causality
demands thermalisation times bigger than $\ell/2$. The current section
and arguments above suggest that in a strongly coupled theory such
states are very fine tuned and more importantly they will still thermalise
fast, in a time close to the bound by causality.

\subsection{Holographic isotropisation simplified}

\begin{figure}
\noindent \begin{centering}
\centerline{\includegraphics[width=14.5cm]{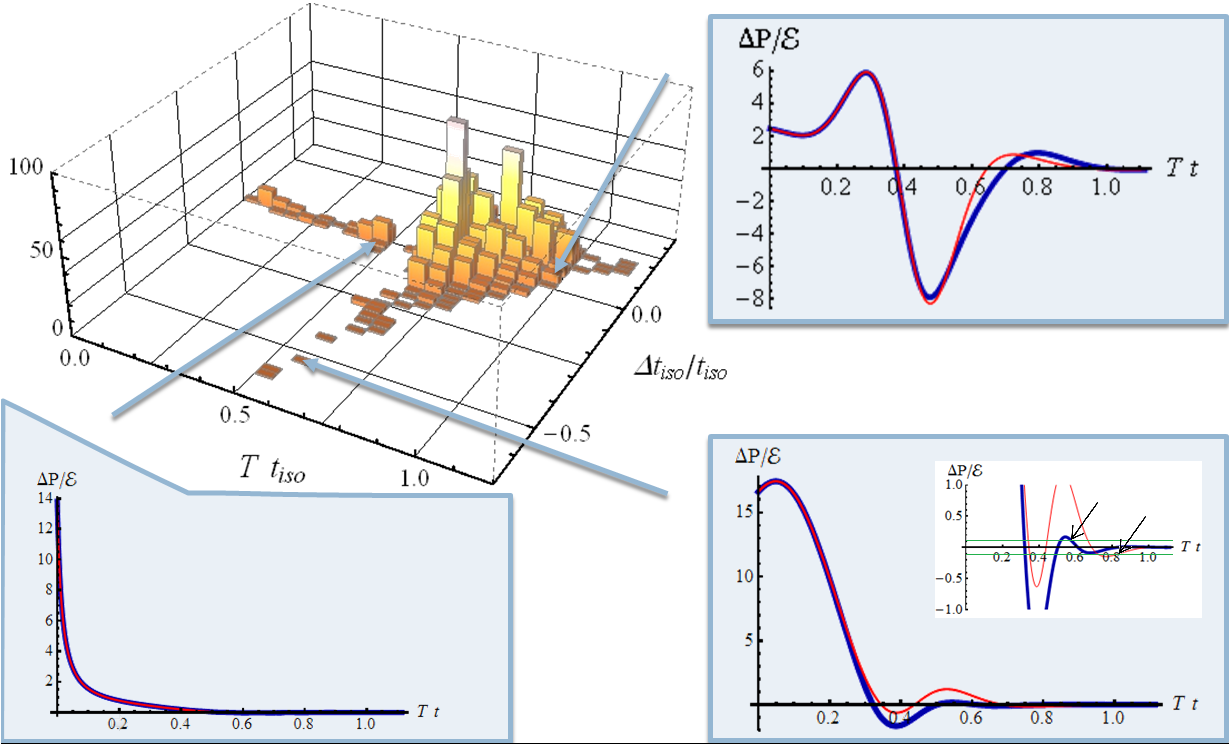}}
\par\end{centering}

\caption{The histogram plots the isotropisation time $t_{\text{iso}}$ versus
the difference between the isotropisation time predicted by the full
and the linear equations, $\Delta t_{\text{iso}}$. The height of
each bar in the histogram indicates the number of initial states for
which the evolution yields values in the corresponding bin. The total
number of initial states is more than 1000. We see both that holographic
isotropisation proceeds quickly, at most over a time scale set by
the inverse temperature, and that the linearised Einstein's equations
correctly reproduce the isotropisation time with a 20\% accuracy in
most cases. A close inspection of one of the few profiles (bottom
right) for which the linearised approximation seemingly fails by more
than 20\% ($\Delta t_{\text{iso}}/t_{\text{iso}}=-0.5$) shows that
it is the imperfect isotropisation criterion which leads to the mismatch
rather than the failure of the linear approximation. Indeed, the plot
shows that, on the scale of the initial anisotropy, the linear result
yields a good approximation. However, the isotropisation criterion
makes no reference to this scale, and results in a 50\% difference
in the isotropisation times, indicated by the arrows on the small
plot. See \cite{Heller:2012je} for a related discussion of subtleties
involved in defining the thermalisation (or more accurately hydrodynamisation)
time in a similar setup. \label{fig:histogram1}}
\end{figure}

The main motivation for studying holographic thermalisation is learning
possible lessons about the way the thermalisation (or rather hydrodynamisation)
process proceeds in relativistic heavy ion collisions at RHIC and
LHC. For this we compare over a 1000 different initial states and
found that the full Einstein equations always lead to an isotropisation
time $t_{\text{iso}}$ less than $1.2/T$, with $T$ the final temperature
of the plasma. Furthermore, we compare all these profiles with their
linearised approximation, and find that the difference in thermalisation
times $\Delta t_{\text{iso}}=t_{\text{iso}}-t_{\text{iso, lin}}$
is almost always less than $20\%$ of $t_{\text{iso}}$. These findings
are summarised in the histogram of figure (\ref{fig:histogram1}).

By replacing QCD by a theory with a gravity dual one only expects
to obtain either qualitative insights or quantitative ball-park estimates
\cite{Mateos:2011bs}. In this sense a $20\%$ accuracy is more than
what is needed in order to understand the phenomena we are interested
in, and at the same time may allow to address otherwise technically
hard-to-tackle questions. Two examples of such problems in the relativistic
heavy ion collisions context are the pre-equilibrium phase of the
elliptic flow and the equilibration of transverse-plane inhomogeneities
following from event-by-event fluctuations. Solving their holographic
analogues in full generality will require complicated simulations
of AdS-black hole spacetimes depending on all five bulk coordinates
and our hope is that a suitably developed linear approximation may
allow us to obtain results with a reasonable accuracy at a much smaller
cost.

The most important open problem is if our linearised simplification
extends to more non-trivial cases where the final state will not be
known in advance. Preliminary results in expanding boost-invariant
plasmas suggest that this is indeed the case, provided one takes care
to chose a proper fiducial state to linearise around. This suggests
interesting opportunities to linearise around more non-trivial backgrounds,
such as presented in the next two chapters.

\chapter{Colliding planar shock waves in AdS\label{chap:Colliding-planar-shock}}

This chapter presents two studies \cite{Casalderrey-Solana:2013aba,Casalderrey-Solana:2013sxa}
of colliding shocks in AdS, which can provide insights in the longitudinal
dynamics in heavy-ion collisions (HIC) in the first far-from-equilibrium
regime. These shocks are constructed such that the field theory stress-tensor
before the collision provides a good model for real heavy-ion collisions.
Although event-by-event fluctuations presumably make transverse dynamics
important, this chapter is limited to shocks homogeneous in the transverse
plane. 

Gravitational shock waves were first studied in \cite{Dray:1984ha},
where they considered boosting a Schwarzschild metric of mass $m$
with a velocity $v$. The shock wave metric can then be obtained by
taking the limit $v\rightarrow1$, keeping the total energy, $\gamma m=(1-v^{2})^{-1/2}m$
fixed. Because of Lorentz contraction, it was found that the metric
was flat everywhere, except on the plane transverse to the direction
of motion.

Our shock waves are somewhat different; they move in 5 dimensional
AdS, are planar in the transverse field theory coordinates, can have
non-trivial structure in the longitudinal direction and do not contain
explicit sources. On the other hand, they can still be thought of
as a boosted source, where the limit is taken of a very energetic
source very far away, such that the profile in this limit is indeed
homogeneous in the transverse plane.

Perhaps more intuitively, one can think of the shock waves from a
field theory perspective. The shock waves are then defined by an energy
density $\mathcal{E}(x_{L},\,\vec{x}_{T})$ at some moment in time
and after which we demand that this energy density moves at the speed
of light. This completely fixes the AdS geometry in the case of pure
gravity without sources.

This chapter is accompanied by the notebook/package \emph{`shockwaves.nb'}.
The notebook contains all details of the computations presented and
for ease of use there are a few sample notebooks to run a simulation
and interpret the results. Note that the package includes the electromagnetic
field and can hence also collide charged shock waves. Details of those
computations are not part of this chapter, which deals with pure gravity
only.

\section{Solving Einstein's equations}

The evolution of Einstein's equations can be conveniently done using
the method described in sections \ref{sec:The-metric-ansatz} and
\ref{sec:Numerics-and-a}, first described in \cite{Chesler:2010bi}.
For this we need to write the metric in the form \ref{eq:metricEF},
which given the planar symmetry in the transverse plane reduces to
\begin{equation}
ds^{2}=dt(2dr-Adt+2Fdz)+S^{2}(e^{B}d\mathbf{x}_{\perp}^{2}+e^{-2B}dz^{2}),\label{eq:metric-shock}
\end{equation}
where all functions now depend on $r,\, t,$ and $z$. Similar to
the homogeneous case initial conditions are given by $B(r,\, t_{0},\, z)$,
$a_{4}(t_{0},\, z)$ and $f_{4}(t_{0},\, z)$, where the latter two
are defined by the near-boundary expansions:
\begin{multline}
\begin{array}{cccccc}
A(r,\, t,\, z) & = & (r+\xi)^{2}-2\partial_{t}\xi+\frac{a_{4}}{r^{2}}+O\left(r^{-3}\right), & B(r,\, t,\, z) & = & \frac{b_{4}}{r^{4}}+O\left(r^{-5}\right),\\
S(r,\, t,\, z) & = & r+\xi+O\left(r^{-5}\right), & F(r,\, t,\, z) & = & \partial_{z}\xi+\frac{f_{4}}{r^{2}}+O\left(r^{-3}\right),
\end{array}\label{eq:near-boundary}
\end{multline}
where also $b_{4}$ and the gauge $\xi$ depend on $t$ and $z$,
and are undetermined by a near-boundary expansion. In the next subsection
we will use the gauge $\xi=0$ to compute the initial conditions,
but during the evolution the gauge freedom will be used to fix the
location of the apparent horizon at $r=1$ (see subsection \ref{subsec:The-apparent-horizon}).

Transforming the near-boundary expansion above to Fefferman-Graham
coordinates (eqn. \ref{eq:metricFG}) we use holographic renormalisation
(section \ref{sub:Holographic-renormalisation-and}) to find the following
energy density, energy flux, longitudinal pressure and transverse
pressure of the boundary field theory:
\begin{equation}
\langle(-T_{t}^{t},\, T_{t}^{z},\, T_{z}^{z},\, T_{\mathbf{x}_{\perp}}^{\mathbf{x}_{\perp}})\rangle\equiv(e,\, s,\, P_{L},\, P_{T})=\frac{1}{4\pi G_{N}}(-\frac{3}{4}a_{4},\, f_{4},\,-\frac{1}{4}a_{4}-2b_{4},\,-\frac{1}{4}a_{4}+b_{4})\label{eq:SEtensor}
\end{equation}
where all functions depend on $t$ and $z$, and for $\mathcal{N}=4$
SU($N_{c}$) SYM we have $G_{N}=\pi/2N_{c}^{2}$.

To find the Einstein equations suitable for the characteristic method
it is essential to understand which function to solve for at each
step. Similar to the homogeneous case (subsection \ref{sub:Solving-Einstein's-equations}),
one starts with $B$, solves for $S$ (2), $F$ (2), $\dot{S}$ (1),
$\dot{B}$ (1) and $A$ (2), where the numbers indicate the order
of the differential equations, which are all ordinary linear differential
equations in $r$. Knowing this, it is straightforward to solve the
Einstein equations for $S''$, $F''$, $\dot{S}'$, $\dot{B}'$, $A''$,
$\ddot{S}$ and $\dot{F}'$, where the latter two are constraints,
only used to find the boundary evolution equations
\begin{equation}
\partial_{t}a_{4}=-\frac{4}{3}\partial_{z}f_{4}\text{ and }\partial_{t}f_{4}=-\frac{1}{4}\partial_{z}a_{4}-2\partial_{z}b_{4},\label{eq:SEconservation-shocks}
\end{equation}
and furthermore as a useful check on numerical accuracy. The full
equations written out can be found either in the Appendix, or the
notebook \emph{shockwaves.nb}.

\subsection{From Fefferman-Graham to Eddington-Finkelstein\label{subsec:FG-to-EF}}

The metric of a single lightlike shock in AdS can be written down
analytically in Fefferman-Graham coordinates \cite{Janik:2005zt}:
\begin{equation}
ds^{2}=\tilde{r}^{2}(-d\tilde{z}_{+}d\tilde{z}_{-}+d\mathbf{x}_{\perp}^{2})+\frac{1}{\tilde{r}^{2}}(d\tilde{r}^{2}+h(\tilde{z}_{\pm})d\tilde{z}_{\pm}^{2})\label{eq:shockFG}
\end{equation}
where $\tilde{z}_{\pm}=\tilde{t}\pm\tilde{z}$ and $h(\tilde{z}_{\pm})$
is arbitrary. Below we restrict to left-moving shocks; right-moving
shocks follow by symmetry. In order to make use of the efficient characteristic
formulation this metric needs to be transformed into the form \ref{eq:metric-shock},
which for a left-moving shock can in general be written as
\begin{eqnarray*}
\tilde{u}(u,\, t,\, z) & = & u+a(u,\, t+z),\\
\tilde{z}_{+}(u,\, t,\, z) & = & t+z+b(u,\, t+z),\\
\tilde{z}_{-}(u,\, t,\, z) & = & t-z+c(u,\, t+z),
\end{eqnarray*}
where we changed $\tilde{r}\rightarrow1/\tilde{u}$. We now have to
demand that the transformed metric satisfies
\begin{equation}
g_{uu}=g_{uz}=0,\qquad g_{ut}=-1/u^{2}.\label{eq:FGtoEF}
\end{equation}
These equations can be solved algebraically order-by-order near the
boundary, leading to the following near-boundary expansion, where
we again fixed the gauge freedom $\xi(t,\, z)=0$:
\begin{equation}
\begin{array}{ccc}
a(u,\, t+z) & = & \frac{1}{6}u^{5}h(t+z)+\frac{1}{10}u^{6}h'(t+z)+\frac{1}{30}u^{7}h''(t+z)+O\left(u^{8}\right),\\
b(u,\, t+z) & = & u+\frac{1}{15}u^{5}h(t+z)+\frac{1}{30}u^{6}h'(t+z)+\frac{1}{105}u^{7}h''(t+z)+O\left(u^{8}\right),\\
c(u,\, t+z) & = & u+\frac{7}{15}u^{5}h(t+z)+\frac{1}{3}u^{6}h'(t+z)+\frac{9}{70}u^{7}h''(t+z)+O\left(u^{8}\right).
\end{array}\label{eq:near-boundaryFGtoEF}
\end{equation}
With these boundary conditions at hand it would be possible to try
and integrate \ref{eq:FGtoEF} into the bulk. Although it is possible
to get rather accurate results using just \emph{Mathematica's }NDSolve,
for the shock waves presented below this method is not good enough.
One of the difficulties in this system is the divergence at the Poincar�
horizon, at $\tilde{r}=0$, which is a non-trivial function of $t+z$,
$u_{hor}(t+z)=-1/a(r,\, t+z)$, in Eddington-Finkelstein coordinates.
Solving for $a,$ $b$, and $c$ on a $u$ and $t+z$ (rectangular)
grid would therefore fail as soon as $a(u,\, t+z)=-1/u$ for some
$t+z$, thereby not giving the complete solution.

Conveniently, eqn. \ref{eq:FGtoEF} can be rewritten such that it
can be solved locally in $t+z$. This is possible by \foreignlanguage{british}{realising}
that paths of varying $u$, keeping $t$ and $z$ constant, have to
satisfy the geodesic equation for light-rays. In the Fefferman-Graham
coordinates the path has the form $x^{\mu}(\lambda)=(\tilde{r}(\lambda,\, t,\, z),\,\tilde{z}_{+}(\lambda,\, t,\, z),\,\tilde{z}_{-}(\lambda,\, t,\, z),\,0,\,0)$,
which leads to second order ordinary differential equations, local
in $t$ and $z$, for the functions $a$, $b$ and $c$. Their explicit
form can be found in the Appendix.

Analogous to subsection \ref{sub:Numerical-implementation--EE} we
redefine $a(\lambda)=\tilde{a}(\lambda)\lambda^{5}$ and similarly
for $b$ and $c$, such that our variables become finite and non-trivial
at the boundary. We solved these modified equations by an $8^{\text{th}}$
order Runge-Kutta stepper, starting at $\lambda=0.03$ with step size
$0.002$, where the boundary expansion \ref{eq:near-boundaryFGtoEF}
(expanded up to order $u^{14}$) provides the boundary conditions.

Having solved the coordinate transformation we can read off $B(t_{0},\, r,\, z)$,
which in this case is the only initial condition depending on the
full AdS geometry:
\begin{eqnarray}
B(t,\,\lambda,\, z) & = & \frac{1}{3}\log(g_{x_{\perp}x_{\perp}}/g_{zz})\\
 & = & -\frac{1}{3}\log\left(\partial_{z}a^{2}+(a+\lambda)^{4}\left(\partial_{z}b+1\right)^{2}h-\left(\partial_{z}b+1\right)\left(\partial_{z}c-1\right)\right).
\end{eqnarray}
This equation depends on space derivatives of $a$, $b$, and $c$,
which are obtained by sampling these functions at 7 points around
the point where $B$ is to be evaluated. As anticipated, this only
works for $\lambda>0.03$, so that for smaller values we used the
near-boundary expansion.

The only other initial conditions needed are $a_{4}(t+z)$ and $f_{4}(t+z)$,
and the initial conditions for right-moving shocks. The former can
again be obtained using the near-boundary expansion, $f_{4}(t+z)=h(t+z)=-\frac{3}{4}a_{4}(t+z)$,
and right-moving shocks are obtained by letting $z\rightarrow-z$
and $f_{4}\rightarrow-f_{4}$.

\subsection{The apparent horizon\label{subsec:The-apparent-horizon}}

Compared to the homogeneous case (section \ref{sec:Numerics-and-a})
the major difference is the non-trivial structure of the (apparent)
horizon as a function of $z$. The apparent horizon is defined as
the outermost surface with a negative outward expansion rate. Although
this definition depends on the time slicing of the spacetime, it has
the large advantage over the event horizon that one can compute the
horizon locally in time. Probably the easiest way to find the apparent
horizon is to compute the outward expansion rate and put it to zero.
Here, however, we present a somewhat more physical derivation.

In this derivation it is assumed that the apparent horizon lies at
constant $r=r_{h}$, which can always be attained using the gauge
freedom $r\rightarrow r+\xi(z)$. The idea is then to shoot outgoing
light rays perpendicular from this surface, so that they maximise
the surface at a time $t+\delta t$. The apparent horizon is then
found by demanding that the surface nevertheless remains constant:
the volume inside cannot expand and is trapped.

In a time $\delta t$ the light rays will in general have traveled
a distance $\delta r(z)$ in $r$ and $\delta z(z)$ in $z$, but
they have to be null:
\begin{equation}
\delta r=\frac{1}{2}\left(\delta t\, A-\frac{\delta z^{2}\, e^{-2B}S{}^{2}}{\delta t}-2\delta z\, F\right).
\end{equation}
The 3-area per transverse 2-area at time $t$ is given by $a(t)\equiv\int dy\sqrt{g_{zz}g_{x_{\perp}x_{\perp}}g_{x_{\perp}x_{\perp}}}=\int_{-\infty}^{z_{f}}S(r,\, t,\, z)^{3}\, dz$
, where we integrate until $z_{f}$ to later find a local formula
by varying $z_{f}$. The difference in area density at time $t+\delta t$
is then given by
\begin{equation}
\delta z(z_{f})S(r,t,z_{f})^{3}+\int_{-\infty}^{z_{f}}3S{}^{2}\left(\delta t\,\partial_{t}S+\frac{1}{2}\partial_{r}S\left(\delta t\, A-\frac{\delta z{}^{2}e^{-2B}S{}^{2}}{\delta t}-2\delta z\, F\right)\right)\, dz,\label{eq:difarea}
\end{equation}
where all functions have been expanded for small $\delta t$ and $\delta z$
and are now evaluated at $r$, $t$ and $z$. Note also that the integration
domain at $t+\delta t$ is changed, which is taken into account by
the first term. It is now possible to maximise this difference by
extremising over $\delta z$, so that we find
\begin{equation}
\delta z=-\frac{\delta t\, e^{2B}F}{S{}^{2}}.
\end{equation}

\noindent Lastly, we put this back in \ref{eq:difarea}, which gives
an apparent horizon if it is zero for all $z_{f}$. The latter is
done by differentiating with respect to $z_{f}$ and letting $z_{f}\rightarrow z$,
giving
\begin{equation}
3S^{2}\dot{S}-\partial_{z}\left(S\, F\, e{}^{2B}\right)+\frac{3}{2}e^{2B}F^{2}S'=0,\label{eq:AH}
\end{equation}
where we furthermore recognised $\dot{S}=\partial_{t}S+\frac{1}{2}AS'$,
and naturally all functions have to be evaluated at $r=r_{h}$. If
the surface is not at constant $r$ the equation remains the same,
replacing $F\rightarrow F+\frac{\partial r_{h}}{\partial z}$, $r_{h}$
being the position of the apparent horizon. For this replacement one
can repeat the whole derivation, but alternatively one can realise
that the horizon is at constant $r_{h}$ for an appropiate gauge $\xi(z)$
(note that eqn. \ref{eq:AH} explicitly does not depend on $\partial_{t}\xi$).
Eqn. \ref{eq:AH} then becomes a second order non-linear equation
for $\xi$ (note that $F$ depends on $\partial_{z}\xi$), after which
it is straightforward to obtain $r_{h}(z)$.

After replacing $F$, equation \ref{eq:AH} becomes a second order
non-linear differential equation in $z$ for $r_{h}(z)$. Typically,
apparent horizons are closed surfaces and hence do not have a boundary,
making \ref{eq:AH} a non-standard boundary value problem \cite{Booth:2005qc}.
In particular, the apparent horizon depends on the curvature on the
entire surface and its time evolution is therefore not necessarily
causal. In the case of shock waves the apparent horizon is planar
and we typically just impose periodic boundary conditions in $z$.
In chapter \ref{chap:Thermalisation-with-radial}, however, there
will be no periodicity, and one has to demand smoothness at the origin
and its location at asymptotic infinity.

Usually an apparent horizon implies the existence of an event horizon
outside the apparent horizon, which can indeed be proven by assuming
the apparent horizon is closed and the spacetime is strongly asymptotically
predictable. In our set-ups these assumptions are not satisfied, but
in the homogeneous setting (section \ref{sec:Numerics-and-a}) we
never found a violation. In the shock wave collision presented in
this chapter the computation of the event horizon is somewhat more
subtle and we did not compute its location, but it was checked that
the region inside the apparent horizon was causally disconnected from
the outside.

In the numerical code it is convenient if the apparent horizon is
at constant $u$, by fixing the gauge $\xi(t,\, z)$. In this way,
one can safely excise the geometry inside the horizon, while still
keeping a rectangular grid and hence keeping the numerics straightforward.
For this, one first needs to find the apparent horizon on the initial
time slice, for which we use Newton's algorithm described in subsection
\ref{subsec.numerics}. While computing this initial apparent horizon,
one necessarily includes part of the geometry behind the horizon,
which can be close to the singularity if the initial $\xi$ is poorly
chosen. To ameliorate this we started the procedure with a large background
energy density, thereby reducing the geometry needed to be covered
numerically, and then used this solution to go to progressively smaller
background energy densities. In this way only a small part behind
the horizon needs to be covered, thereby avoiding the singularity.

During the evolution there are several ways to keep the apparent horizon
at constant $u$, basically by choosing the right $\xi(t,\, z)$.
One approach, used in \cite{vanderSchee:2012qj}, is to evolve the
geometry for a small time, determine the new apparent horizon, and
adapt $\xi(t,\, z)$ accordingly. More efficient, it is also possible
to determine $\partial_{t}\xi(t,\, z)$ by demanding that the time
derivative of eqn. \ref{eq:AH} vanishes:
\begin{equation}
\partial_{t}(3S^{2}\dot{S}-\partial_{z}\left(S\, F\, e{}^{2B}\right)+\frac{3}{2}e^{2B}F^{2}S')=0.\label{eq:AHdt}
\end{equation}
This is relatively straightforward by rewriting time derivatives as
dotted derivatives, and then eliminating $\ddot{S}$ using the Einstein's
equations. After that the only function which depends on $\partial_{t}\xi$
is $A$, through eqn. \ref{eq:near-boundary}, which is present only
linearly in eqn. \ref{eq:AHdt}. During the evolution we work with
$\tilde{A}$, which has $\partial_{t}\xi$ subtracted, and hence does
not depend on $\partial_{t}\xi$. Rewriting eqn. \ref{eq:AHdt} in
this way the dependence on $\partial_{t}\xi$ is explicit and can
be solved as a linear second order differential equation in $z$.
During a time step $\partial_{t}\xi$ is then used in computing $\partial_{t}\tilde{B}$
and to evolve $\xi$ forward in time.

\subsection{Some technical tricks\label{sub:Some-technical-tricks}}

As already mentioned, compared to a homogeneous metric the evolution
of shock waves is considerably harder to \foreignlanguage{british}{stabilise}.
The first reason is the smallness of the horizon far away from the
collision, whereby a horizon could absorb numerical errors. This absorbing
power can be increased by including a regulator energy density. Naturally
it has to be checked that this regulator energy density is not so
big as to affect the physics involved. For this, we ran all simulations
with 2 or 3 different regulators, and then combined those into a single
simulation by a first or second order extrapolation to a simulation
with zero regulator. Doing this it is also straightforward to check
if one has indeed converged.

Although the technique of computing $\partial_{t}\xi$ is very accurate,
it presupposes that the apparent horizon is located at $r=1$. Some
small errors in $\partial_{t}\xi$ will violate this assumption, thereby
possibly causing the apparent horizon to slowly drift away from $r=1$.
To oppose this problem we computed the apparent horizon every 100
steps and performed the transformation $r\rightarrow r+\delta\xi(z)$
in order to fix the horizon again at $r=1$. Another problem can be
a growing asymmetric mode of $\xi(z)$, even though the original collision
is symmetric. For a symmetric shock wave collision we therefore explicitly
\foreignlanguage{british}{symmetrised} $\partial_{t}\xi$. For asymmetric
collisions this is obviously not possible, thereby making these collisions
more challenging to perform.

A well-known problem in the analysis of non-linear differential equations
is aliasing, where short wavelength modes can be artificially excited
\cite{boyd2001chebyshev,Chesler:2013lia}. To ameliorate this problem
one can include some artificial damping in the equations (numerical
viscosity, implemented as regulator energy density), or filter out
these small-wavelength modes. A convenient way to filter these modes
is to interpolate the results on a grid with 2/3 the number of gridpoints
and then interpolate this new function again on the original grid%
\footnote{For the to be presented thin shocks it turns out that it works better
to remove one third of the highest Fourier modes in the longitudinal
direction, which is what is used in those computations.%
}. In addition to filtering all time derivatives we sometimes used
`smoothing', where every point is recomputed by a quadratic least-square
fit using the nearest 5 points in the longitudinal direction. This
smoothing looses precision and is almost never used in computations
presented in this thesis.

Lastly, when naively computing the modified Einstein equations there
will be terms which are large near the boundary, such as $S\sim1/u$.
Although all terms together will not diverge near the boundary, individual
terms may do so. Subtracting two such large numbers can therefore
lead to a large round-off error near the boundary. In chapter \ref{chap:Thermalisation-with-radial}
we resolved this by using high-precision numbers (up to 100 digits),
but here we found it easiest to expand all equations. This leads to
longer equations, but this outweighs the computational ease of using
double precision numbers by far.

Some of the issues above can be improved upon by using the technique
of spectral elements instead of the current pseudospectral implementation
\cite{Chesler:2013lia}. In that case it is even possible to have
a stable code without regulator energy density at all.

\subsection{Hydrodynamisation\label{sub:Hydrodynamisation}}

The shocks and the first far-from-equilibrium stage of the collision
are not well described by hydrodynamics. Shortly afterwards, however,
hydrodynamics does become a good description, which can be seen as
a manifestation of the fluid/gravity duality (see section \ref{sec:Relativistic-hydrodynamics-and}).
It is important to stress that at this moment it is not necessary
that the fluid is (locally) thermalised, which would imply equal pressures
in all directions. Viscous hydrodynamics, on the other hand, can be
very anisotropic, where the size of the anisotropy is governed by
the viscous corrections.

One of the important lessons in thermalisation, perhaps mainly learnt
through holographic studies \cite{Chesler:2010bi,Heller:2011ju,vanderSchee:2012qj},
is that viscous hydrodynamics can be applicable when these viscous
corrections are still large, sometimes as large as the pressure itself.
This came as a surprise, since hydrodynamics can be thought of as
a gradient expansion, so that this expansion was not expected to converge
if the first order viscous correction is big. If one is very precise
one should therefore say that the process of thermalisation is \foreignlanguage{british}{characterised}
by an initial far-from-equilibrium phase and a second phase described
by hydrodynamics with a large anisotropy. The first process towards
hydrodynamics is then called `hydrodynamisation'.

In order to study this hydrodynamisation we compare the full stress-tensor
obtained by the gravitational simulation with the hydrodynamic stress
tensor (eq. \ref{eq:hydro-constituive}), for which we need the local
energy density $e_{\text{loc}}$ and the velocity field $u_{\mu}$,
defined as the timelike eigenvector and corresponding eigenvalue of
the stress tensor: $T_{\mu}^{\nu}u_{\nu}=-e_{loc}u_{\mu}$ (called
the Landau frame). For the stress-tensor \ref{eq:SEtensor} this gives:
\begin{eqnarray}
4\pi G_{N}e_{loc} & = & \frac{1}{4}a_{4}-b_{4}-\frac{1}{2}\sqrt{\left(a_{4}+2b_{4}\right){}^{2}-4f_{4}^{2}},\\
v\equiv u_{z}/u_{t} & = & \frac{a_{4}+2b_{4}+\sqrt{\left(a_{4}+2b_{4}\right){}^{2}-4f_{4}^{2}}}{2f_{4}},
\end{eqnarray}
where we note that $u_{z}$ vanishes in the $f_{4}\rightarrow0$ limit,
since $a_{4}+2b_{4}$ will generally be negative%
\footnote{Both in this chapter and chapter \ref{chap:Thermalisation-with-radial}
there will be far-from-equilibrium regions where a local rest frame
cannot be found, usually indicated by complex eigenvalues of the stress-tensor.
These regions are not described by hydrodynamics.%
}. This leads us to the shear tensor in first-order hydrodynamics,
$\pi_{\mu\nu}=-\eta[e_{loc}]\,\sigma_{\mu\nu}$, with $\sigma$ from
equation \ref{eq:sigma}:
\begin{equation}
\sigma_{tt}/v^{2}=\sigma_{tz}/v=\sigma_{zz}=\frac{4}{3}\gamma^{5}\left(v\frac{\partial v}{\partial t}+\frac{\partial v}{\partial z}\right)\;\text{and}\;\sigma_{x_{\perp}x_{\perp}}=-\frac{2}{3}\gamma^{3}\left(v\frac{\partial v}{\partial t}+\frac{\partial v}{\partial z}\right),
\end{equation}
where $\gamma=(1-v^{2})^{-1/2}$. For strongly coupled $\mathcal{N}=4$
SYM theory the shear viscosity equals $\eta=s/4\pi=(e_{loc}/6){}^{3/4}\sqrt{N_{c}/\pi}$,
where $s$ is the entropy density \cite{Policastro:2001yc}. This
hydrodynamic stress tensor then allows a direct comparison with the
full stress tensor obtained from AdS, by comparing the longitudinal
and transverse pressure $P_{L}$ and $P_{T}$ (eq. \ref{eq:SEtensor})
with the hydro pressures predicted by eq. \ref{eq:hydro-constituive}:
\begin{eqnarray}
P_{L,hydro} & = & \frac{1}{3}\left(\left(4\gamma^{2}-3\right)e_{loc}-4\gamma^{5}\eta\left(v\frac{\partial v}{\partial t}+\frac{\partial v}{\partial z}\right)\right),\nonumber \\
P_{T,hydro} & = & \frac{1}{3}\left(e_{loc}+2\gamma^{3}\eta\left(v\frac{\partial v}{\partial t}+\frac{\partial v}{\partial z}\right)\right).\label{eq:hydroshocks}
\end{eqnarray}
Analogously with eqn. \ref{eq.tiso} we therefore define $t_{hyd}$
by
\begin{equation}
\Delta P(t>t_{hyd})\leq0.05\, e_{loc},
\end{equation}
where we choose $\Delta P=P_{T}-P_{T,hydro}$ since the transverse
pressure is equal in both lab-frame and local rest-frame. Tracelessness
furthermore then implies that the local longitudinal pressures agree
within 10\%, just as in the homogeneous case. We usually limit ourselves
to $z=0$, but generally find that hydrodynamisation occurs to a good
approximation at constant proper time.

In chapter \ref{chap:Thermalisation-with-radial} a somewhat different
way of studying the transition towards hydrodynamics is presented.
There, for several starting times we converted the full stress tensor
to initial conditions for hydrodynamics, which in the second-order
formalism chosen there equals the local energy density $e_{\text{loc}}$,
the velocity field $u_{\mu}$ and the shear tensor $\pi_{\mu\nu}$.
Using hydrodynamical evolution one can then obtain the future stress-tensor,
which can be directly compared with the stress-tensor obtained through
the full gravitational evolution. If those agree for the full future
it is clear that hydrodynamics was applicable at the starting time.

\section{A dynamical cross-over\label{sec:A-dynamical-cross-over}}

\subsection{Shock profiles and physical units\label{sub:Shock-profiles-and}}

The functions $h_{+}(z_{+})$ and $h_{-}(z_{-})$ for left and right
moving shocks in eqn. \ref{eq:shockFG} can freely be chosen, specifying
the energy density of the shock in the longitudinal direction. In
this analysis these functions are restricted to be equal%
\footnote{It is possible to collide asymmetric shock collisions, as was done
in \cite{Casalderrey-Solana:2013sxa}. This is however technically
more challenging and does not provide a significantly better understanding
of the physics involved.%
} and composed of one or two Gaussians: 
\begin{equation}
h(z)=\frac{\mu^{3}}{w\sqrt{8\pi}}\left\{ \exp\left[\frac{\left(z-\frac{1}{2}\ell\right)^{2}}{2w^{2}}\right]+\exp\left[\frac{\left(z+\frac{1}{2}\ell\right)^{2}}{2w^{2}}\right]\right\} ,\label{eq:choiceforh}
\end{equation}
where $\ell$ is the distance between the Gaussians ($\ell=0$ represents
a single Gaussian), $w$ the width, and $\mu^{3}=\int_{-\infty}^{\infty}h(z)dz$
the total energy per transverse area in the field theory, divided
by $N_{c}^{2}/2\pi^{2}$. Following from conformal symmetry in the
field theory, eqn. \ref{eq:shockFG} is invariant under $z_{\pm}\rightarrow\lambda z_{\pm}$,
$\mathbf{x}_{\perp}\rightarrow\lambda\mathbf{x}_{\perp}$, $\tilde{r}\rightarrow\tilde{r}/\lambda$
and $h(z)\rightarrow h(z)/\lambda^{4}$, as is expected from the interpretation
of $h(z)$ as energy density. This, however, means that the physics
we are about to find only depends on the dimensionless products $\mu w$
and $\mu\ell$.

In previous works \cite{Chesler:2010bi,Casalderrey-Solana:2013aba,Casalderrey-Solana:2013sxa}
it was always chosen to show results in terms of dimensionless ratios,
such as $e/\mu^{4}$. In this thesis, the goal is to make contact
with heavy-ion collisions, and we therefore usually adopt physical
units. Since all calculations are in AdS using pure gravity the results
are best thought of as being collisions in $\mathcal{N}=4$ SYM, with
an $SU(N_{c})$ gauge group. To make contact with QCD one could naively
use the same gauge group and set $N_{c}=3$. It is possible to do
slightly better, following Gubser, Pufu and Yarom \cite{Gubser:2008pc},
by equating the equation of state, $e/T^{4}$, of both theories, which
for QCD can be accurately computed in lattice QCD \cite{Cheng:2007jq},
giving $e/T^{4}\approx12$. For $\mathcal{N}=4$ SYM the equation
of state follows from AdS/CFT, and reads $e/T^{4}=3N_{c}^{2}\pi^{2}/8$,
so that we find $N_{c}\approx1.8$. With this method the degrees of
freedom approximately match, although the field content is of course
still different.

Secondly, we have to model the energy density of colliding ions. In
this thesis most plots will be made for LHC collisions, but extensions
to RHIC collisions are straightforward. For the energy density per
transverse area we will integrate the Wood-Saxon distribution:
\begin{equation}
T_{A}(x,\, y)=\epsilon_{0}\int_{-\infty}^{\infty}dz'\left[1+e^{(\sqrt{x^{2}+y^{2}+z'^{2}}-R)/a}\right]^{-1}\,,\label{eq:Wood-saxon}
\end{equation}
where for lead ions $R=6.62$ fm and $a=0.546$ fm \cite{Alver:2008aq},
and $2\epsilon_{0}$ is chosen such that the total energy equals $A\sqrt{s_{NN}}=207*2.76\,\text{TeV}=571$
TeV. Finally, we then match $N_{c}^{2}\mu^{3}/2\pi^{2}$, the total
energy density per transverse area, with the thickness function $T_{A}(x,\, y)$,
giving $\mu=44.6\,\text{fm}^{-1}$ at the centre of a central Pb-Pb
collision, which will be used in all plots in this chapter%
\footnote{Note that energy and inverse length can be converted using $\hbar$
and $c$, implying that $0.197\,\text{GeV\,\ fm}=\hbar c=1$ when
using natural units. This also shows that the characteristic size
of the plasma, around $15$ fm, is of the same order as the energyscale.%
}. For non-central collisions the asymmetry gives a different energy
locally for the left and right moving nucleus, such that one has to
determine $\mu$ in the \foreignlanguage{british}{centre-of-mass}
frame, which depends on the transverse plane; one finds $\mu(x,\, y)^{3}\, N_{c}^{2}/2\pi^{2}=\sqrt{T_{A}(x-b/2,\, y)T_{A}(x+b/2,\, y)}$,
with $b$ the impact parameter.

The longitudinal distribution \ref{eq:choiceforh} differs in two
respects from the width of a real nucleus; firstly we will typically
take one constant $w$ representing the full transverse plane and
secondly we have shocks of finite thickness, even though they move
at exactly the speed of light. On the other hand, our results suggest
that in the right regime (see subsection \ref{sub:From-full-stopping})
the final results are relatively independent of the longitudinal profile.
More importantly, the shocks do not take transverse dynamics into
account, which will be commented on in chapter \ref{chap:Thermalisation-with-radial}.

We stress, however, that these results are obtained in a theory not
like QCD in many respects \cite{Gubser:2009fc} and therefore should
not be taken as a prediction for realistic values of real-world collisions.
On the contrary, they may serve to give intuition and guidance how
close AdS/CFT results come to realistic values and/or compare them
with similar computations at weakly coupled QCD, such as the colour
glass condensate \cite{McLerran:1993ni,Gelis:2012ri}. That said,
little is known about the initial stage of heavy-ion collisions, and
it would therefore be interesting to see whether AdS/CFT results can
provide a realistic initial stage for heavy-ion collisions, such as
attempted in chapter \ref{chap:Thermalisation-with-radial} and more
thoroughly discussed in section \ref{sec:A-comparison-with}.

\subsection{The performance of the numerical code}

As already outlined in subsection \ref{sub:Some-technical-tricks}
it is not straightforward to achieve a stable code. Most importantly
we added a regulator energy density, to ensure there is a large enough
horizon throughout the longitudinal domain. This horizon has a dissipative
and therefore stabilising effect.

An important condition for a stable code in an explicit scheme is
the CFL condition, by Courant, Friedrichs and Lewy \cite{courant1928partiellen}.
They proved that stability for propagating waves requires the analytical
dependence of a point to be contained in the numerical domain of dependence.
In practice this usually translates into timesteps being smaller than
the distance between grid points, if the velocity is one. For pseudospectral
methods the numerical domain of dependence comprises the full domain,
so one may be tempted to conclude that the CFL condition is trivially
satisfied. The CFL condition, however, is not sufficient, and one
can show that a CFL-like condition is also present when using pseudospectral
methods \cite{gottlieb1991cfl}. This led us to use a timestep of
$\delta t=1/7n_{z}^{2}$, as the smallest distance between two gridpoints
equals $\delta x\approx2.6/n_{z}^{2}$, but we stress there is some
degree of trial and error to find the right constant of proportionality.

\begin{figure}
\begin{centering}
\includegraphics[width=1\textwidth]{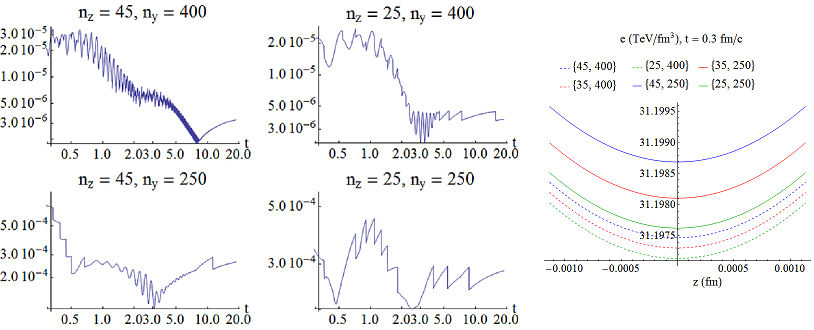}
\par\end{centering}

\centering{}\caption{The maximum value of the constraint violation (last line of equation
\ref{eq:EEshock}) as a function of time for various number of grid
points ($n_{z}$ AdS direction, $n_{y}$ longitudinal direction).
The constraint decreases when increasing $n_{y}$, but the $n_{z}$
dependence is weak (however, for narrower shocks one needs $n_{z}\approx40$
for stable evolution). Note the discontinuities in the constraint
violation, which are at points where the grid is changed to put the
horizon at $r=1$, and/or smoothing is applied (see subsection \ref{sub:Some-technical-tricks}).
This plot is for a shock with $\mu w=0.75$, $\ell=0$ and $e_{reg}/e_{shock}=0.05$,
with $e_{reg}$ the regulator energy density and $e_{shock}$ the
peak of the ingoing shock energy density. These are the standard values
of the package \emph{`shockwaves.nb'}. The right plot compares the
various solutions at the end of the evolution, around mid-rapidity.
There is only a very small difference, mainly due to smoothing effects.\label{fig:stabilitycheck} }
\end{figure}

In order to study the stability and performance of the code we found
it crucial to monitor the constraint violation, analogously to what
was done in figure \ref{fig:constraint}. Here we just use the maximum
of the difference in $\ddot{S}$ computed directly from $\dot{S}$
or by using the constraint Einstein equation for $\ddot{S}$ (equation
\ref{eq:EEshock}). Figure \ref{fig:stabilitycheck} illustrates this
constraint violation for several simulations with different number
of grid points, using the `standard' values of the numerical code.
One can see that in this case the constraint violation is mainly sensitive
to $n_{y}$, and more importantly that all resulting stress-energy
tensors are practically identical.

\subsection{From full stopping to transparency \cite{Casalderrey-Solana:2013aba}\label{sub:From-full-stopping}}

\begin{figure}
\centering{}%
\begin{tabular}{cc}
\includegraphics[width=0.5\textwidth]{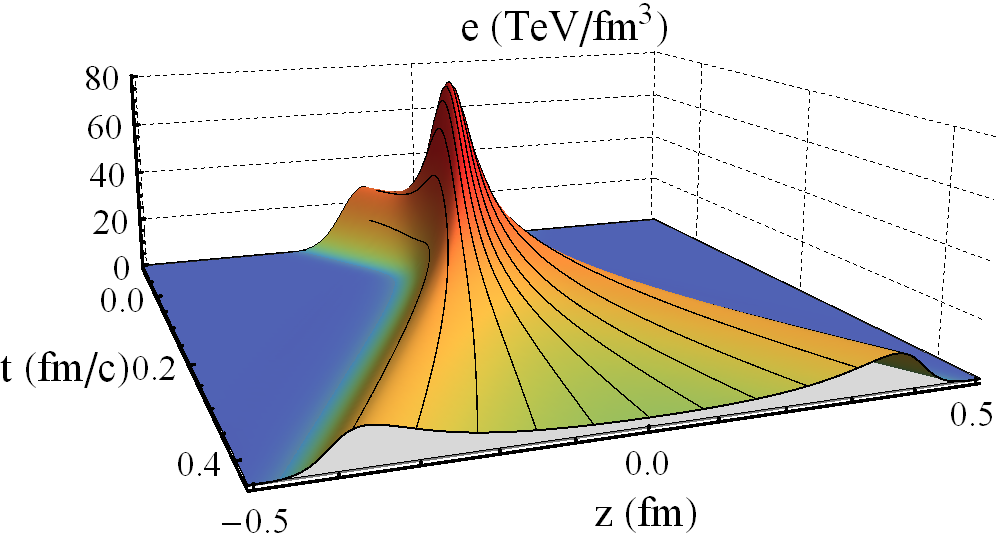}  & \includegraphics[width=0.5\textwidth]{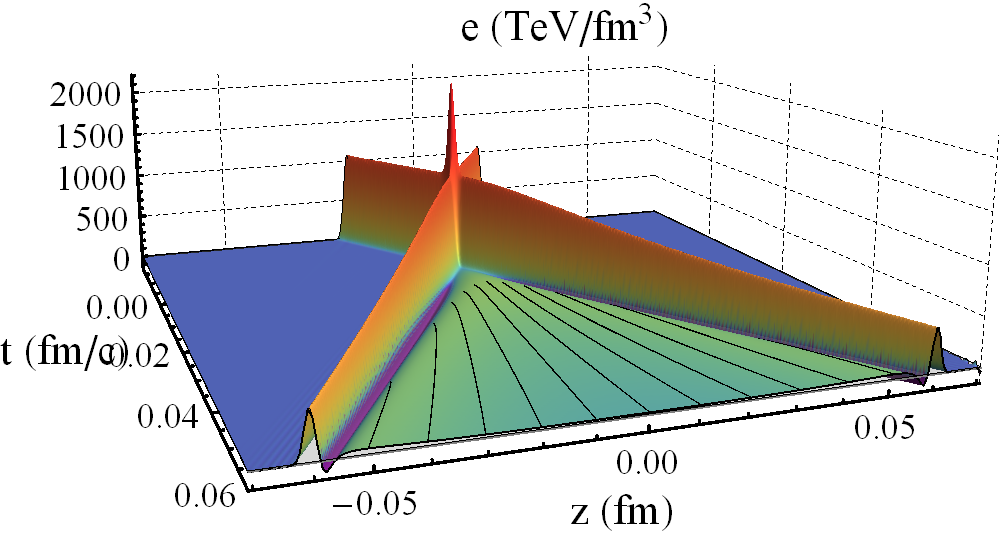}{\large{} }\tabularnewline
\includegraphics[width=0.5\textwidth]{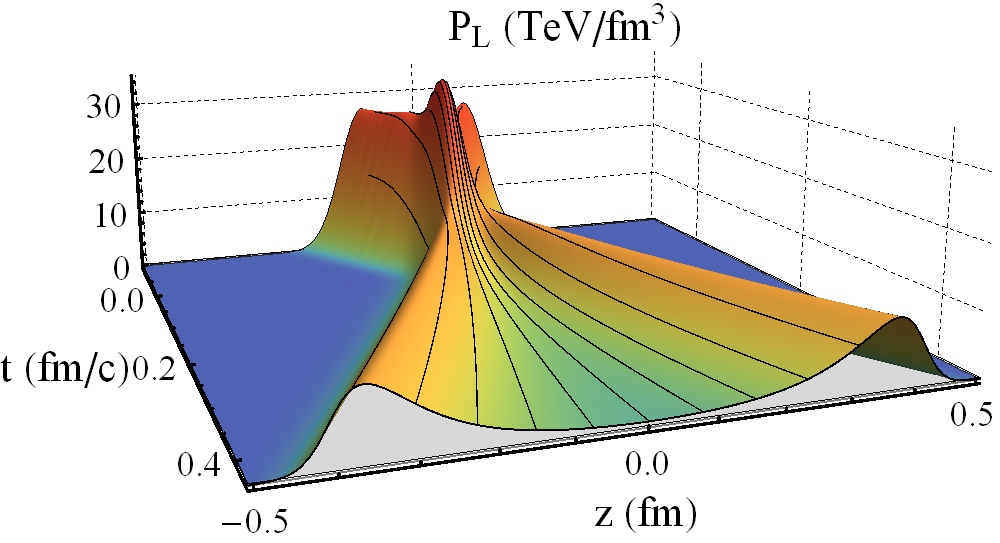} & \includegraphics[width=0.5\textwidth]{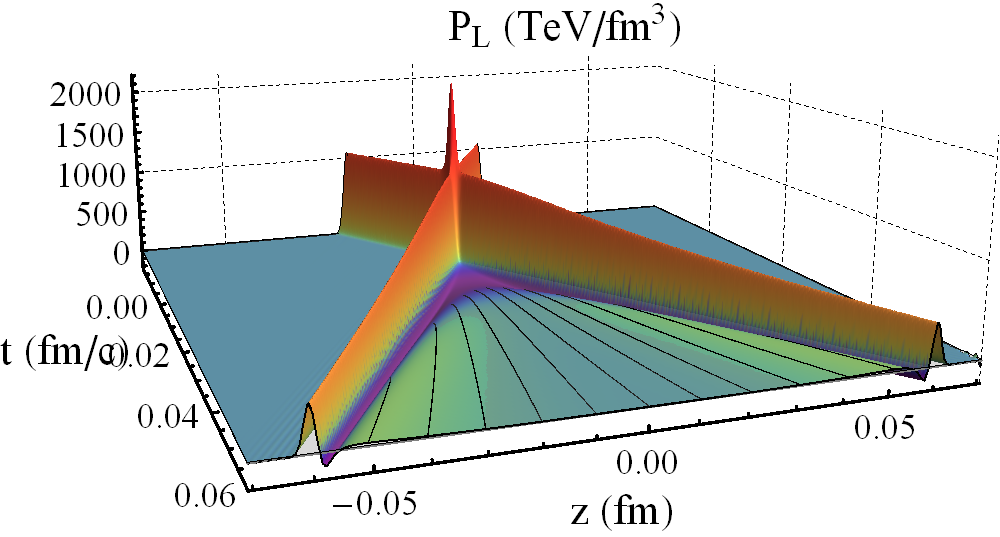} \tabularnewline
\includegraphics[width=0.5\textwidth]{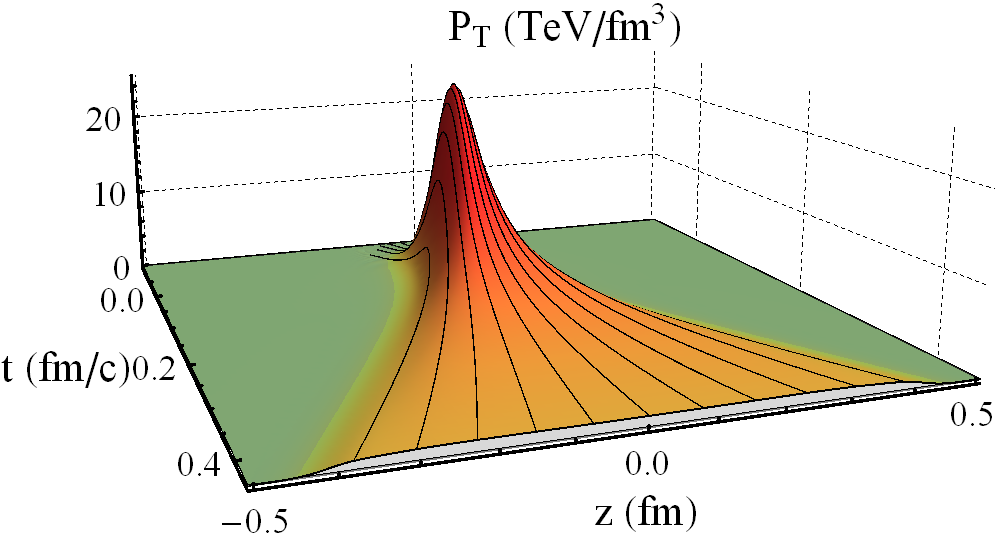}  & \includegraphics[width=0.5\textwidth]{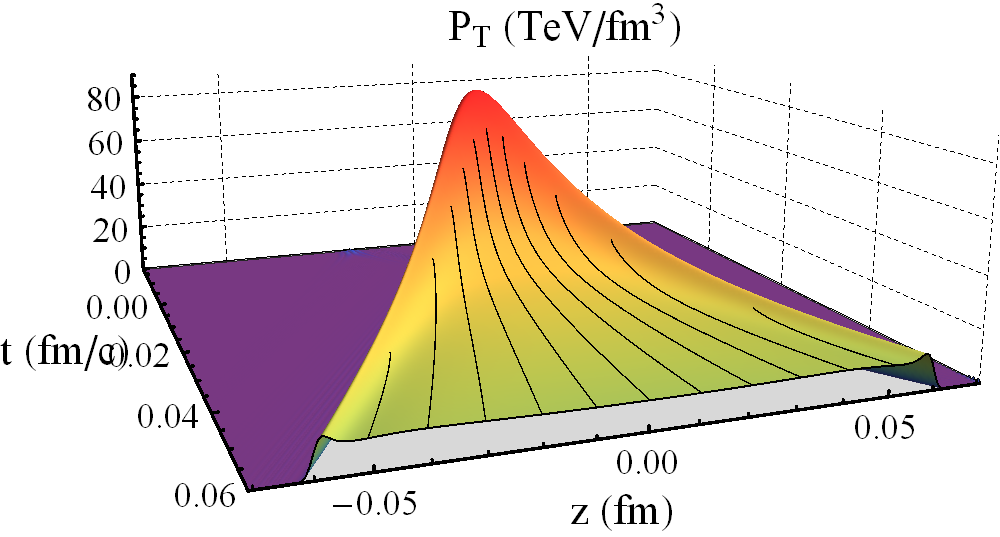} \tabularnewline
\end{tabular}\caption{Energy density and pressures for collisions of thick (left row, $\mu w=1.9$
or $w=0.04$ fm) and thin (bottom row, $\mu w=0.05$ or $w=0.001$
fm) shocks as a function of time and longitudinal coordinate $z$.
The \foreignlanguage{british}{grey} planes lie at the origin of the
vertical axes. As expected, the initial shocks only carry longitudinal
pressure. Note that the transverse pressure for the narrow shock is
surprisingly flat as a function of $z$, we will come back to this
in figure \ref{fig:eloc}. \label{fig:EnergyDensity} }
\end{figure}

An interesting cross-over occurs between what we call wide shocks
(eqn. \ref{eq:choiceforh} with $\ell=0$ and $\mu w\gtrsim0.5$)
and thin shocks ($\mu w\lesssim0.25$). These are represented in figure
\ref{fig:EnergyDensity} by the energy densities and pressures of
simulations with $\mu w=1.9$ (thick) and $\mu w=0.05$ (thin). It
is clear that the thin shocks do not have time to thermalise during
the collision, and hence they pass right through each other, forming
a plasma only later on. On the other hand, for a bigger width the
shocks can sufficiently thermalise already during the collision.

\begin{figure}
\begin{centering}
\includegraphics[width=0.4\textwidth]{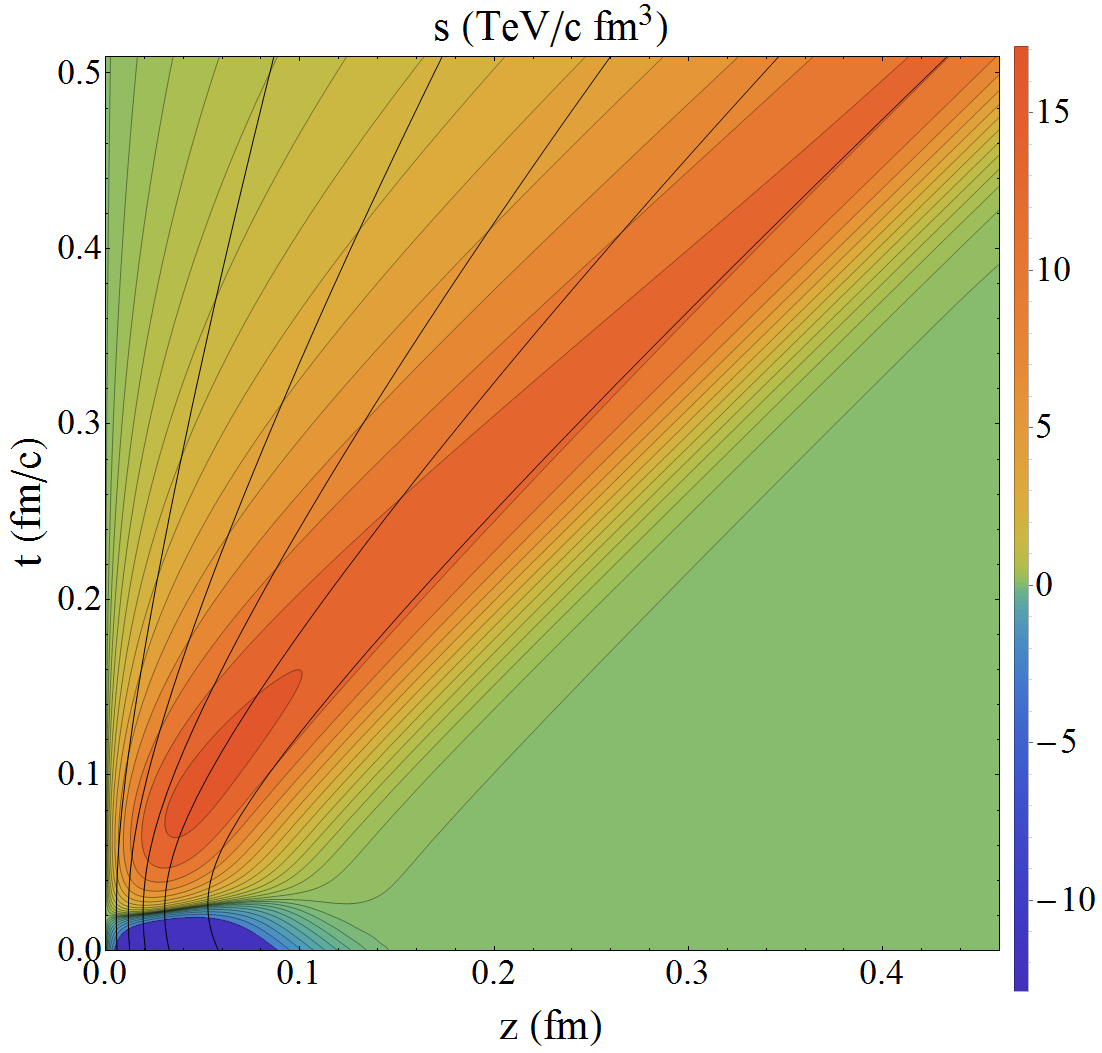} $\quad$\includegraphics[width=0.4\textwidth]{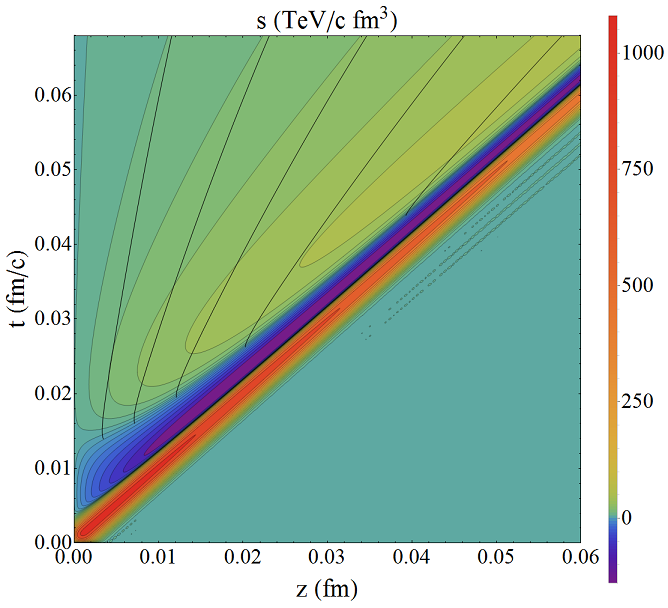}
\par\end{centering}

\centering{}\caption{Energy flux for collisions of thick (left) and thin (right) shocks.
The black lines are streamlines of the produced plasma. \label{fig:flux}}
\end{figure}

\begin{figure}
\begin{centering}
\includegraphics[width=0.45\textwidth]{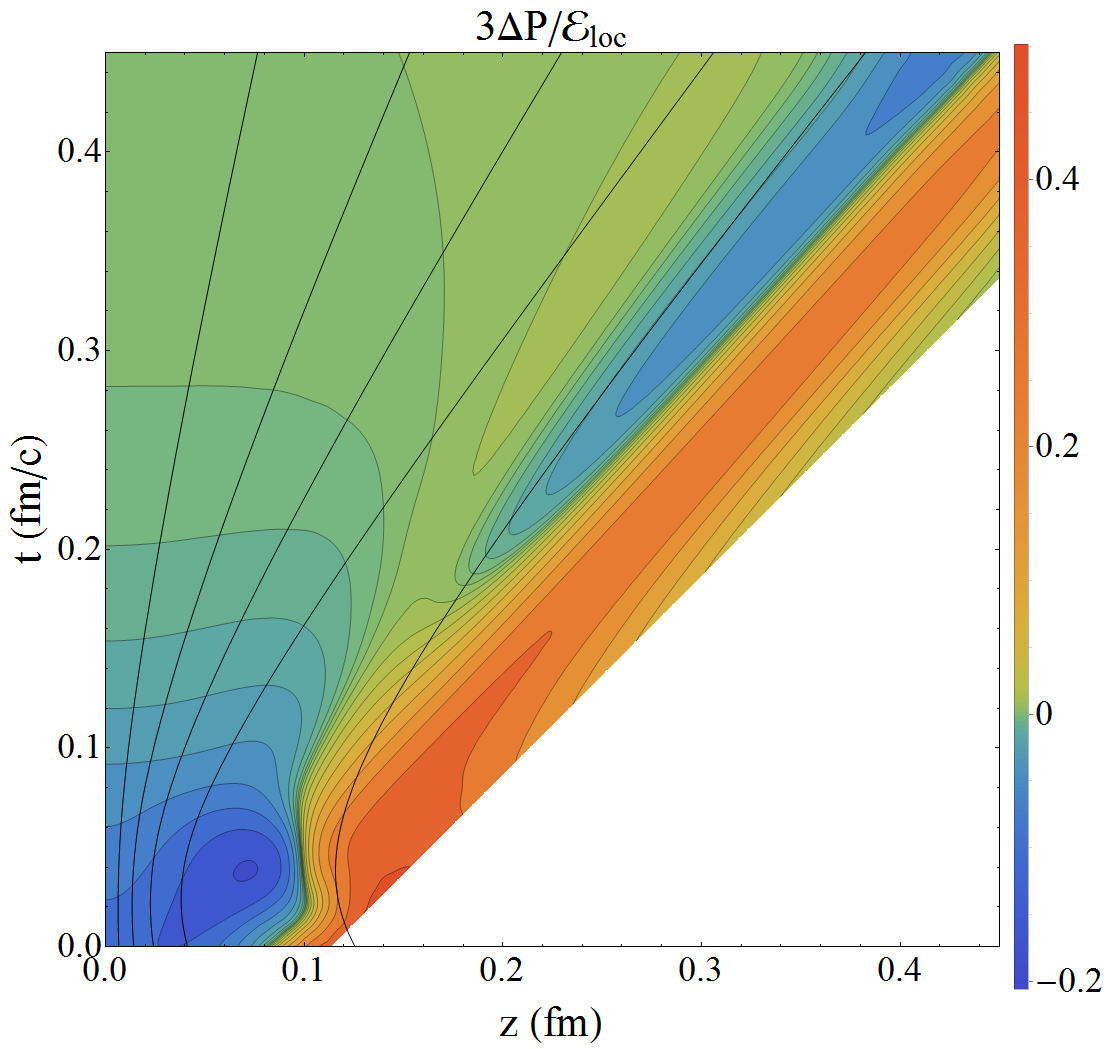}\includegraphics[width=0.45\textwidth]{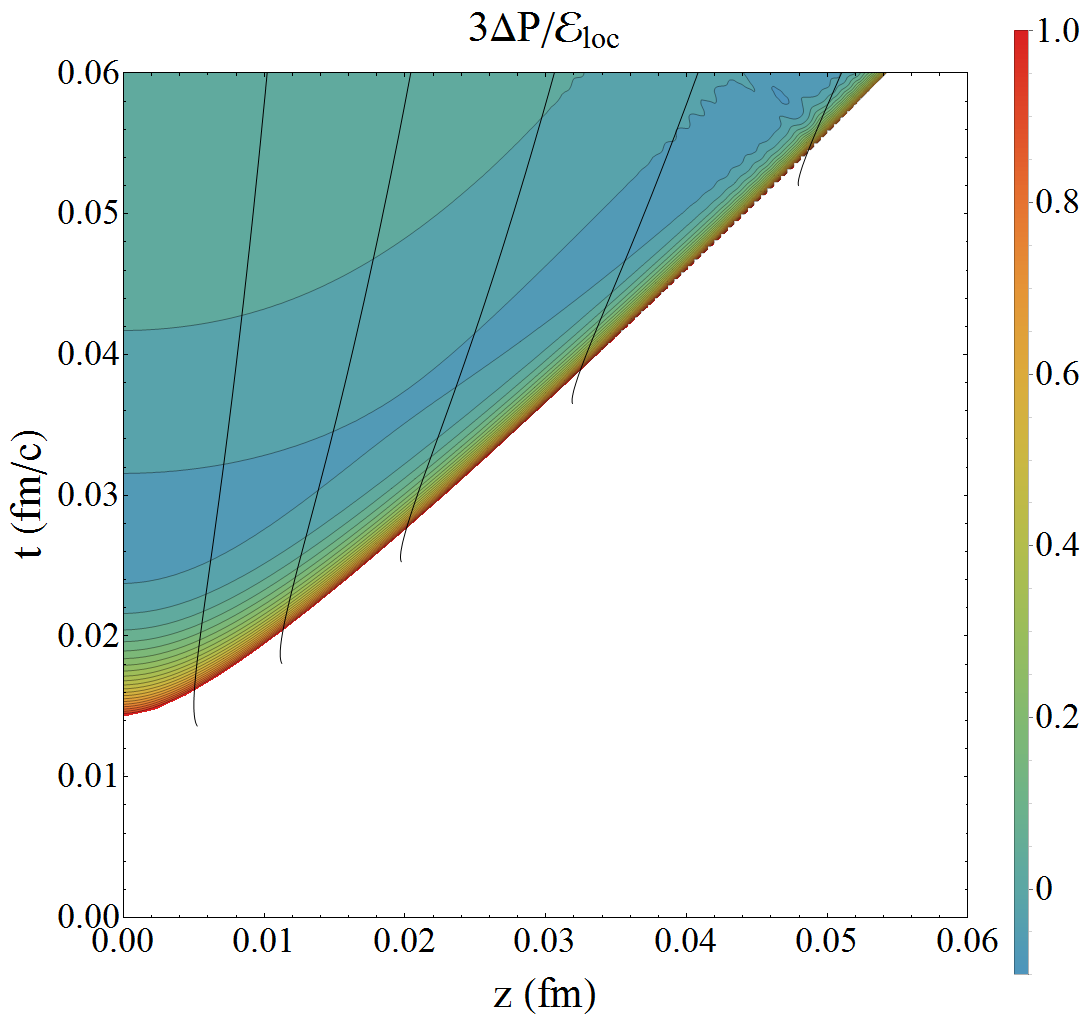}
\par\end{centering}

\centering{}\caption{$3\Delta{\cal P}_{L}^{\textrm{loc}}/\mathcal{E}_{\text{loc}}$ for
thick (left) and thin (right) shocks. The white areas indicate regions
outside the lightcone or where hydrodynamics deviates by more than
100\%. The black lines are again stream lines, whereby it can be seen
that for narrow shocks a local restframe is only defined right before
the region where hydrodynamics becomes applicable. As opposed to all
other plots, these plots are evaluated using a single evolution without
correcting for the regulator energy density, as this displays the
transition to hydrodynamics most clearly.\label{fig:hydro-shocks}}
\end{figure}

\begin{figure}
\centering{}%
\begin{tabular}{cc}
\includegraphics[width=0.42\textwidth]{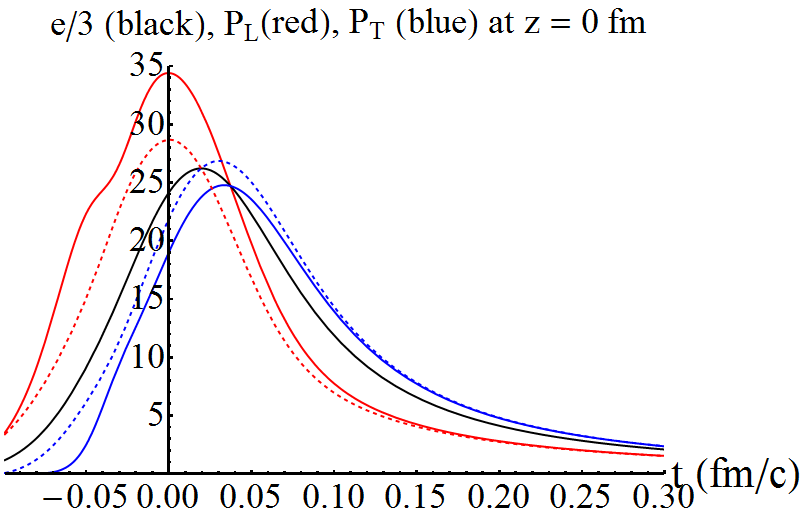}  & \includegraphics[width=0.42\textwidth]{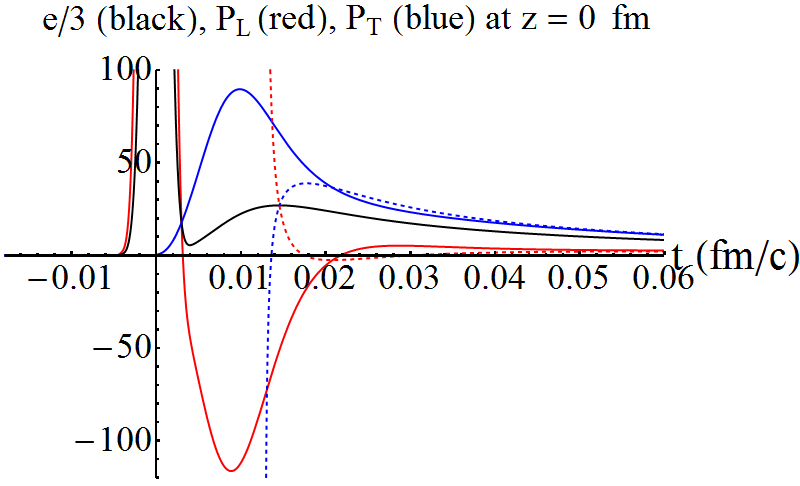}\tabularnewline
\includegraphics[width=0.42\textwidth]{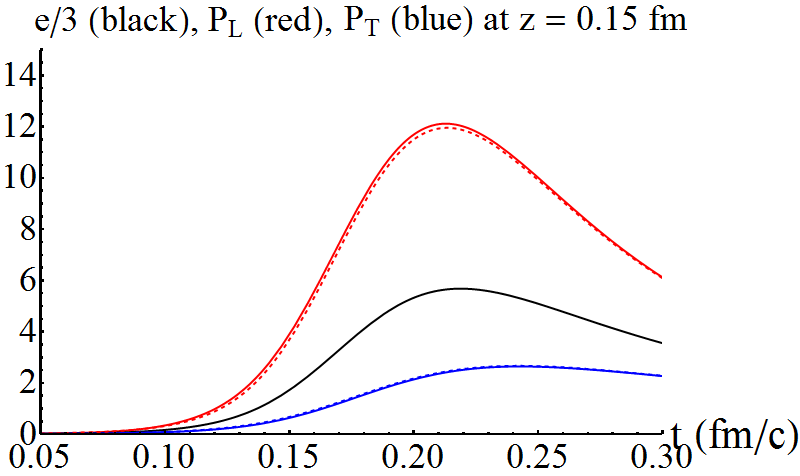}  & \includegraphics[width=0.42\textwidth]{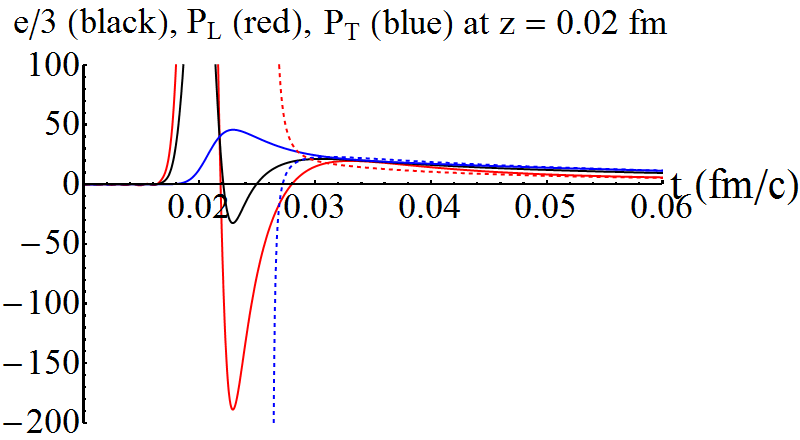}\tabularnewline
\includegraphics[width=0.42\textwidth]{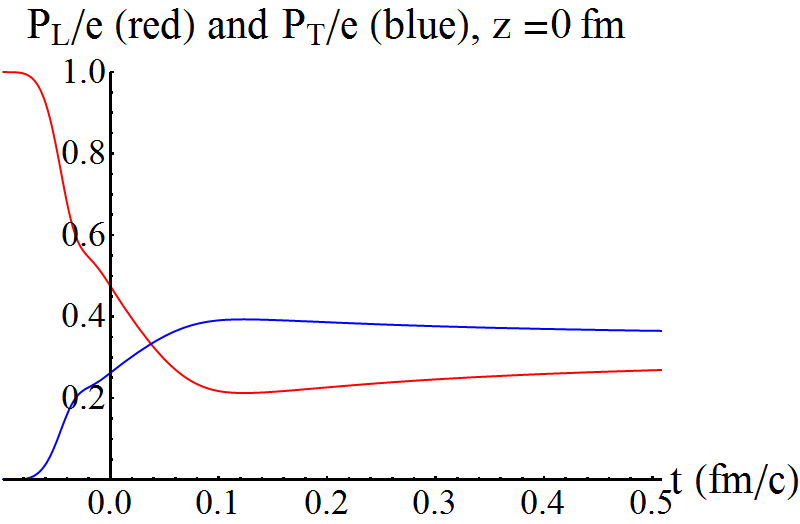} & \includegraphics[width=0.42\textwidth]{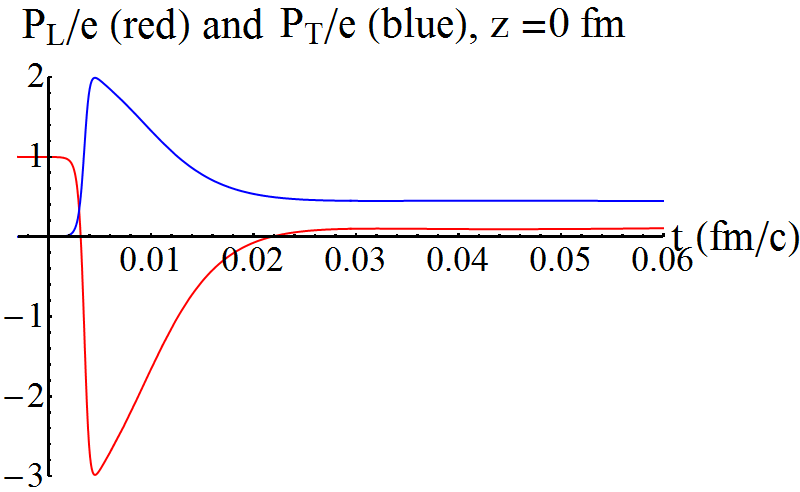}\tabularnewline
\end{tabular} \caption{Energy density and pressures of wide (left) and narrow (right) shock
wave collisions at the centre and off-centre (middle). The dashed
lines indicate the prediction by first order hydrodynamics (eqn. \ref{eq:hydroshocks}).
The last row plots the pressure over the energy density; the narrow
shock ratios directly after the shock almost reach two and minus three,
which agrees with analytic computations with delta-shocks \cite{Grumiller:2008va}.\label{fig:2Dhydro}}
\end{figure}

The thick shocks hence illustrate a full-stopping scenario. As the
shocks start to interact the energy density gets compressed and `piles
up', comes to an almost complete stop, and subsequently explodes hydrodynamically.
Indeed, at the time $t_{\text{max}}\simeq0.020\,\text{fm}/c$ at which
the energy density reaches its maximum in the top-left plot, the energy
density profile is approximately a rescaled version of one of the
incoming Gaussians, with about three times its height (see table \ref{tableshocks})
and {\small 2/3} its width. At this time, $90\%$ of the energy is
contained in a region of size $\Delta z\simeq2.4w$ in which the flow
velocity is everywhere $|v|\lesssim0.1$. Similarly, the energy flux
in this region is less than $10\%$ of the maximum incoming flux,
as illustrated by figure \ref{fig:flux}(left). Importantly, it can
also be checked that hydrodynamics is applicable, as can be seen in
figure \ref{fig:hydro-shocks}(left) and figure \ref{fig:2Dhydro}(left),
where it is seen that hydrodynamics becomes applicable even before
$t_{\text{max}}$.

The thin shocks illustrate a transparency scenario. In this case the
shocks pass through each other and, although their shape gets altered,
they keep moving at $v\simeq1$, as seen in figure \ref{fig:flux}(right).
The most dramatic modification in their shape is a region of negative
$e$ and $P_{L}$ that trails right behind the receding shocks. While
the negative $e$ only develops away from the centre of the collision,
the negative $P_{L}$ is already present at $z=0$, as shown more
clearly in the bottom-right plot of Fig.~\ref{fig:2Dhydro}. These
features are compatible with the general principles of Quantum Field
Theory \cite{Ford:1999qv}, since the `negative region' is far from
equilibrium and highly localised near a bigger region with positive
energy and pressure. In the case of thin shocks, we see from figure
\ref{fig:flux}(right) and figure \ref{fig:2Dhydro}(right) that there
is a clear separation between non-hydrodynamic receding maxima and
a plasma in between them that is described by hydrodynamics only at
sufficiently late times. At sufficiently late times it is also visible
from figure \ref{fig:EnergyDensity} that the receding maxima suffer
significant attenuation. We therefore emphasize that our use of the
term `transparency' refers to time scales longer than $t_{\text{hyd}}$
but shorter than the attenuation time. Furthermore, this `transparency'
is not necessarily related to the similarly transparent model used
in heavy-ion collisions. In particular, our model is not necessarily
boost-invariant (see section \ref{sec:Rapidity-profile:-Bjorken})
and all energy does end up in the plasma at late times, which suggests
that these collisions are not necessarily transparent for baryon number
density \cite{Bearden:2003hx}.

Several quantities of interest are given in table \ref{tableshocks}.
We see that $t_{\text{max}}>0$ for thick shocks, whereas for thin
shocks $t_{\text{max}}\simeq0$, as it would be in the absence of
interactions. Similarly, the maximum energy density $e_{\text{max}}$
is just the sum of the incoming energies $e_{\text{start}}$ for thin
shocks, indicating that, unlike for thick shocks, there is no compression
or piling up for thin shocks. The minimum energy density $e_{\text{min}}$
is negative for sufficiently thin shocks, as expected. The fact that
$t_{\text{hyd}}<0$ is negative for thick shocks simply means that
hydrodynamics becomes applicable even before the shocks fully overlap.
The temperature at the moment of hydrodynamisation , $T_{\text{hyd}}$,
is surprisingly constant. As in other models \cite{Chesler:2010bi,Chesler:2008hg},
the product $t_{\text{hyd}}T_{\text{hyd}}$ is smaller than unity
and fairly constant, which leads to hydrodynamisation times (significantly)
shorter than 1 fm. 

The anisotropy $P_{T}/P_{L}$ at the times where hydrodynamics is
applicable increases as the width decreases, reaching values as large
as $\sim15$. It is remarkable that such strong anisotropies can be
well described by first-order hydrodynamics. On the other hand hydrodynamics
fails to describe the regions with negative pressure, as expected
on thermodynamic grounds. It is therefore interesting that hydrodynamics
really seems applicable `as fast as possible'.

\begin{figure}
\begin{centering}
\includegraphics[width=1\textwidth]{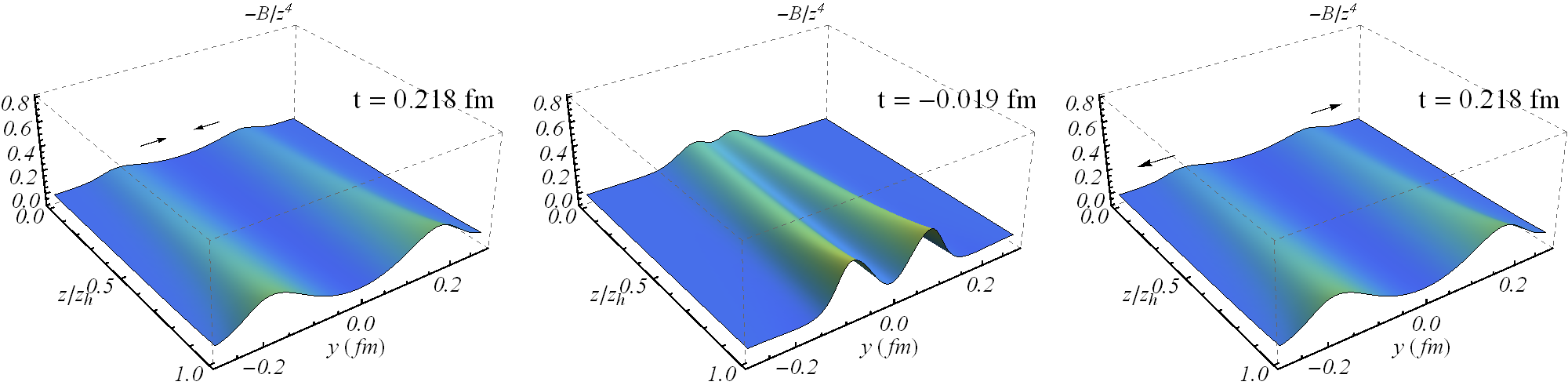} 
\par\end{centering}

\begin{centering}
\includegraphics[width=1\textwidth]{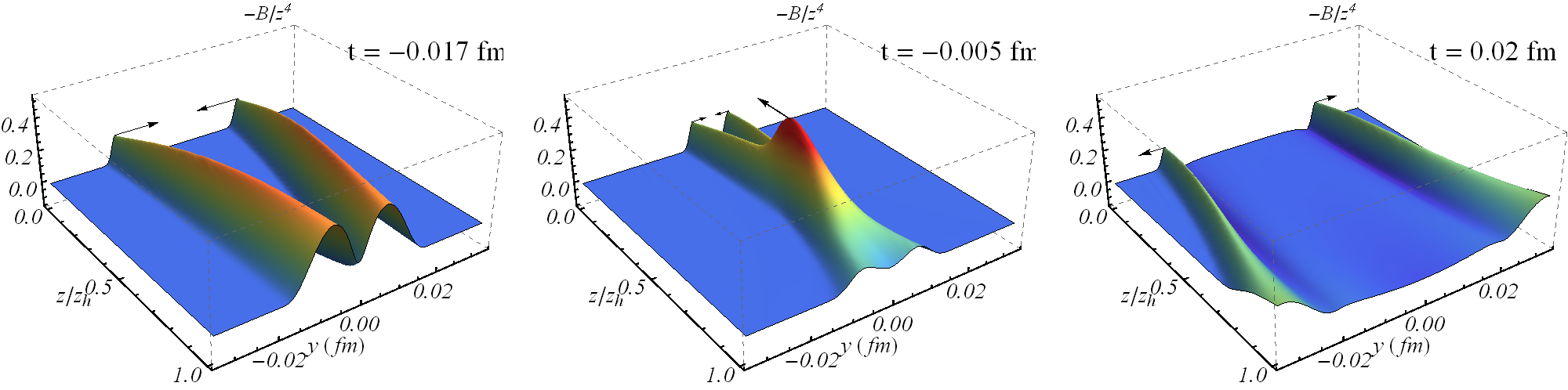} 
\par\end{centering}

\centering{}\caption{Here we illustrate the bulk dynamics by plotting $B/z^{4}$ between
boundary and horizon at various moments for wide (top) and narrow
(bottom) shocks. While $B$ is gauge dependent the figure still allows
to understand the basic difference between both shocks. Firstly, one
notices that in Eddington-Finkelstein coordinates the shocks collide
earlier deep in the bulk. The non-linear dynamics there, however,
arrives only at the boundary a time $\sim1/2T$ after the shocks have
collided. More importantly, it is apparent that the narrow shocks
are a superposition for a larger time, as indicated by the `blob'
hitting the boundary. The wide shocks, on the other hand, thermalise
continuously during the collision.\label{fig:bulkmovie}}
\end{figure}

\begin{figure}
\begin{centering}
\includegraphics[width=0.9\textwidth]{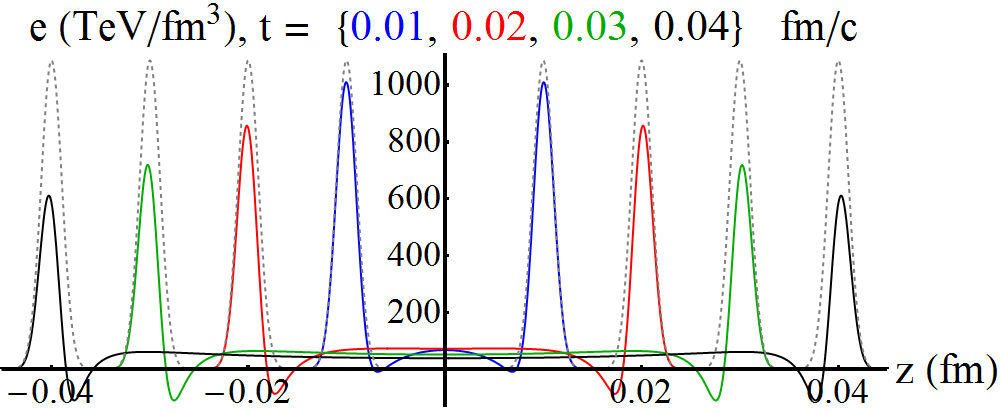} 
\par\end{centering}

\centering{}\caption{We plot snapshots of the energy density of figure \ref{fig:EnergyDensity}
(right) at different times, including snapshots of the original shocks
propagating without any interaction (dashed). The figure confirms
our bulk interpretation that the shocks decay due to non-linear dynamics
in the bulk, which in particular implies that the front of the shock
decays latest, as it it takes bulk dynamics longest to `catch up'
with the front of the shock.\label{fig:energysnapshots}}
\end{figure}

The crossover can be heuristically understood on the gravity side
(figure \ref{fig:bulkmovie}). Since each of the colliding shock waves
is a normalisable solution in the bulk, the metric near the AdS boundary
is a small deviation from AdS. Consequently, the gravitational evolution
is linear near the boundary for some time $t_{\text{lin}}$. The deviation
becomes of order one at $u\sim\mu^{-1}$, with $u$ the usual Fefferman-Graham
holographic coordinate. At this depth gravity becomes strong and the
evolution is non-linear. This non-linearity takes $t_{\text{lin}}\sim u\sim\mu^{-1}$
to propagate to the boundary. If $w\ll t_{\text{lin}}$, i.e.~if
$\mu w\ll1$, there is a clear separation between the linear and the
non-linear regimes. For thin shocks, this is illustrated by e.g.~figure
\ref{fig:2Dhydro}(left, top), where the energy density exhibits two
maxima around $\mu t\sim0$ and $\mu t\sim1$. The former corresponds
to the two shocks passing through each other; the latter corresponds
to the arrival to the boundary of the non-linear pulse from the bulk.
In this sense the pulse is responsible for the `creation' of the plasma
in between the thin receding shocks. Another clear manifestation is
present near the light cone; the non-linear pulse takes longest to
`catch up' with the front of the shocks, which hence decay the latest,
as is clear in figure \ref{fig:energysnapshots}. We note that in
our simulations we always work with Gaussian shocks of finite size,
so that the lightcone is also smoothed out; it is still an interesting
question what happens in the true delta-limit \cite{Grumiller:2008va}.

In contrast, for thick shocks $\mu w\gg1$, meaning that $t_{\text{lin}}\ll w$.
In this case the pulse reaches the boundary before the shocks have
passed through each other and essentially all the evolution is non-linear.
In figure \ref{fig:bulkmovie} this can be seen as a continuous thermalisation
during the collision. The metric is therefore not never the sum of
the incoming shocks, in contrast to the narrow shocks.

This analysis suggests that we have identified all the qualitatively
different dynamical regimes. Presumably we have also considered values
of $\mu w$ sufficiently representative of the asymptotic regimes
$\mu w\gg1$ and $\mu w\ll1$. For thick shocks this is suggested
by the fact that they come very close to a complete stop and subsequently
evolve hydrodynamically. For thin shocks this is suggested by comparison
of figure \ref{fig:2Dhydro}(bottom right) with \cite{Grumiller:2008va}.
This reference studied the delta-function limit $w\to0$ with $\mu$
fixed and found that the pressure/energy ratios are $P_{L}/e=-3$
and $P_{T}/e=2$ at $t\to0^{+}$. Figure \ref{fig:2Dhydro}(right)
shows that these are also the extremum values attained by our thin
shocks. 

\begin{table}[tp]
\begin{centering}
\begin{tabular}{|c|c|c|c|c|c|c|c|}
\hline 
$w$(am)  & $\mu w$  & $t_{\text{max}}$(am/c)  & $\frac{{\textstyle e_{\max}}}{{\textstyle e_{\text{start}}}}$  & $\frac{{\textstyle e_{\text{min}}}}{{\textstyle e_{\text{start}}}}$ & $t_{\text{hyd}}$(am/c)  & $T_{\text{hyd}}$(GeV)  & $t_{\text{hyd}}T_{\text{hyd}}$ \tabularnewline
\hline 
43.0  & 1.89  & 19.4 & 2.9  & 0.  & -2  & 2.6 & -0.02 \tabularnewline
17.0  & 0.75  & 3.5 & 2.3  & 0.  & 34  & 2.6 & 0.45 \tabularnewline
6.7  & 0.30  & 0.6 & 2.0  & 0.  & 23  & 2.7 & 0.32 \tabularnewline
2.7  & 0.12  & 0 & 2.0  & 0.  & 20  & 2.6 & 0.27 \tabularnewline
1.8  & 0.08  & 0 & 2.0  & -0.01  & 20  & 2.6 & 0.27 \tabularnewline
1.1  & 0.05  & 0 & 2.0  & -0.1  & 20  & 2.6 & 0.26 \tabularnewline
\hline 
\end{tabular}
\par\end{centering}

\caption{Numerical values of several quantities of interest for single shocks
($\ell=0$).\label{tableshocks}}
\end{table}

\section{Longitudinal coherence\label{sec:Longitudinal-coherence}}

It is an interesting question if the plasma created in the collisions
of section \ref{sec:A-dynamical-cross-over} depends on the longitudinal
structure of the colliding objects, which was previously restricted
to be a Gaussian. In heavy-ion collisions this structure would be
more complicated, but perhaps even more interesting are recent proton-lead
collisions. These are inherently a-symmetric, where in the \foreignlanguage{british}{centre-of-mass}
frame the energy density of the proton would be much more concentrated
than the energy density of the nucleus.

\begin{figure}[H]
\centering{}%
\begin{tabular}{cc}
\includegraphics[width=0.49\textwidth]{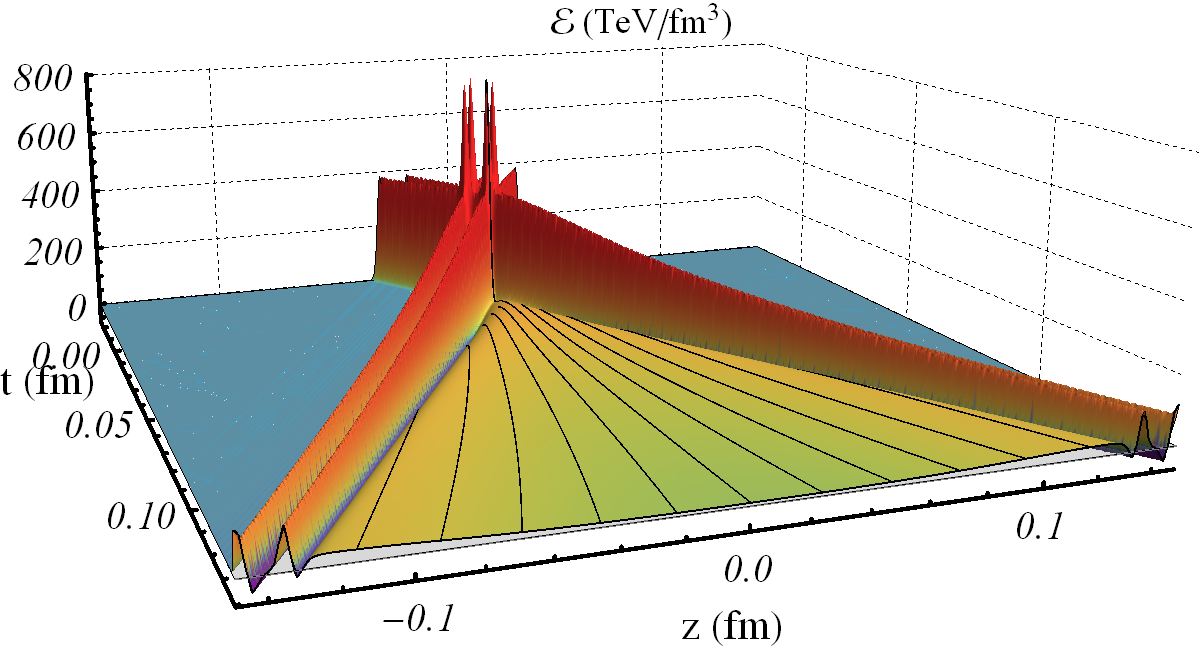}  & \includegraphics[width=0.49\textwidth]{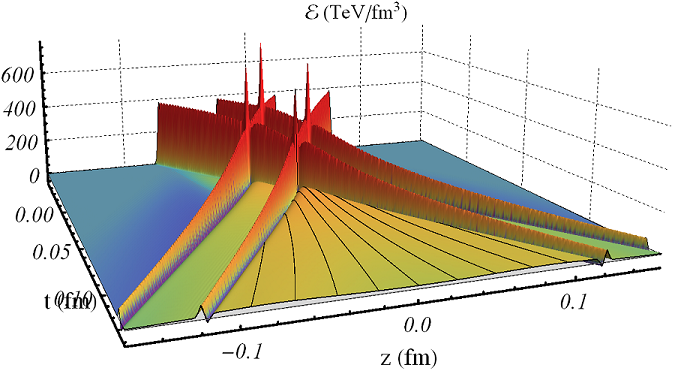}\tabularnewline
\end{tabular} \caption{Energy density of two collisions composed of narrow shocks with separation
$\ell=8w$ (left) and $\ell=32w$ (right), with $w=0.05/\mu=0.001$
fm. The black lines are streamlines of the produced plasma. \label{fig:EnergyDensitypPb}}
\end{figure}

To study the dependence on the longitudinal structure figure \ref{fig:EnergyDensitypPb}
shows the energy density for the two collisions in the second row
of Table \ref{TableofShocks}: a coherent collision with (eqn. \ref{eq:choiceforh}with
$\ell=8w$ and $\mu w=0.05$) (left) and an incoherent collision with
$\ell=32w$ and $\mu w=0.05$ (right). These are shocks composed of
two Gaussian `thin' constituents (as in subsection \ref{sub:From-full-stopping})
which are separated by $\mu\ell=8\mu w=0.4$ (left) and $\mu\ell=32\mu w=1.8$
(right). As expected from subsection \ref{sub:From-full-stopping},
the thin constituents pass through each other virtually undisturbed
and then start to attenuate. Close to the light-cone, both figures
show the initial shock profiles after the collision, indicating that
in both cases the high-rapidity region is sensitive to the initial
structure of the shocks.

\begin{figure}[H]
\centering{}%
\begin{tabular}{cc}
\includegraphics[width=0.49\textwidth]{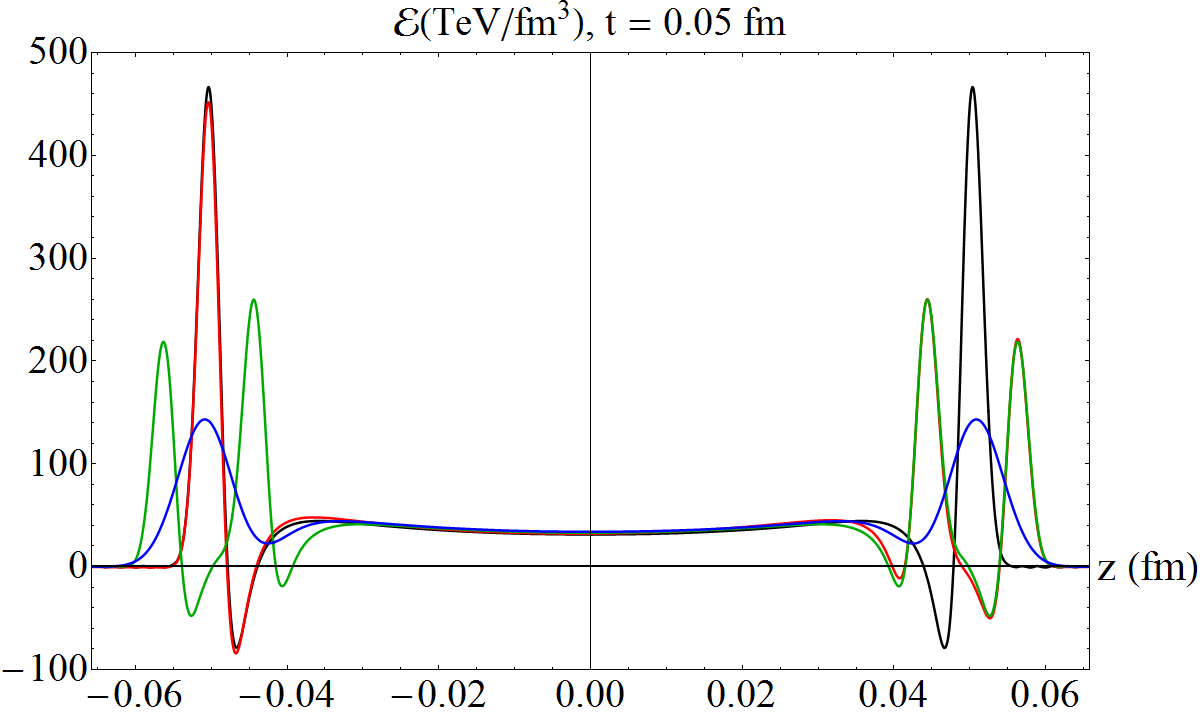}  & \includegraphics[width=0.49\textwidth]{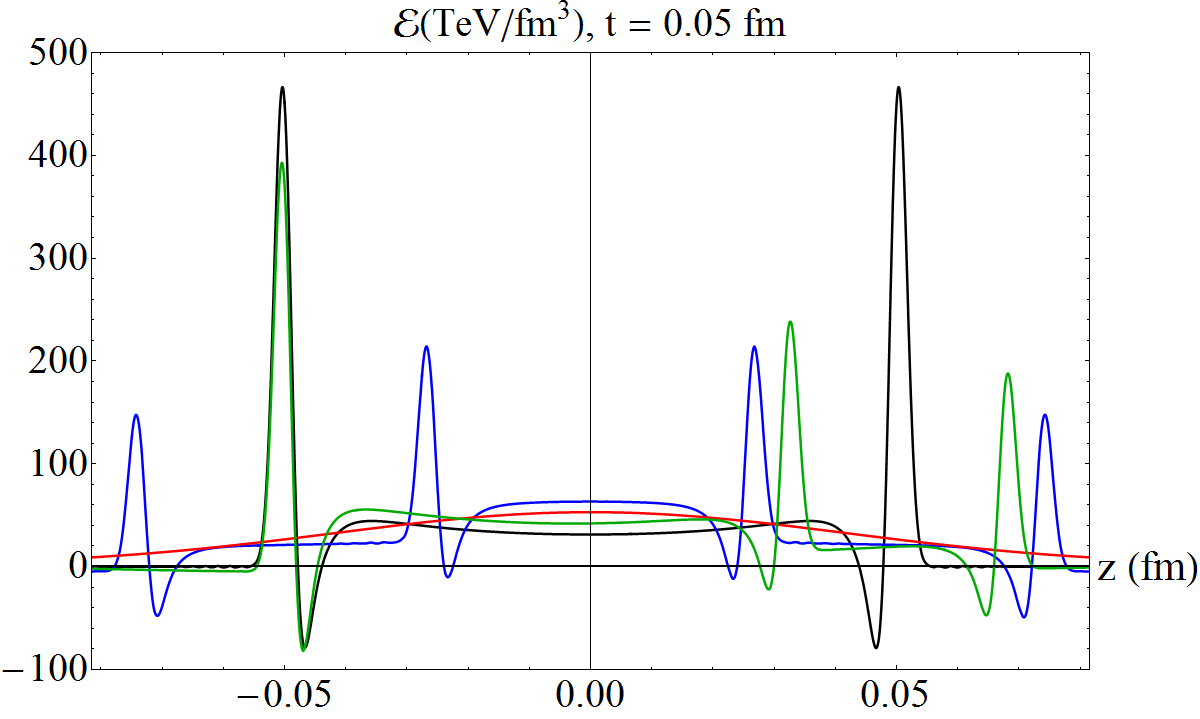}{\large{} }\tabularnewline
\end{tabular} \caption{Energy density at $t=0.05$ fm for different shock collisions \foreignlanguage{british}{characterised}
by the parameters displayed in Table \ref{TableofShocks}. \label{fig:ShockComparison} }
\end{figure}

In contrast, the mid-rapidity region of figure \ref{fig:EnergyDensitypPb}
(left) keeps no memory of the initial structure of the shocks. This
is illustrated in figure \ref{fig:ShockComparison} (left), which
shows snapshots of the energy density at a fixed time after hydrodynamisation,
$t=0.05$ fm, for the several collisions with different initial shock
structures but with the same total energy listed in the left part
of Table \ref{TableofShocks}. We see that the energy density around
mid-rapidity for the single-double collision of \ref{fig:EnergyDensitypPb}
(left) is identical to that for a single-single or a double-double
collision with constituents of the same width, and for a single-single
collision with twice-as-thick constituents. In all these cases the
hydrodynamisation time and the hydrodynamisation temperature are independent
of the initial structure of the shocks. For single shocks this is
consistent with section \ref{sec:A-dynamical-cross-over}, where it
was found that the hydrodynamisation properties of the plasma are
independent of the widths of the initial shocks provided these satisfy
$\mu w\lesssim0.2$.

\begin{table}[H]
\caption{\label{TableofShocks} Parameters of the shocks displayed in figure
\ref{fig:ShockComparison}. The 8/0 indicates an asymmetric collision
with different $\ell$ for left and right moving shocks.}

\centering{}%
\begin{tabular}{|l||c|c|c|c|c||c|c|c|c|c|}
\hline 
 & \multicolumn{5}{c||}{Left} & \multicolumn{5}{c|}{Right}\tabularnewline
\hline 
 & $\mu w$  & $\ell/w$  & $t_{\text{hyd}}$  & $T_{\text{hyd}}/\mu$  & $\ell T_{\text{hyd}}$  & $\mu w$  & $\ell/w$  & $\mu t_{\text{hyd}}$ & $T_{\text{hyd}}/\mu$  & $\ell T_{\text{hyd}}$ \tabularnewline
\hline 
Black & 0.05  & 0  & 0.88  & 0.30  & 0.05  & 0.05  & 0  & 0.88  & 0.30  & 0.05\tabularnewline
\hline 
Red & 0.05  & 8/0 & 0.88  & 0.30  & 0.12  & 1.9 & 0  & 0.95  & 0.31  & 0.36 \tabularnewline
\hline 
Green & 0.05  & 8 & 0.88  & 0.30  & 0.12  & 0.05  & 32/0 & 1.20  & 0.33  & 0.48 \tabularnewline
\hline 
Blue & 0.10  & 0  & 0.88  & 0.30  & 0.1  & 0.05 & 32  & -0.08  & 0.30  & 1.9 \tabularnewline
\hline 
\end{tabular}
\end{table}

Figure \ref{fig:ShockComparison}(right) shows analogous snapshots
for the collisions listed on the right part of table \ref{TableofShocks},
which again have the same total energy but differ in the initial structure
of the shocks. One of the curves is the same single-single collision
of thin shocks from figure \ref{fig:ShockComparison} (left), which
is included for comparison. The other three curves all have $\ell>0.26/T_{\text{hyd}}$
and they illustrate the incoherent regime, namely the fact that the
energy density around mid-rapidity, as well as the hydrodynamisation
time and the hydrodynamisation temperature, are sensitive to the initial
structure of the shocks. Note that the different hydrodynamisation
temperatures would translate into about a 30\% difference in the energy
density at mid-rapidity (which scales roughly as $T_{\text{hyd}}^{4}$)
even if each of these curves were plotted at its corresponding hydrodynamisation
time.

From the gauge theory viewpoint, these results imply that the smallest
longitudinal structure that the fields in the mid-rapidity region
can resolve is set by the inverse temperature at hydrodynamisation,
which in the coherent regime is $T_{\text{hyd}}=0.3\mu$. Clearly,
the plasma will be sensitive to the structure of the initial shocks
if their characteristic size, $\ell_{\text{char}}$, is larger than
the formation time of the hydrodynamised plasma, $t_{\text{hyd}}$.
By inspection of table \ref{TableofShocks} we see that the transition
between the coherent and the incoherent regimes takes place at a scale
$\ell_{\text{coh}}$ such that $0.12<\ell_{\text{coh}}T_{\text{hyd}}<0.36$.
Since this transition is smooth, $\ell_{\text{coh}}$ is not sharply
defined. Motivated by the considerations above, we therefore choose
to define it as the hydrodynamisation time for single-single collisions
of thin shocks, which yields $\ell_{\text{coh}}=0.26/T_{\text{hyd}}$.

\begin{figure}
\centering{}%
\begin{tabular}{cc}
\includegraphics[width=0.49\textwidth]{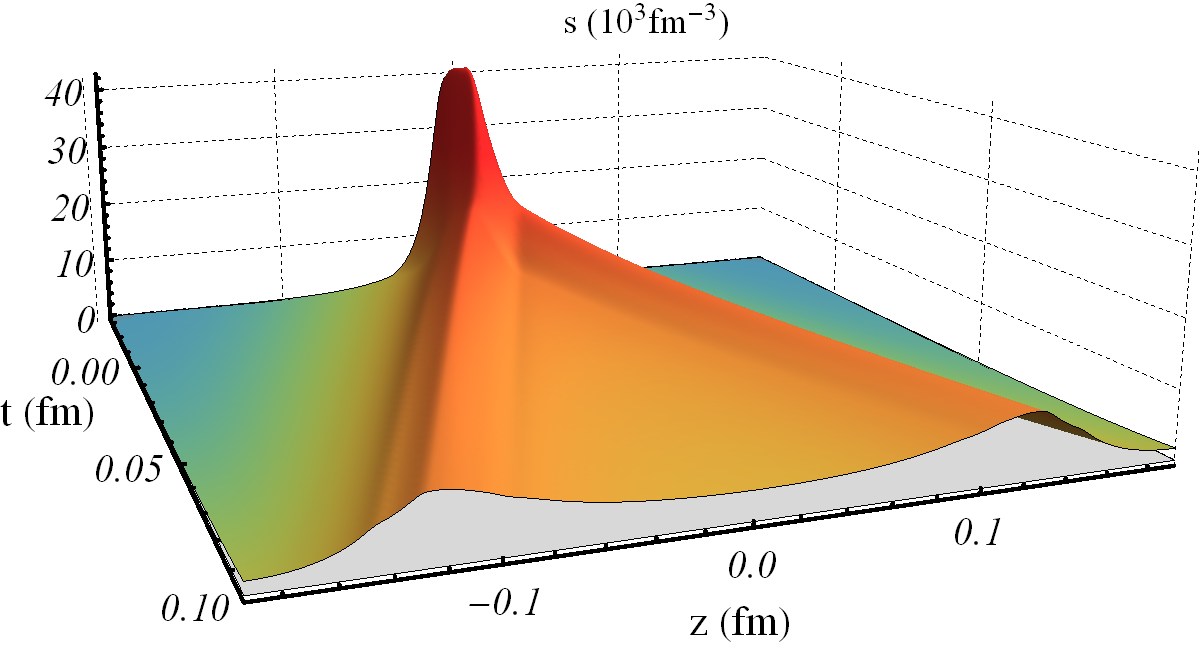}  & \includegraphics[width=0.49\textwidth]{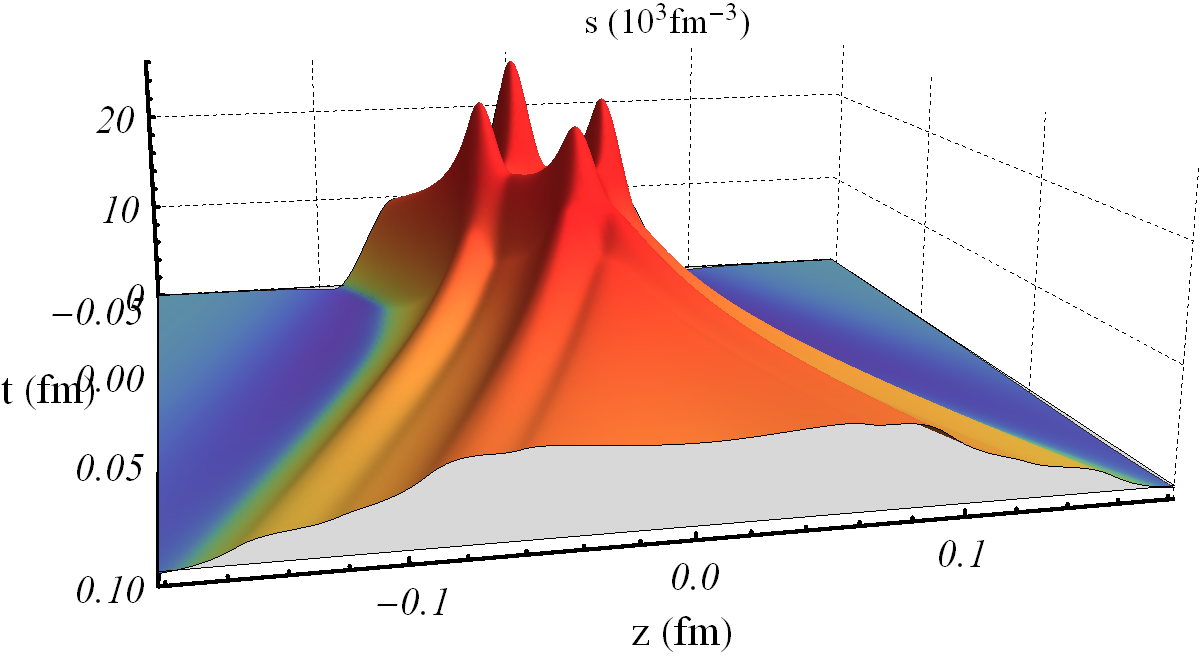}\tabularnewline
\end{tabular} \caption{Entropy density $s$ found by computing the volume element on the
apparent horizons formed in the two collisions depicted in figure
\ref{fig:EnergyDensitypPb}. Note that the entropy is only defined
near equilibrium, so that outside this region $s$ should be interpreted
as a lower bound on the entropy which is going to be produced. \label{fig:horizonp-Pb}}
\end{figure}

This picture is supported by the gravitational description. In figure
\ref{fig:horizonp-Pb} we show the entropy density from the apparent
horizon formed in the two collisions displayed in figure \ref{fig:EnergyDensitypPb}.
Although this quantity depends on the slicing of the space-time, close
to equilibrium it provides a lower bound for the entropy density \cite{Figueras:2009iu}.
According to the gauge/gravity duality, the horizon encodes the physics
at the thermal scale. Heuristically, one may say that figure \ref{fig:horizonp-Pb}
provides an effective picture of figure \ref{fig:EnergyDensitypPb}
in which all length scales shorter than the thermal scale have been
integrated out. It is therefore suggestive that in figure \ref{fig:horizonp-Pb}
(left) there is no trace of the microscopic structure of the shocks
even at the time $t=0$ of the collision. In contrast, for the further-separated
colliding shock constituents of figure \ref{fig:EnergyDensitypPb}
(right), the corresponding apparent horizon in figure \ref{fig:horizonp-Pb}
(right) reflects the initial configuration, albeit with a significant
smoothing due to the integration of scales.

\subsection{Consequences for HIC\label{sub:Consequences-for-HIC}}

\noindent Since longitudinal coherence only depends on the inability
of the horizon to resolve sub-thermal length scales, we expect this
coherence to occur in holographic high-energy collisions more general
than the simple model considered here. These may include collisions
of shocks with profiles more general than eqn. \ref{eq:choiceforh}
and collisions with non-trivial transverse dynamics, at least if the
transverse expansion rate is slower than the longitudinal one. In
the following we take this as an assumption and explore interesting
consequences for high-multiplicity (p/d)+A collisions. Furthermore,
we consider the limits in which the physics of bulk-particle production
is assumed to be exclusively strongly or weakly coupled, the hope
being that these limits bracket the production dynamics at the energies
of present colliders.

In the strong coupling limit our results, together with the large
Lorentz contraction of the colliding projectiles at RHIC and LHC,
suggest that most of the participating nucleons act coherently in
the formation of the plasma. As a consequence, the momentum rapidity
of the plasma's c.o.m., $y_{\text{plasma}}$, should coincide with
the momentum rapidity of the c.o.m.~of all the participating nucleons,
$y_{\text{part}}$. Since the local energy density at fixed proper
time is maximal at $y_{\text{plasma}}$ \cite{Casalderrey-Solana:2013aba,Chesler:2013lia},
the maximum in the rapidity distribution of particles, $y_{\text{max}}$,
also coincides with $y_{\text{part}}$. For a generic collision with
$N_{\text{A}}$ ($N_{\text{B}}$) right-moving (left-moving) participating
nucleons moving at rapidity $y_{\text{A}}$ ($y_{\text{B}}$), we
have that $y_{\text{part}}=\tfrac{1}{2}\,\log(N_{\text{A}}/N_{\text{B}})+y_{\text{NN}}$,
where $y_{\text{NN}}=\tfrac{1}{2}(y_{\text{A}}+y_{\text{B}})$ is
the rapidity of the nucleon-nucleon c.o.m. As a consequence, event-by-event
fluctuations in the number of participating nucleons in A+A collisions
lead to fluctuations in $y_{\text{max}}$ according to $y_{\text{part}}$,
as was also studied in \cite{Bzdak:2012tp}. Similarly, in p(d)+A
collisions $y_{\text{max}}$ shifts to the A side due to the asymmetric
collision geometry. Taking $N_{A}=15-30$ as representative values
for central p(d)+A collisions at the LHC (RHIC) we find $y_{\text{max}}=0.9\,(1.3)-1.2\,(1.7)$.
An additional result of the strong-coupling model is that the plasma
is $y$-reflection-symmetric around $y_{\text{plasma}}$. Interestingly,
particle production in d+A collisions at RHIC \cite{Back:2003hx}
seems consistent with both of these features, as already noted in
\cite{Steinberg:2007fg}.

At weak coupling we may determine $y_{\text{max}}$ via perturbative
QCD. For nuclei moving at large rapidities, $\left|y_{\text{A}}-y_{\text{B}}\right|\gg1$,
this can be estimated by equating the squared saturation scales of
both colliding objects \cite{Xiao:2005pc}, $Q_{s}^{2}(N_{\text{A}},y_{\text{max}})=Q_{s}^{2}(N_{\text{B}},y_{\text{max}})$.
Far from its own rapidity $y_{\text{C}}$, the saturation scale of
a nucleus with $N_{\text{C}}$ participating nucleons evolves as $Q_{s}^{2}(N_{\text{C}},y)\sim N_{\text{C}}\exp\left(\bar{\lambda}\left|y-y_{\text{C}}\right|\right)$
\cite{Kharzeev:2001gp,Kharzeev:2004if}. The coupling-dependent exponent
$\bar{\lambda}$ can be extracted from fits to HERA data within the
saturation framework \cite{Stasto:2000er} and is given by $\bar{\lambda}\simeq0.25$
\cite{Kharzeev:2001gp,Kharzeev:2004if}, reflecting the fact that
in perturbative QCD the fraction of energy available for particle
production decreases with energy. Substituting in the equation for
$y_{\text{max}}$ we find $y_{\text{max}}=\frac{1}{2\bar{\lambda}}\,\log(N_{\text{A}}/N_{\text{B}})+y_{\text{NN}}$.

\begin{figure}
\centering{}\includegraphics[width=0.95\textwidth]{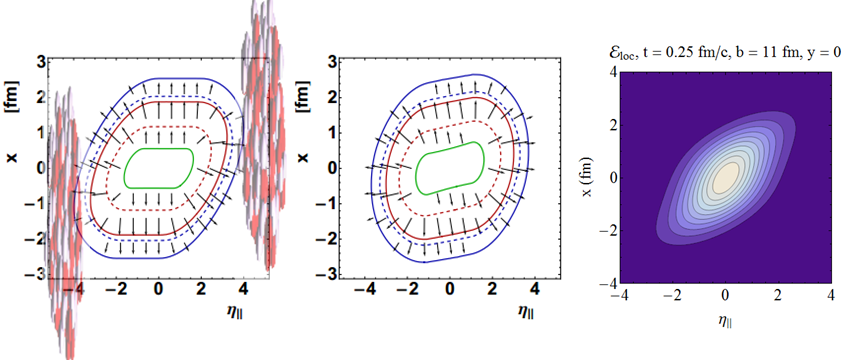} \caption{The figure on the left (adapted from \cite{Bozek:2010bi}) represents
constant pressure contours for initial conditions for an off-central
collision computed by assuming coherence and a plateau in rapidity.
This initial condition turns out not to produce the correct direct
flow $v_{1}(\mbox{\ensuremath{\eta})}$, whereas a phenomenologically
inspired tilt (middle figure, from \cite{Bozek:2010bi}) does produce
the correct $v_{1}(\mbox{\ensuremath{\eta})}$. On the other hand,
if one uses a Gaussian rapidity distribution with width 0.95 (see
section \ref{sec:Rapidity-profile:-Bjorken}) the initial condition
looks similar to the tilted one, with using coherence (right). It
is interesting that planar shock waves give such interesting consequences
for these off-central collisions.\label{fig:directflow}}
\end{figure}

Another interesting consequence applies to off-central nucleus-nucleus
collisions. There the rapidity dependence of the direct flow $v_{1}$
depends crucially on the longitudinal deposition of the energy \cite{Bozek:2010bi}%
\footnote{We thank Piotr Bozek for bring this work to our attention.%
}. In this work the `firestreak model' shifts rapidity to the centre-of-mass
just like our proposal above. Quite surprisingly reference \cite{Bozek:2010bi}
found that this does not produce the correct direct flow as a function
of rapidity. Interestingly, if one replaces their boost-invariant
rapidity profile with the rapidity profile of section \ref{sec:Rapidity-profile:-Bjorken}
the direct flow seems to be closer to the experimental value (see
figure \ref{fig:directflow}).

Lastly, our coherent picture will mean that at a fixed position in
the transverse plane the correlations in the longitudinal direction
will be basically constant over the complete range. This is because
the plasma will only be sensitive to the total energy per transverse
area, whereby more energy will just increase the energy density over
the total longitudinal range. Of course, there will still be thermal
fluctuations ($1/N_{c}$ supressed in our set-up), such as interestingly
studied in \cite{Springer:2012iz}. This therefore means that we expect
fluctuations in the longitudinal direction to be solely due to thermal
fluctuation, as opposed to transverse fluctuations, which originate
both from initial state fluctuations and thermal fluctuations. To
measure thermal transport coefficients it can therefore be beneficial
to focus on longitudinal dynamics.

We thus conclude that longitudinal coherence has consequences for
off-central nucleus-nucleus collisions and for proton-nucleus collisions,
whereby strong- and weak-coupling predictions are considerably different.
Especially the value for $y_{\text{max}}$ in p+A collisions differs
by about a factor of four. This makes the possible experimental extraction
of $y_{\text{max}}$ from RHIC \cite{Abelev:2008ab} or LHC \cite{ALICE:2012xs}
d/p+A data extremely interesting, since the result may help constrain
the mechanism of bulk-matter production.

\section{Rapidity profile: Bjorken vs Landau?\label{sec:Rapidity-profile:-Bjorken}}

In heavy-ion collisions there are two interesting models to describe
the initial longitudinal dynamics. The first one is described by Landau
\cite{Landau:1953gs} in 1953, where he assumes the two nuclei to
be completely equilibrated and at rest at the moment they completely
overlap. Due to all the energy being concentrated in an extremely
small volume, this model will lead to a violent, but hydrodynamic,
explosion afterwards. An impressive success of the Landau model is
its prediction that the total number of particles scales as $s_{NN}^{1/4}$,
which holds experimentally for a very large range in collision energies
$\sqrt{s_{NN}}$ \cite{Back:2006yw,Steinberg:2004vy,Wong:2008ta},
albeit being violated at LHC \cite{Chatrchyan:2012mb} (see also subsection
\ref{sub:Multiplicities-and-a}).

Importantly, while the Landau model seems to reproduce the total particle
number (see however subsection \ref{sub:Multiplicities-and-a}), it
is not generally believed to be an accurate description of heavy-ion
collisions. Partly, this is because at such high energies the coupling
of QCD is generally not assumed to be strong enough to cause the needed
stopping. More importantly, it is experimentally found that the conserved
baryon number (the number of protons minus antiprotons) ends up at
high rapidities \cite{Bearden:2003hx}, which may be unnatural in
the Landau model. In the current AdS model there is no such conserved
charge, but we plan to report on this in the near future.

The second model was developed much later, in a famous paper by Bjorken
\cite{Bjorken:1982qr}. Building on previous intuition \cite{Feynman:1969ej,Cooper:1974qi}
he assumed interactions to be sufficiently weak such that the nuclei
could pass through each other, virtually unperturbed. In the middle
a plasma would form, which would be invariant under boosts, or equivalently
shifts in rapidity. Here, proper time $\tau$ and rapidity $y$ are
defined by $t=\tau\cosh y$ and $z=\tau\sinh y$. Of course it is
not possible to have a boost-invariant plasma for all rapidities (it
would require infinite energy), but one may imagine that the range
of approximate boost-invariance will grow with growing collision energy.

The basic assumptions of these two models cannot be derived from (strongly
coupled) QCD. It is therefore of great interest to see how our results,
at infinite coupling, fit into these two models. Indeed, our full-stopping
scenario is in close similarity with the Landau model, as was already
clear in \cite{Chesler:2010bi}. Perhaps the only difference is that
we find a very specific energy ($\mu w=0.75$, also used in \cite{Chesler:2010bi})
to \foreignlanguage{british}{realise} the Landau model, whereas the
Landau model is supposed to be correct for a wide range of energies.
At lower energies%
\footnote{Note that the energy per transverse area, $\mu^{3}$, scales as the
gamma factor $\gamma=\sqrt{s_{NN}}/(2m_{p})$, with $m_{p}$ the proton
mass, and that Lorentz contraction makes $w$ scale as $\gamma^{-1}$,
such that $\mu w\sim\gamma^{-2/3}$. This means that lower energy
collisions are best thought of as collisions with large $\mu w$.%
}, there is a significant `piling up' of energy. This will reduce the
scaling $s_{NN}^{1/4}$ somewhat, but probably too little to be easily
detectable experimentally.

At high energies we find a transparent regime, showing that infinite
coupling does not necessarily lead to full stopping and is compatible
with receding shocks moving at the speed of light. Indeed, a simple
field theory scaling argument shows that this is inevitable. In the
Landau model the inverse temperature of the equilibrated plasma scales
as $e^{-1/4}\sim(\mu^{3}/w)^{-1/4}\sim\gamma^{-1/2}$, and can be
thought, both thermodynamically and quantum mechanically, as the minimum
time scale or distance required to interact. The width, however, scales
as $w\sim\gamma^{-1}$, such that it is easy to see that for large
enough $\gamma$ the width is smaller than the minimum time to interact,
and hence the shocks necessarily pass through each other: they do
not have time to thermalise.

\begin{figure}
\centering{}\includegraphics[width=0.47\textwidth]{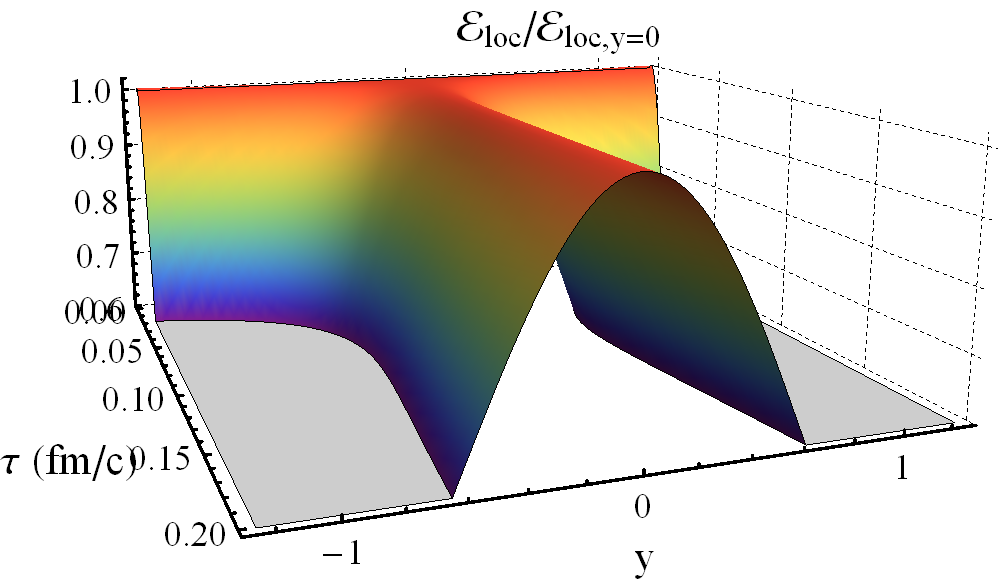}$\,\,$
\includegraphics[width=0.48\textwidth]{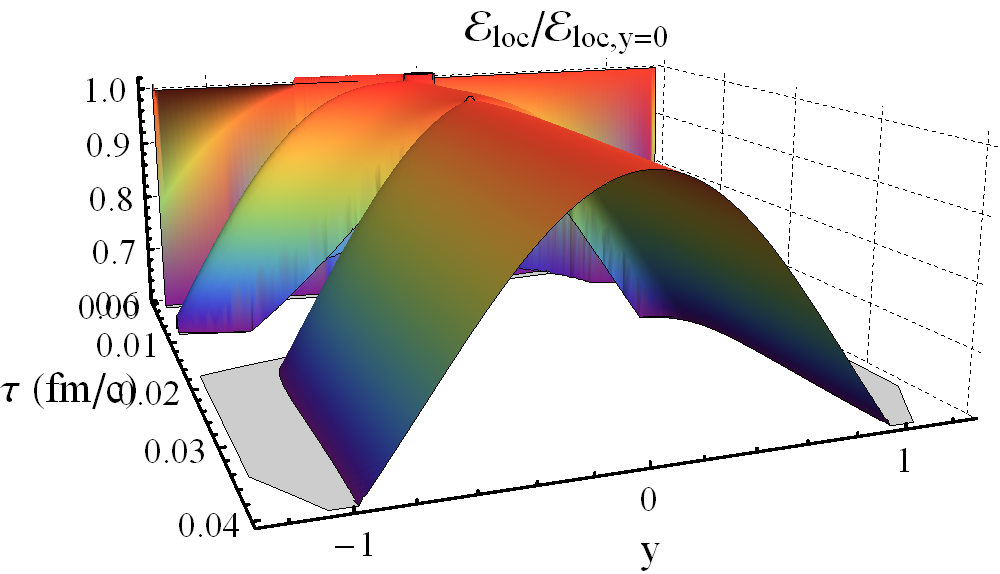}\caption{Energy density in the local rest frame around mid-rapidity as a function
of spacetime rapidity $y$ and proper time $\tau$ for thick (left)
and thin (right) shocks (the same shocks as figure \ref{fig:EnergyDensity}).
In the right case we have excluded from the plot the region in which
the local rest frame is not defined because $2|s|>|e+P_{L}|$. \label{fig:rapidity} }
\end{figure}

The plasma formed by the narrow shocks, however, does not obey the
boost-invariance of the Bjorken model. This is most easily seen in
figure \ref{fig:rapidity}, where we have changed to proper-time and
spacetime-rapidity coordinates. The `tubes' at late times show that
the local energy density at mid-rapidity is not rapidity-independent
but has an approximately Gaussian profile of width about 0.95 in the
transparent regime.

It is sensible to ask how figure \ref{fig:rapidity} would look like
in limit of infinite collision energy. This can be answered by \foreignlanguage{british}{realising}
that longitudinal coherence (section \ref{sec:Longitudinal-coherence})
shows a plasma at mid-rapidity which is independent of $\mu w$, provided
$\mu w$ is small enough. We can therefore say that figure \ref{fig:rapidity}(right)
is already very close to the delta-limit, or infinite energy limit,
where $\mu w=0$. We therefore believe to have arrived at a universal
high energy rapidity profile for conformal theories at infinite coupling,
which is notably different from the Bjorken model above.

\begin{figure}
\centering{}\includegraphics[width=0.65\textwidth]{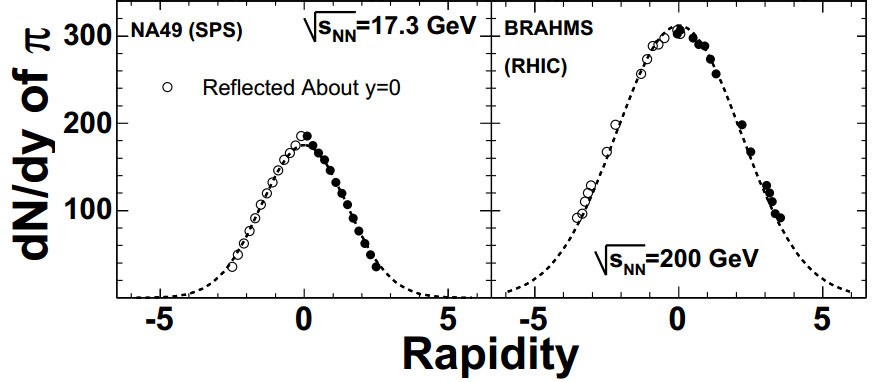}\caption{Particle multiplicities of $\pi^{-}$ and $\pi^{+}$ particles at
$\sqrt{s_{NN}}=17.3$ GeV (SPS, \cite{Afanasiev:2002mx}) and $\sqrt{s_{NN}}=200$
GeV (RHIC, \cite{Bearden:2004yx}). The dashed lines are fitted Gaussians
of width $1.42\pm0.02\,\text{(stat)}$ and $2.25\pm0.02\,\text{(stat)}$
respectively, the figure is taken from \cite{Back:2004je,Romatschke:2009im}.
It can be seen that the Gaussians describe the distribution well,
which was our main motivation when comparing figure \ref{fig:rapidity}
with a Gaussian. It should be kept in mind, however, that these experimental
data represent the final rapidity distribution, and hence do not necessarily
agree with the initial distribution, such as we try to compute. Also,
measuring rapidity distributions necessitates measuring particle masses,
which at LHC is unfortunately not possible over the full range of
rapidities.\label{fig:RHICrapidity} }
\end{figure}

\subsection{Local energy density in real time}

\begin{figure}
\centering{}\includegraphics[width=0.55\textwidth]{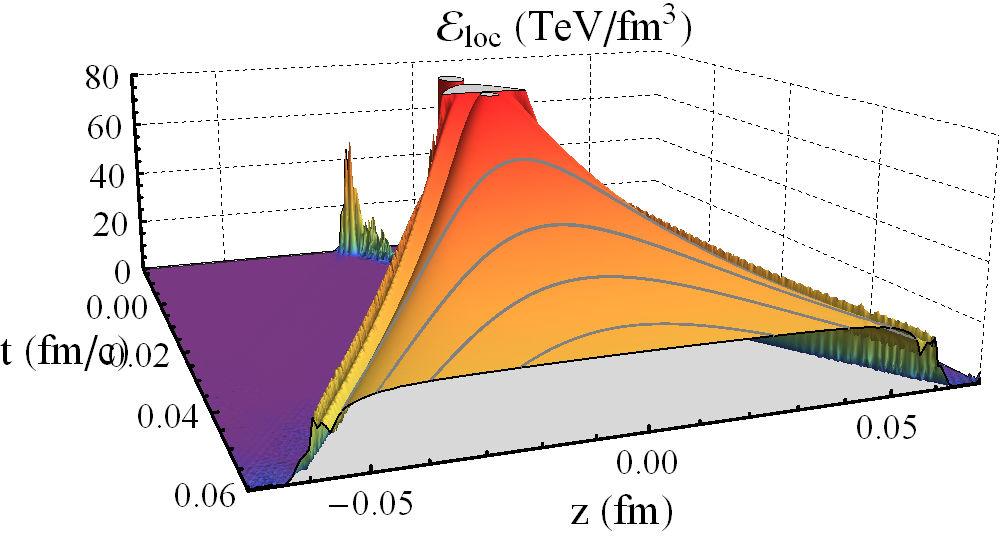}$\,\,$\includegraphics[width=0.42\textwidth]{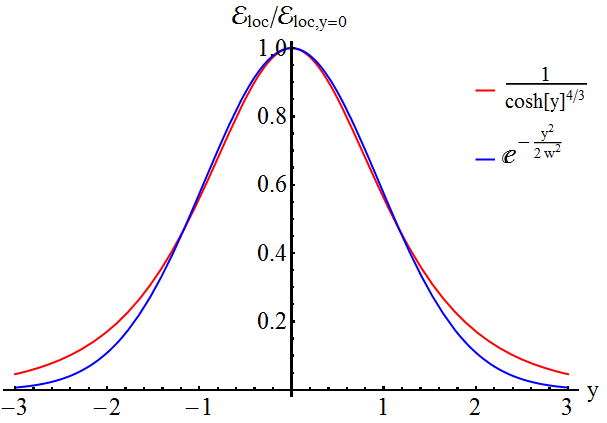}
\caption{Energy density in the local rest frame as a function time and space
for thin shocks with constant proper time curve (\foreignlanguage{british}{grey}).
The curve is remarkably flat shortly after the applicability of hydrodynamics
It is an open problem to understand this flatness, but it can be used
to estimate the rapidity profile plotted in figure \ref{fig:rapidity}
being $\cosh(y)^{-4/3}$. At the right this shape is compared to a
Gaussian of width 0.95, and they fit remarkably well in the region
we can compare with our theoretical computation (figure\ref{fig:rapidity})
or even with experimental data (figure\ref{fig:RHICrapidity}). Current
simulations for LHC collisions would start with a completely flat
profile for $y\in(-5,\,5)$ \cite{Schenke:2012hg}. \label{fig:eloc} }
\end{figure}

Interestingly, the rapidity shape described above may have a more
natural interpretation in real space, instead of rapidity space. When
plotting the local energy density as a function of time and space
for narrow shocks (figure \ref{fig:eloc}, left), one notices that
shortly after the applicability of hydrodynamics (at around 0.02 fm)
the local energy density is remarkably flat as a function $z$ at
fixed $t$ (real time!). This could already be noticed by looking
at the transverse pressure (figure \ref{fig:EnergyDensity} lower
right), which is already in the local rest frame. We stress that the
constancy of $e_{\text{loc}}$ should come as a surprise; in fact,
we could not think of an argument independent of our numerical simulation.
In particular, a boost-invariant $e_{\text{loc}}$ would look very
differently, being constant along the grey lines in figure \ref{fig:eloc}
or flat in figure \ref{fig:rapidity}. There is no real translation
symmetry in $z$ either, as the velocity profile approximates the
expected $z/t$ well. Note also that $e_{\text{loc}}$ will not remain
constant at later times; the hydrodynamic expansion will widen the
rapidity distribution, which will change $e_{\text{loc}}$. This,
however, is just due to hydrodynamics and therefore of lesser interest
in our context of thermalisation.

With this extra understanding it is then only natural that at constant
proper time $\tau$ (\foreignlanguage{british}{grey} lines in figure
\ref{fig:eloc}) the energy density decreases with rapidity $y$.
We can write $e_{\text{loc}}(t,\, z)=f(t)$, where $f(t)$ will decrease
due to the expansion and longitudinal pressure of the plasma. In the
regime plotted this can be fairly well approximated as $f(t)\sim t^{-4/3}$,
which is also the late time boost-invariant result%
\footnote{It was also noted in \cite{Chesler:2013lia} that even though the
plasma is not boost-invariant, one can still succesfully use boost-invariant
hydrodynamics at fixed rapidity for a limited time. This simply means
that the rapidity gradients can be neglected during the short time
scales considered here. For longer evolutions of order of the lifetime
of the plasma this is not expected to hold anymore.%
}. As a function of proper time this gives 
\begin{equation}
e_{\text{loc}}\sim(\tau\,\cosh(y))^{-4/3}.
\end{equation}
Comparing this rapidity shape with the Gaussian conjectured above
leads to very similar curves until $\mathcal{E}_{\text{loc}}/\mathcal{E}_{\text{loc,y=0}}\approx0.3$
(figure \ref{fig:eloc}, right), below which it is difficult to make
either theoretical calculations (figure \ref{fig:rapidity}) or experimental
measurements (figure \ref{fig:RHICrapidity}). So perhaps the rapidity
profile is more similar to $\cosh(y)^{-4/3}$ instead of a Gaussian.
This would also provide a natural explanation why the Gaussian has
width 0.95.

\subsection{Multiplicities and a comparison with experiments\label{sub:Multiplicities-and-a}}

\begin{figure}
\centering{}\includegraphics[width=0.46\textwidth]{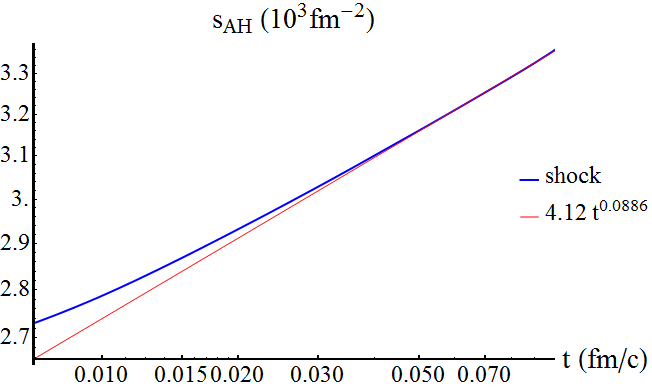}$\,\,$\includegraphics[width=0.42\textwidth]{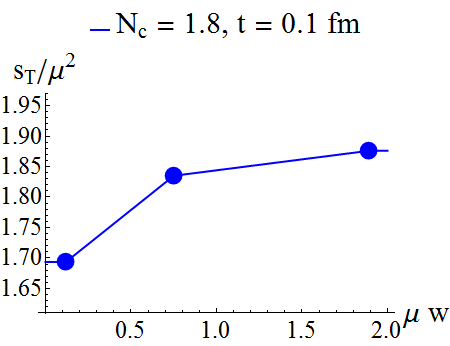}
\caption{Using figure \ref{fig:horizonp-Pb} we can obtain the total entropy
per transverse area for thin shocks as a function of time. We find
that most entropy is already produced before $t=0$, and that the
late-time increase is well described by a slow power of $t$. On the
right we evaluated this total entropy for several shock simulations
at $t=0.1$ fm, showing a mild dependence on the width. \label{fig:AHshocks} }
\end{figure}

Given that the computations above suggest a universal rapidity profile
(figure \ref{fig:rapidity}) in the high energy limit, it is interesting
to ask if this kind of shape may be \foreignlanguage{british}{realised}
in real heavy-ion collisions. It is challenging to compare directly
with experiments, firstly because the plasma evolution can significantly
change the profile, and secondly it is only possible to measure the
pseudo-rapidity (related to the angle of flight $p_{z}/|p|$), which
can only be converted to real rapidity (related to $p_{z}/E$) if
the particle mass is known.

An indication that a Gaussian-like initial profile can be realistic
comes from RHIC, where it is possible to convert to rapidity and an
approximately Gaussian shape was found (see figure \ref{fig:RHICrapidity}).
The width of this Gaussian grows with energy, but it is unclear how
much of this growth is due to evolution or due to the initial profile. 

To make such an estimate we can compute the initial entropy, which
is approximately conserved during the evolution if the viscosity is
small enough. This entropy $S$ directly translates into the total
number of charged particles produced: $N_{charged}=S/7.5$ \cite{Pal:2003rz,Muller:2005en,Gubser:2008pc}.
While the shock waves only simulate collisions which are homogeneous
in the transverse plane, it is not unreasonable to assume that the
entropy density in the first moments (where the longitudinal size
is much smaller than the transverse size) does not depend much on
transverse gradients. If most of the fluid is then well described
by hydrodynamics, one can obtain a good estimate of the total entropy
of the collision debris. Figure \ref{fig:AHshocks} indicates that
this is indeed the case.

\begin{figure}
\centering{}\includegraphics[width=0.48\textwidth]{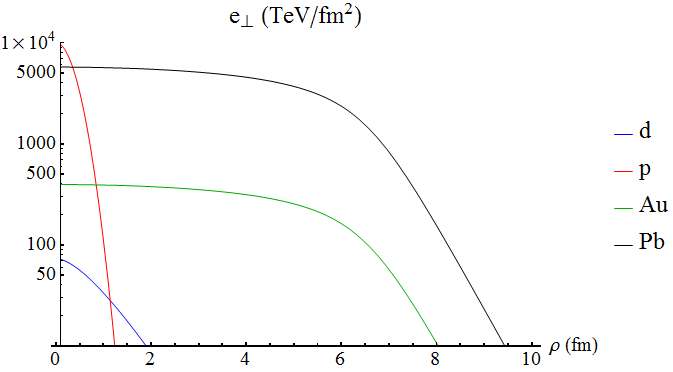}$\,\,$\includegraphics[width=0.42\textwidth]{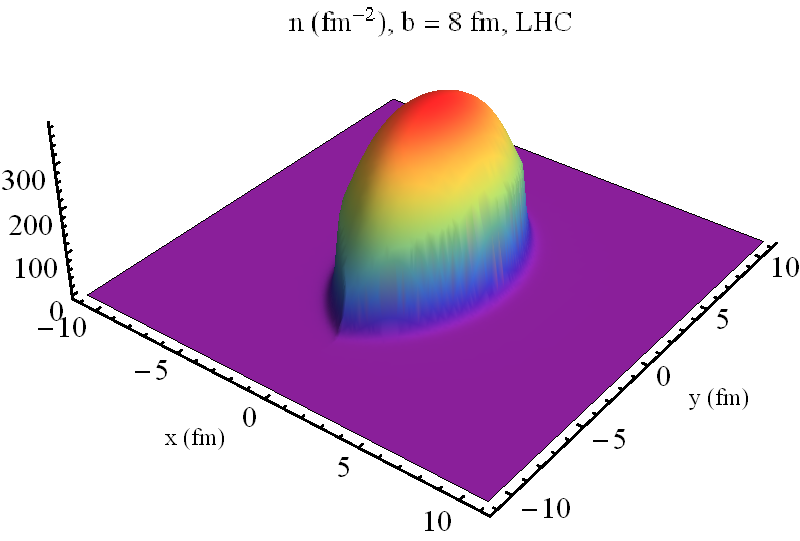}
\caption{Using the total entropy per transverse area as a function of the energy
scale $\mu$ and the width $w$ (figure \ref{fig:AHshocks}) we can
use this to compute the total entropy for various transverse energy
densities (left, values from \cite{Alver:2008aq}). As an example
we plot the multiplicity per transverse area for an LHC lead-lead
collision (right).\label{fig:models} }
\end{figure}

\begin{figure}
\begin{centering}
\includegraphics[width=0.48\textwidth]{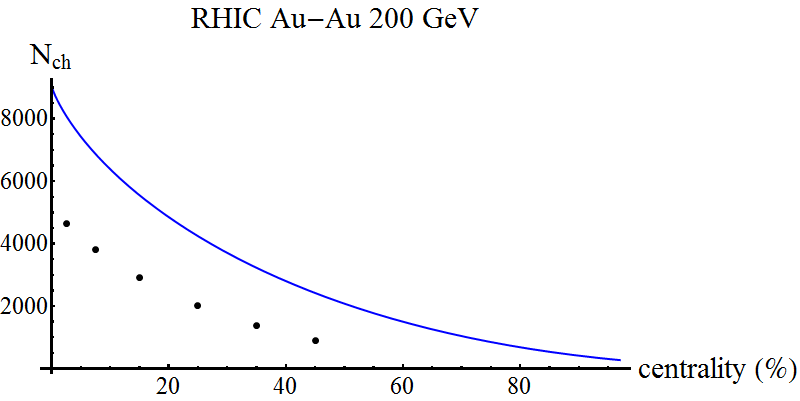}\includegraphics[width=0.48\textwidth]{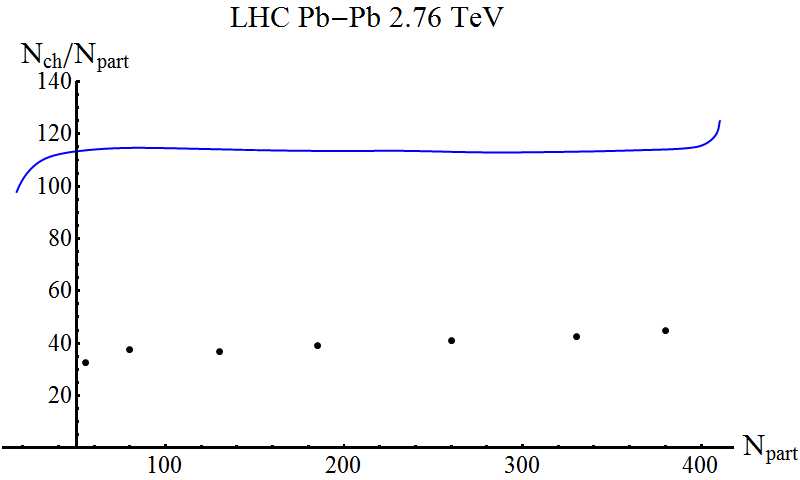}
\par\end{centering}

\centering{}\includegraphics[width=0.48\textwidth]{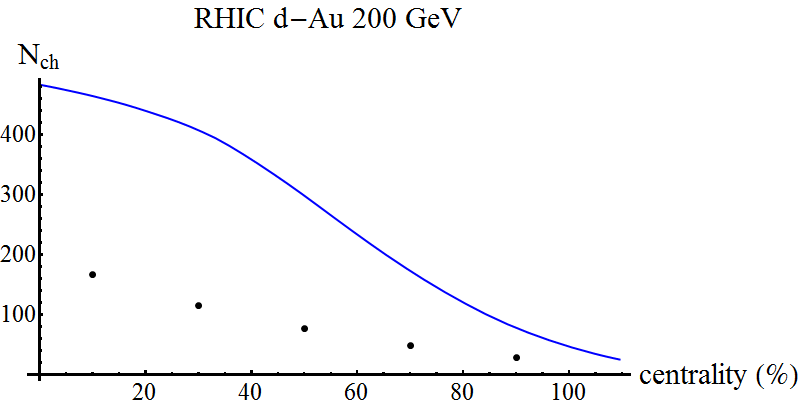}\includegraphics[width=0.48\textwidth]{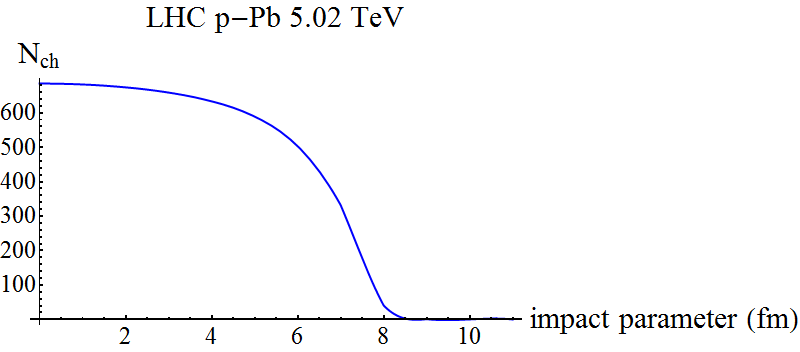}\caption{We plot the total number of charged particles for the models of figure
\ref{fig:models}. The centrality is computed using formula \ref{eq:centrality},
the number of participants by integrating the energy in the region
where the energy density of the left and right moving shock differs
by less than a factor of 15. The mismatch at RHIC is slightly surprising,
but can be traced back to mainly viscous effects. The mismatch at
LHC confirms previous work \cite{Gubser:2008pc,Lin:2009pn,Kiritsis:2011yn},
and is generally considered to be due to using infinite coupling.
Nevertheless, the qualitative trends are well reproduced and furthermore
we would like to stress that our simple model does not have any free
parameters and one can therefore optimistically say that the result
is rather close to the experimental results.\label{fig:Multiplicities} }
\end{figure}

The only other input needed is the \foreignlanguage{british}{centre-of-mass}
energy density as a function of the transverse plane, which we take
from an optical Glauber model (eqn. \ref{eq:Wood-saxon}). We then
integrate the total entropy over the transverse plane, depending on
the impact parameter $b$, to get an estimate of the total entropy
and multiplicity. Figure \ref{fig:Multiplicities} plots the resulting
multiplicities for gold-gold, deuteron-gold, lead-lead and proton-lead
collisions, starting with the energy profiles of figure \ref{fig:models}. 

Even at RHIC the multiplicity is higher than the experimental result,
which contrasts slightly with previous claims that the Landau model
could give the right particle multiplicity \cite{Steinberg:2004vy,Gubser:2008pc}.
This is largely due to a significant entropy production by viscous
effects, which were previously neglected. On the other hand we should
note that we did not take into account that a real nucleus is shaped
irregularly, and that therefore some nucleons will not collide, which
will reduce the multiplicity by perhaps $10-15\%$. For deuteron collisions
the fit is also worse, but it should be kept in mind that the uncertainty
for the deuteron shape is considerable.

Nevertheless, the agreement is much worse at LHC (consistent with
previous results \cite{Gubser:2008pc,Gubser:2009sx,Lin:2009pn,Kiritsis:2011yn}).
As the total energy of a collision is conserved this means that real
central collisions have only part of their energy deposited in the
plasma (unlike our results), or the energy per particle in real collisions
is considerably higher. As our initial conditions are unlikely to
produce less radial flow this energy per particle would most likely
be in the longitudinal direction. This would imply that real collisions
have a broader rapidity profile than the profile found in section
\ref{sec:Rapidity-profile:-Bjorken}, which can perhaps be thought
of as an effect of the infinite coupling approximation.

Although the mismatch between experimental data might seem worrisome,
we do not think this is the case. Firstly, it is reassuring that the
qualitative trends in figure \ref{fig:Multiplicities} are well respected.
Secondly, it is natural to expect that taking the infinite coupling
limit produces too much stopping. Lastly, we stress that our model
basically has no free parameters at all. Conventional models usually
have the normalisation of the initial energy density, the coupling
constant or other parameters which are more or less free, or fitted
to related experiments. \pagebreak{}In the AdS/CFT model the only
real freedom is the number of colours $N_{c}$, which has little effect
on the total multiplicity. In this light the presented mismatch is
not surprising, but an opportunity to improve the model by including
weak-coupling effects in systematic or less systematic ways, which
is partly the topic of the next chapter. 

\chapter{Thermalisation with radial flow\label{chap:Thermalisation-with-radial}}

\begin{figure}
\begin{centering}
\includegraphics[width=8cm]{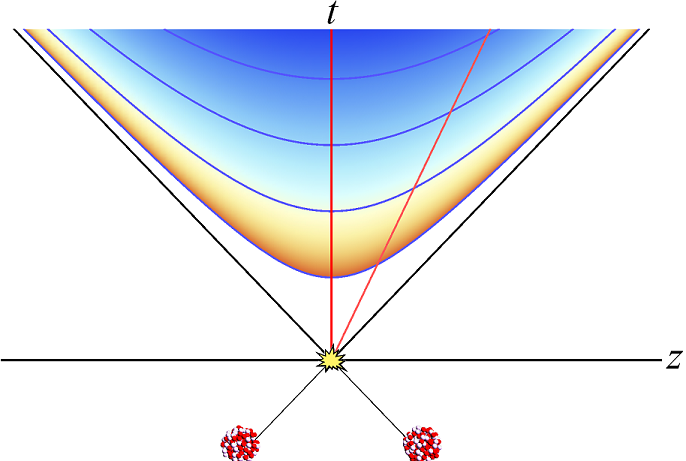}
\par\end{centering}

\caption{A simple and often used model of a heavy-ion collision was proposed
by Bjorken \cite{Bjorken:1982qr}, where he assumed that a heavy-ion
collision is approximately boost-invariant, at least near $z=0$.
This means that all physics only depends on proper time $\tau=\sqrt{t^{2}-z^{2}}$,
and hence that all physics experienced by observers, such as a temperature
field illustrated here, is independent of its frame. The two red lines
would illustrate two such frames, which indeed has equal temperatures
at equal$\tau$.\label{fig:boost-invariance}}
\end{figure}

Most AdS/CFT studies, including the previous chapters, have completely
neglected dynamics of heavy-ion collisions in the transverse plane.
This is unfortunate, since the build-up of momentum in the transverse
plane is directly related to experimentally measurable quantities.
Including transverse dynamics in the shock wave collisions of chapter
\ref{chap:Colliding-planar-shock} is possible, but numerically more
involved. Here instead, we will approximate the longitudinal dynamics
as boost-invariant (see figure \ref{fig:boost-invariance}), as is
usually done in studies of heavy-ion collisions%
\footnote{Currently there are full 3+1 dimensional simulations of heavy-ion
collisions available, see for instance \cite{Schenke:2010rr}. On
the other hand, these simulations are still approximately boost-invariant
around mid-rapidity for the full evolution.%
}. Furthermore we assume rotational symmetry, restricting ourselves
to head-on collisions, thereby keeping the numerical code effectively
2+1 dimensional.

Firstly, we present two simple initial conditions; the first starts
with a blob of energy with a diameter of approximately 14 fm in vacuum,
whereas the second has a blob of about 1 fm in a bath of half the
peak energy density. These initial states can model the overall thermalisation
of a central collision and the evolution of an initial fluctuation
in such a collision. Fluctuations are caused by the random distribution
of protons and neutrons in nuclei will lead to large fluctuations
in the (local) distribution of energy, which can is measured \cite{Timmins:2013hq,Aad:2013xma}
and interesting to study \cite{Stephanov:1999zu,Teaney:2010vd,Luzum:2013yya,Heinz:2013th,Schenke:2012hg}.
For the bulk metric we started with vacuum AdS, but adapted the near-boundary
coefficients for the energy density and the pressures according to
the Glauber model.

Thereafter a more ambitious project is presented, where the initial
data is inspired from a small-time expansion of shock wave collisions
\cite{Romatschke:2013re}. The resulting stress-tensor is then evolved
using a state-of-the-art hydrodynamics solver \cite{Luzum:2008cw},
including a final hadronic cascade code \cite{Pratt:2010jt}. This
allows a direct comparison with experimental data, which fits surprisingly
well. We will explicitly show the main advantage of this AdS/CFT approach,
being the dynamical transition from far-from-equilibrium to hydrodynamics,
which in other models usually has to be assumed.

\section{The holographic set-up with two examples\label{sec:The-holographic-set-up}}

\noindent As our coordinates in the field theory it is natural to
use proper time $\tau$ and rapidity $y$, defined by $t=\tau\cosh y$
and $z=\tau\sinh y$, and angular coordinates $\rho$ and $\theta$
in the transverse plane. The assumptions of boost-invariance and rotational
symmetry then imply that all functions are independent of $y$ and
$\theta$. In these coordinates the flat metric of the field theory
reads 
\begin{equation}
ds_{B}^{2}=-d\tau^{2}+d\rho^{2}+\rho^{2}d\theta^{2}+\tau^{2}dy^{2}.\label{eq:metric-boundary}
\end{equation}
 Given these symmetries and using generalized Eddington-Finkelstein
coordinates, we can write the dual AdS metric \ref{eq:metricEF} as
\begin{equation}
ds^{2}=d\tau(-Ad\tau+2dr+2Fd\rho)+S^{2}(e^{-B-C}dy^{2}+e^{B}d\rho^{2}+e^{C}d\theta^{2}),\label{eq:metricradial}
\end{equation}
where $A$, $B$, $C$, $S$ and $F$ are all functions of $\tau$,
$\rho$ and the AdS radial coordinate $r$. Note that the absence
of homogeneity in the transverse plane now leaves two non-trivial
functions, $B$ and $C$, as part of the spatial metric, $h_{ij}$
in eqn. \ref{eq:metricEF}, which are both needed as an initial condition.
Also, due to the flat, but non-trivial boundary metric \ref{eq:metric-boundary},
the near-boundary expansion is somewhat more complicated:{\small 
\begin{eqnarray}
A & = & r^{2}+\frac{a_{4}(\tau,\rho)}{r^{2}}+O\left(r^{-3}\right),\nonumber \\
B & = & -\frac{2}{3}\log(\tau\rho)+\frac{3r\tau(1-2r\tau)-2}{9r^{3}\tau^{3}}+\frac{b_{4}(\tau,\rho)}{r^{4}}+O\left(r^{-5}\right),\nonumber \\
C & = & -\frac{2}{3}\log(\tau/\rho^{2})+\frac{3r\tau(1-2r\tau)-2}{9r^{3}\tau^{3}}+\frac{c_{4}(\tau,\rho)}{r^{4}}+O(r^{-5}),\nonumber \\
S & = & \rho^{1/3}\frac{3r\tau(9r\tau(3r\tau(3r\tau+1)-1)+5)-10}{243r^{3}\tau^{11/3}}+O\left(r^{-4}\right),\nonumber \\
F & = & \frac{f_{4}(\tau,\rho)}{r^{2}}+O\left(r^{-3}\right),\label{eq:near-boundary-radial}
\end{eqnarray}
w}here in these expressions we again fixed the residual gauge freedom
$\xi(\tau,\,\rho)=0$. The normalisable modes of the metric, $a_{4}$,
$b_{4}$, $c_{4}$ and $f_{4}$, depend on the full bulk geometry
and cannot be determined from a near-boundary expansion. Using holographic
renormalisation we determine the stress tensor of the dual field theory
(subsection \ref{sub:Holographic-renormalisation-and}), which has
five independent non-zero components:
\begin{align}
\varepsilon & \equiv-T_{\tau}^{\tau}=-\frac{3N_{c}^{2}}{8\pi^{2}}a_{4},\nonumber \\
s & \equiv T_{\rho}^{\tau}=\frac{N_{c}^{2}}{2\pi^{2}}f_{4},\nonumber \\
p_{\rho} & \equiv T_{\rho}^{\rho}=\frac{N_{c}^{2}}{2\pi^{2}}\left(-\frac{1}{6\tau^{4}}-\frac{1}{4}a_{4}+b_{4}\right),\nonumber \\
p_{\theta} & \equiv T_{\theta}^{\theta}=\frac{N_{c}^{2}}{2\pi^{2}}\left(-\frac{1}{6\tau^{4}}-\frac{1}{4}a_{4}+c_{4}\right),\nonumber \\
p_{y} & \equiv T_{y}^{y}=\varepsilon-p_{\rho}-p_{\theta},\label{eq:stress-tensor-radial}
\end{align}
all functions of $\tau$ and $\rho$. Note that in our actual code
we fix $\xi(\tau,\,\rho)$ by the apparent horizon, such that both
\ref{eq:near-boundary-radial} and \ref{eq:stress-tensor-radial}
will change. The final field theory stress-tensor, on the other hand,
naturally does not depend on this gauge choice. The conservation of
the stress tensor implies that
\begin{eqnarray}
\partial_{\tau}a_{4} & = & -\frac{12\tau^{4}\left(\rho\left(\tau\partial_{\rho}f_{4}+a_{4}+b_{4}+c_{4}\right)+\tau\, f_{4}\right)-4\rho}{9\rho\tau^{5}},\nonumber \\
\partial_{\tau}f_{4} & = & -\frac{1}{4}\partial_{\rho}a_{4}+\partial_{\rho}b_{4}+\frac{b_{4}-c_{4}}{\rho}-\frac{f_{4}}{\tau}.\label{eq:SEconservation-radial}
\end{eqnarray}

\noindent 
\begin{figure}
\begin{centering}
\includegraphics[width=8cm]{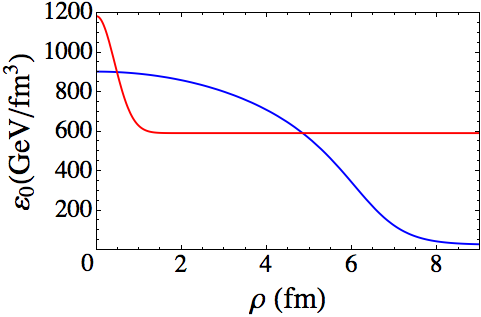} 
\par\end{centering}

\caption{The initial energy density profiles at $\tau_{in}=0.12$ fm as a function
of the distance to the origin. The blue curve models a central heavy-ion
collision; the red curve models a fluctuation in such a collision.\label{fig:The-initial-energy}}
\end{figure}

Our model basically contains two scales: the initial energy density
and the characteristic scale in the radial direction. We can, however,
make use of the scale invariance of the field theory to rescale our
coordinates such that at $\tau=0.6$ fm the energy density at the
origin equals $\varepsilon_{0}=187\,\text{GeV/fm}^{3}$. We choose
this combination to reproduce the final multiplicities of central
heavy-ion collisions at LHC \cite{Niemi:2011ix}. For the radial profile
we then consider two types of initial conditions, specified at some
small time $\tau_{in}\approx0.12$ fm %
\footnote{In principle, this provides an extra scale, but this initial time
seems small enough not to have a large influence.%
}. The first is a model for a head-on collision, where the shape of
the energy density is provided by the Glauber model, having an approximate
radius of 6.5 fm. The second energy density profile models one fluctuation
in the initial state of such a collision. We take a Gaussian of width
0.5 fm for this profile (see figure \ref{fig:The-initial-energy}).
For both initial conditions we assume that initially there is no radial
momentum, such that $f_{4}(\tau_{in},\,\rho)=0$.

Importantly, we must also specify the metric functions $B(r,\,\tau_{in},\,\rho)$
and $C(r,\,\tau_{in},\,\rho)$ on a full time-slice of the bulk AdS
geometry. These two functions, together with $a_{4}$, $f_{4}$ and
the Einstein equations, specify the complete metric and its time derivative
on a time-slice. In principle, these functions should follow from
a model describing the very first less strongly coupled stage after
the collision, such as the Glauber model or the Color Glass Condensate.
However, these models themselves contain significant uncertainties
and, more importantly, it is not clear how to map them to this gravitational
setting. Refining our initial conditions in the next section, we will
restrict ourselves here with a simple choice, where $B$ and $C$
are the same functions as in vacuum AdS, but with modified $b_{4}$
and $c_{4}$, such that the longitudinal pressure $p_{y}$ vanishes
initially:
\begin{equation}
B(r,\,\tau_{in},\,\rho)=C(r,\,\tau_{in},\,\rho)=-\frac{2}{3}\log((\tau+1/r)\rho)-\frac{1}{8}a_{4}(\tau_{in},\,\rho)/r^{4}.
\end{equation}

Having specified the initial and boundary conditions we can solve
Einstein's equations (see Appendix A) numerically %
\footnote{The numerical code, results and a movie of the radial velocity can
be downloaded at \href{http://www.staff.science.uu.nl/~schee118/}{www.staff.science.uu.nl/$\sim$schee118/}%
}, using essentially the same scheme as in chapter \ref{chap:Colliding-planar-shock}.
One difference is the required boundary conditions in the $\rho$
direction, which in this case means smoothness at the origin and at
infinity. Also the condition for the apparent horizon changes slightly,
becoming
\begin{equation}
3S^{2}\dot{S}-\partial_{z}\left(S\, F\, e{}^{-B}\right)+\frac{3}{2}e^{-B}F^{2}S'=0,\label{eq:AHradial}
\end{equation}
and lastly we used a grid in the $\rho$ direction parametrised by
$\rho=L\frac{x}{(1-x^{2})^{1/\#}}$, where $x$ ranges from 0 to 1,
and $\#$ is 20 and 4 for the nucleus and fluctuation model respectively.
Typically around 35 grid points in both directions are used, and we
choose $L=18$ or $L=2$ for the nucleus and fluctuation models respectively.
Complicating the implementation somewhat, all quantities were again
modified to be finite and non-trivial at the boundary, where in this
case also $\rho=0$ is included as a boundary.

\begin{figure}
\begin{centering}
\includegraphics[width=8cm]{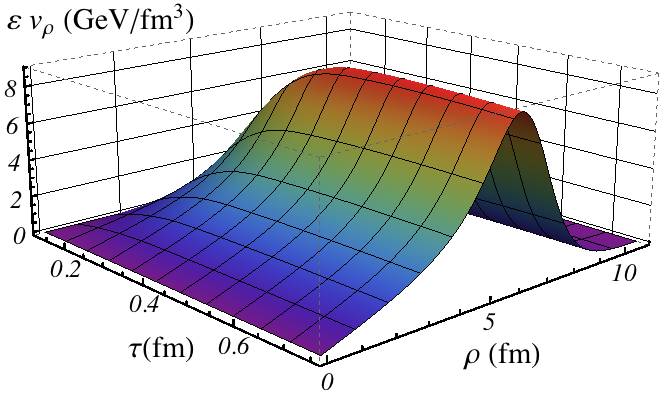}
\par\end{centering}

\caption{The radial velocity times the energy density as a function of proper
time $\tau$ and distance to the origin $\rho$ for our model of a
nucleus. Note that at late times the increasing velocity is almost
exactly compensated by the decreasing energy density (which is due
to the longitudinal expansion). The slope at the origin at the end
of our simulation equals 0.66 GeV/$\text{fm}^{4}$.\label{fig:momentum-flow}}
\end{figure}

\begin{figure}
\begin{centering}
\includegraphics[width=8cm]{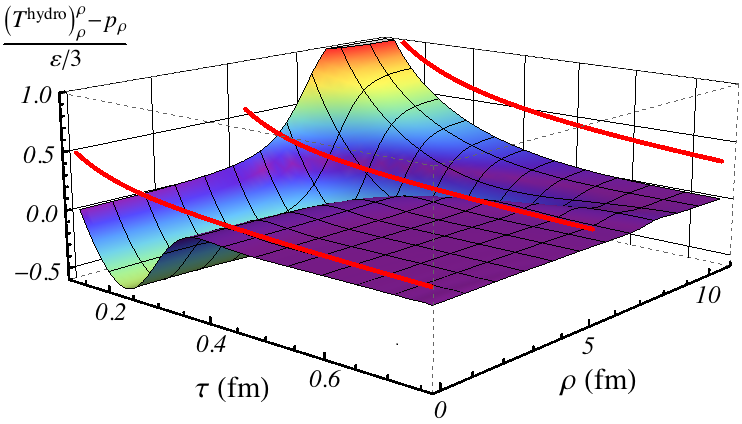}
\par\end{centering}

\caption{The difference between the full non-equilibrium $p_{\rho}$ and the
pressure given by first order hydrodynamics. Although hydrodynamics
applies very quickly, the viscous contribution is still large (shown
by a red lines). The relatively high values for $\rho>7$ fm are a
consequence of the very small energy density. For the model of a fluctuation
the graph is similar, with equally quick thermalisation.\label{fig:hydro}}
\end{figure}

\begin{figure}
\begin{centering}
\begin{tabular}{cc}
\includegraphics[width=0.48\textwidth]{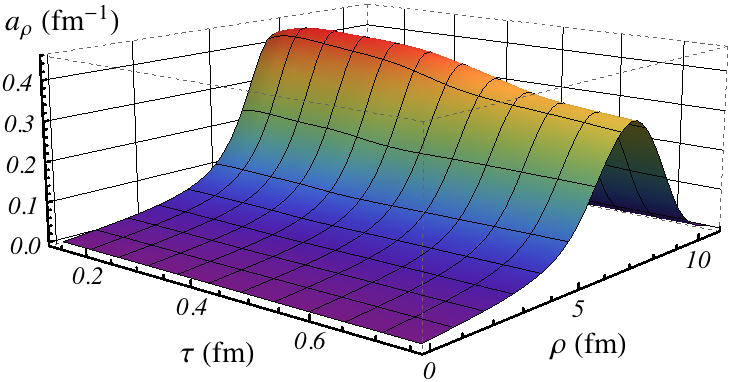} & \includegraphics[width=0.48\textwidth]{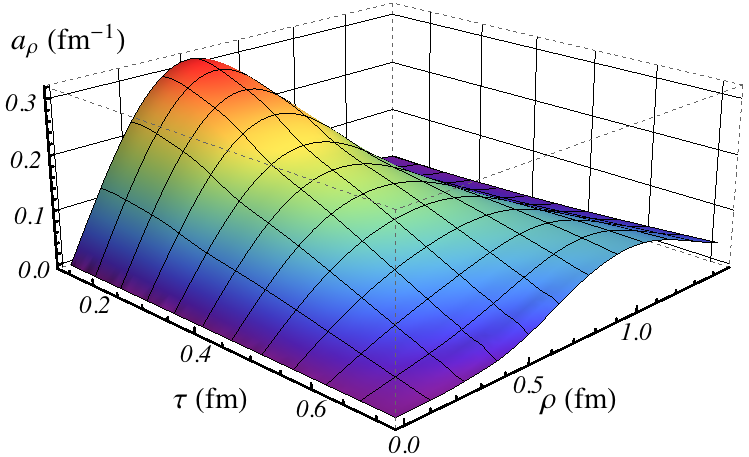} \tabularnewline
(a)  & (b)\tabularnewline
\end{tabular}
\par\end{centering}

\caption{ (a) The radial acceleration of our nucleus model. The acceleration
decreases after some time, which is mainly a consequence of the decrease
in radial pressure, due to the isotropisation. Thereafter the acceleration
is quite steady and mainly localised near the boundary of the nucleus.
(b) The radial acceleration of our fluctuation model. Since the bump
of energy is much smaller one can clearly see the spreading out and
the decrease in acceleration. As will also be clear from figure \ref{fig:velocity-gradient},
this model reaches a lower radial speed than the model for the nucleus.
\label{fig:acceleration}}
\end{figure}

After determining the stress tensor one can extract the radial velocity,
defined again by the boost after which there is no momentum flow.
Figure \ref{fig:momentum-flow} shows this velocity times the energy
density, which gives a good measure of the momentum flow. The radial
velocity, together with the stress tensor in the local rest frame,
can be used to compute the stress tensor according to hydrodynamics.
Although initially there will not be local equilibrium, at late times
a hydrodynamic expansion is expected to be valid. It is therefore
interesting to compare the actual pressures with the pressures which
follow from a hydrodynamic expansion \cite{Baier:2007ix,Janik:2005zt},
analogously to subsection \ref{sub:Hydrodynamisation}.

In figure \ref{fig:hydro} we plot the difference of $p_{\rho}$ and
the corresponding first order hydrodynamic prediction of our model
of a nucleus. The stress tensor is excellently described by hydrodynamics
as soon as $\tau=0.35$ fm. At the border of our nucleus this is slightly
subtler, since the stress tensor is rather small there, and it becomes
comparable to our regulator energy density. We therefore cannot say
too much about this, but the agreement with hydrodynamics is also
there encouraging. We note that in previous studies \cite{Heller:2011ju,Heller:2012km}
somewhat larger thermalisation times (with respect to the local temperature)
were found, so we expect more exotic initial conditions in our bulk
AdS to give somewhat later thermalisation.

In figure \ref{fig:acceleration}b we plot the radial acceleration
of our model of a fluctuation. We notice the acceleration already
decreases considerably during our simulation, in contrast with the
model for the nucleus (figure \ref{fig:acceleration}a). Also, the
acceleration increases rapidly near the origin, whereas for the nucleus
it is rather narrowly peaked near the boundary of the nucleus. This
means that fluctuations are expected to spread out rather quickly.
Perhaps surprisingly, also the stress tensor for the fluctuation is
governed by hydrodynamics within 0.35 fm.

The main motivation for the two examples above is to provide a description
of the far-from-equilibrium stage of heavy-ion collisions, including
non-trivial dynamics in the transverse plane. While we kept rotational
symmetry in the transverse plane, we believe our study can be used
more generally. One reason for this is an old result in asymptotically
flat space \cite{Price:1994pm}, recently studied in asymptotically
AdS (\cite{Heller:2012km} and section \ref{sec:Numerics-and-a}),
that during black hole formation gravity can be well approximated
by linearising around the final state. We therefore believe that an
initial energy profile with many fluctuations could be well approximated
by superposing the result of our fluctuation presented above.

\begin{figure}
\begin{centering}
\includegraphics[width=8cm]{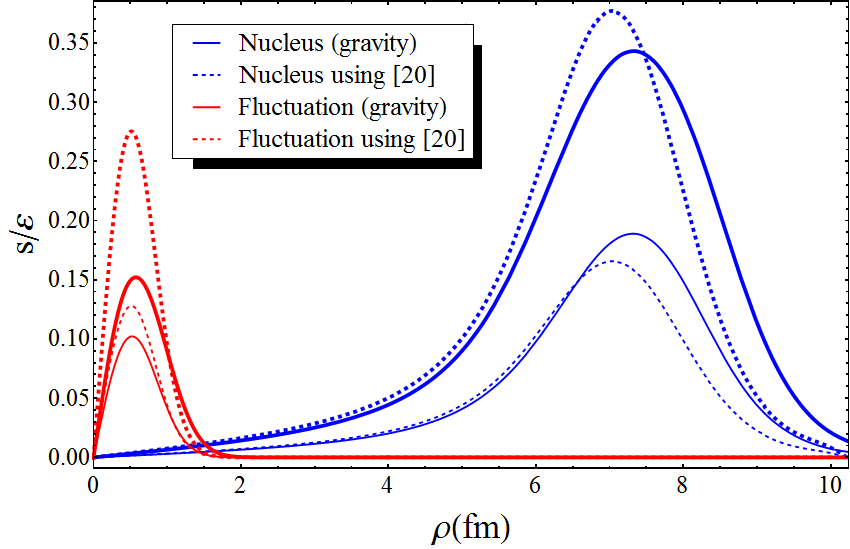} 
\par\end{centering}

\caption{Here we plot the momentum flow $s$ divided by the energy density,
at time $\tau=0.75$ fm (thick lines) and $\tau=0.4$ fm (thin lines),
as a function of $\rho$. The plots compare our gravitational results
with formula \ref{eq:velocitygradient}, which was found in \cite{Vredevoogd:2008id}.
The two results are remarkably similar, especially at earlier times,
and considering the dynamics takes place at very different scales.
We suggest formula \ref{eq:velocitygradient} as an initial condition
for non-symmetric hydrodynamic simulations, where a simulation should
start at about $\tau=0.4$ fm if fluctuations are present. \label{fig:velocity-gradient}}
\end{figure}

Also, it should be possible to use our results for non-central collisions.
This can be seen by comparing with a formula for universal initial
flow \cite{Vredevoogd:2008id}. There, they assume that the anisotropy
is independent of $\rho$, the transverse pressures are equal and
that the velocity is approximately linear in time. Without using any
hydrodynamics, they used the conservation of the stress tensor to
arrive at the following local formula for the transverse momentum
of the stress tensor: 
\begin{equation}
\vec{s}/\varepsilon\approx-\frac{\vec{\nabla}_{\perp}\varepsilon_{0}}{2\varepsilon_{0}}(\tau-\tau_{in}),\label{eq:velocitygradient}
\end{equation}
 where $\varepsilon_{0}$ is the initial energy density. This formula
(see fig. \ref{fig:velocity-gradient}) works remarkably well at early
times and also later on for the nucleus model. At later times the
transverse velocities of fluctuations are smaller, which is due to
the decreasing acceleration (displayed in figure \ref{fig:acceleration}b).
This result therefore increases confidence in the result of \cite{Vredevoogd:2008id},
which can be used in less symmetric situations. When including fluctuations,
however, one should hydrodynamics as soon as $\tau=0.4$ fm to get
more accurate results.

\noindent 

\section[A dynamical simulation of central nuclear collisions]{A fully dynamical simulation of central nuclear collisions}

In this section we attempt to obtain initial conditions by including
the far-from-equilibrium stage obtained from colliding shock waves
in AdS. After the matter has equilibrated, we match the AdS/CFT results
onto a standard viscous hydrodynamics code which, once the matter
has cooled below the QCD phase transition temperature $T_{c}$, is
itself matched onto a standard hadronic cascade code, thereby achieving
a fully dynamical simulation of a boost-invariant heavy-ion collision.

As in eqn. \ref{eq:Wood-saxon} and section \ref{sec:The-holographic-set-up},
the main physical input for our simulation will be the energy density
of a highly boosted and Lorentz contracted nucleus, $T_{tt}=\delta(t+z)T_{A}(\rho)$,
with the {}``thickness function'' 
\begin{equation}
T_{A}(\rho)=\epsilon_{0}\int_{-\infty}^{\infty}dz\left[1+e^{(\sqrt{\rho^{2}+z^{2}}-R)/a}\right]^{-1}\,,\label{Fermi}
\end{equation}
 where $R=6.62$ fm, $a=0.546$ fm for a $^{208}{\rm Pb}$ nucleus\cite{Alver:2008aq}.
The normalization $\epsilon_{0}$ is a measure of the energy of the
nucleus, and given the simplicity of our model, we will use this constant
to match the experimentally observed number of particles ({}``multiplicity'',
$dN/dY$). It is noteworthy that analogously to subsection \ref{sub:Shock-profiles-and}
we could match $\epsilon_{0}$ to energies used at LHC. This, however,
would lead to $\epsilon_{0}$ being more than a hundred times bigger
than what we will find, which is perhaps unsurprising in our simple
model with boost-invariance. 

Here we again describe a relativistic nucleus as a gravitational shockwave
in AdS, whereby the stress-energy tensor of a nucleus can be exactly
matched to the thickness function. For a head-on (central) collision
this shockwave collision has been written down and solved near the
boundary of AdS in Ref.~\cite{Romatschke:2013re}, resulting in the
stress-energy tensor at early times to leading order in $t$ that
reads
\begin{eqnarray}
e=2T_{A}^{2}(\rho)\tau^{2}\,,\ u^{\rho}=-\frac{T_{A}^{\prime}(\rho)}{3T_{A}(\rho)}\tau\,,\ \frac{P_{L}}{P_{T}}=-\frac{3}{2},\quad\label{earlytimeTmunu}
\end{eqnarray}
where in the local rest frame $T_{\nu}^{\mu}={\rm diag}(-e,\, P_{T},\, P_{T},\, P_{L})$
and $u^{\mu}$ is defined as in \ref{eq:hydro-constituive}. The velocity
dependence and pressure anisotropy are consistent with our numerical
computations (figure \ref{fig:2Dhydro}) and the universal flow formula
\ref{eq:velocitygradient}. The early time result (\ref{earlytimeTmunu})
fixes the first few near-boundary series coefficients of $B$ and
$C$, but does not fix the metric functions deep in the bulk, leading
to an unstable time evolution. In order to have a stable time evolution,
we introduce a function with one bulk parameter $\sigma$ to extend
the metric functions to arbitrary $r$, specifically choosing 
\begin{equation}
B(r,\tau,\rho)\rightarrow B_{0}(r,\tau,\rho)+\sum_{i=0}^{6}\frac{b_{i}(\tau,\rho)r^{-i}}{1+\sigma^{7}r^{-7}}\,,\label{sigma}
\end{equation}
and analogously for $C$. Here $B_{0}(r,\,\tau,\,\rho)=-\frac{2}{3}\log((\tau+1/r)\rho)$
would give vacuum AdS, and the $b_{i}(\tau,\rho)$ are choosen such
that the stress tensor \ref{eq:stress-tensor-radial} equals eqn.~\ref{earlytimeTmunu}.
Having $B$, $C$, $a_{4}$ and $f_{4}$ at a time $\tau_{{\rm init}}$
and choosing a value for $\sigma$, the future metric is completely
determined by the same method as in section \ref{sec:The-holographic-set-up}.

From the metric we again extract the full $T^{\mu\nu}$ (eqn. \ref{eq:stress-tensor-radial})
and in particular observe the transition from early-time, far-from
equilibrium dynamics to a fluid described by viscous hydrodynamics.
At some value of proper time $\tau_{{\rm hydro}}$, we stop the evolution
using Einstein equations and extract $e,\, u^{\mu},\,\pi^{\mu\nu}$
from eqn. \ref{eq:hydro-constituive}. These functions provide the
initial conditions for the well-tested relativistic viscous hydrodynamic
code vh2 (version 1.0) \cite{Luzum:2008cw}, which uses an equation
of state (EoS) inspired by lattice QCD and has, for simplicity, $\eta/s=\frac{1}{4\pi}$.
Since this EoS differs from the conformal EoS of our AdS model there
will be a discontinuity in the pressure. At high temperatures, however,
QCD is approximately conformal and in our simulations the discontinuity
at the center was never more than 15\%.

The hydrodynamic code simulates the evolution from $\tau=\tau_{{\rm hydro}}$
until the last fluid cell has cooled down below $T_{{\rm sw}}=0.17$
GeV. The hydrodynamic variables along the hypersurface defined by
$T=T_{{\rm sw}}$ are stored and converted into particle spectra using
the technique from Ref.~\cite{Pratt:2010jt}. The subsequent particle
scattering is treated using a hadron cascade \cite{Novak:2013bqa}
for resonances with masses up to $2.2$ GeV by simulating $500$ Monte-Carlo
generated events. Once the particles have stopped interacting, and
particles unstable under the strong force have decayed, light particle
transverse momentum spectra are analyzed and can be compared to data.

From the hydrodynamic evolution onward our model uses techniques and
parameters which are fairly standard. The initial conditions for hydrodynamics,
however, are now determined using a far-from-equilibrium evolution.
We modeled this phase as a strongly coupled CFT, described by gravity
in AdS. This introduces new parameters and functions, namely the initialization
time $\tau_{{\rm init}}$, the normalization $\epsilon_{0}$, the
bulk function with parameter $\sigma$ and the AdS/hydro switching
time $\tau_{{\rm hydro}}$. We will explore the effects of changing
these parameters below.

\begin{figure}[t]
\centering{}\includegraphics[width=10cm]{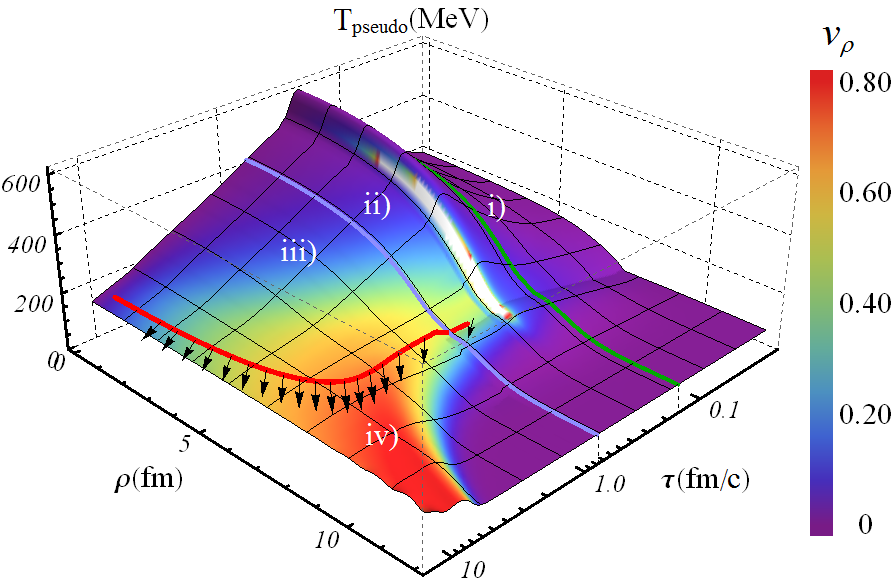} \caption{Assuming (\ref{eq:hydro-constituive}) applied, a {}``pseudo'' temperature
(defined by using Eq.~(\ref{eq:hydro-constituive}) with $e=e(T_{{\rm pseudo}})$)
and radial velocity $v_{\rho}=u_{\rho}/u_{\tau}$ are extracted for
a representative simulation. The plot illustrates four physical tools
used: i) early time expansion, ii) numerical AdS evolution, iii) viscous
hydrodynamics until $T=0.17\,\text{GeV}$, iv) kinetic theory after
conversion into particles (indicated by arrows). The (white) region
close to $\tau\sim0.2$ fm/c, $\rho\sim5$ fm/c indicates a far-from-equilibrium
domain where a local rest frame cannot be found.\label{fig:Temp1} }
\end{figure}

\subsection{Resulting particle spectra }

\begin{figure}[t]
\centering{}\includegraphics[width=8cm]{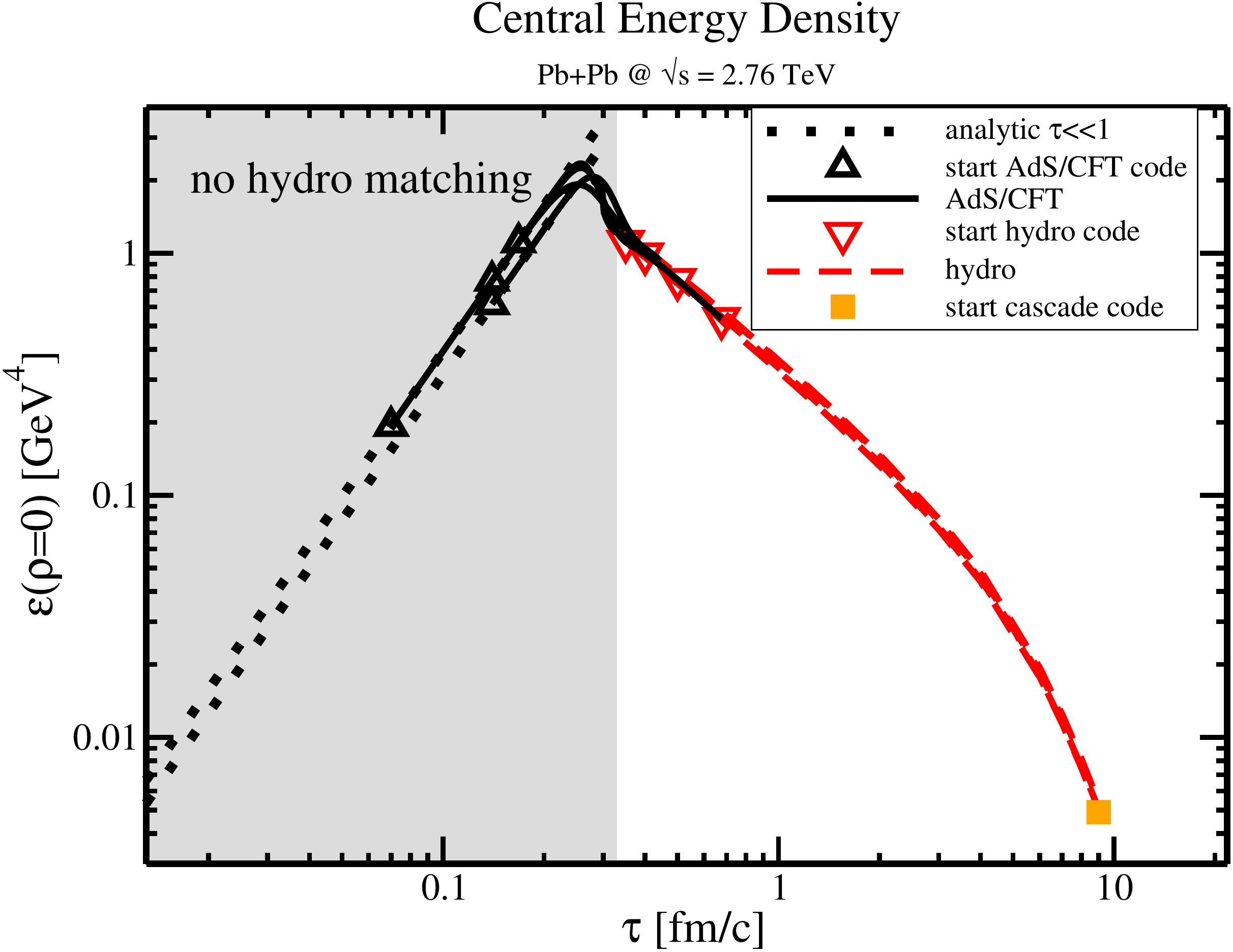} \caption{ Time evolution of the energy density at the center of the fireball
for different values of the regulator $\sigma$, different AdS/CFT
starting times $\tau_{{\rm init}}$ and different AdS/hydro switching
times $\tau_{{\rm hydro}}$. Shown are the analytic early time result
(dotted), the numerical AdS/CFT evolution (full lines), the numerical
hydro evolution (dashed lines) and the conversion point to the hadron
cascade. For $\tau\lesssim0.35$ fm/c, no sensible matching from AdS/CFT
to a hydrodynamic evolution is possible ({}``no hydro matching''),
cf.~Fig.~\ref{fig:aniso1}.\label{fig:ed1} }
\end{figure}

\begin{figure}[t]
\centering{}\includegraphics[width=7cm]{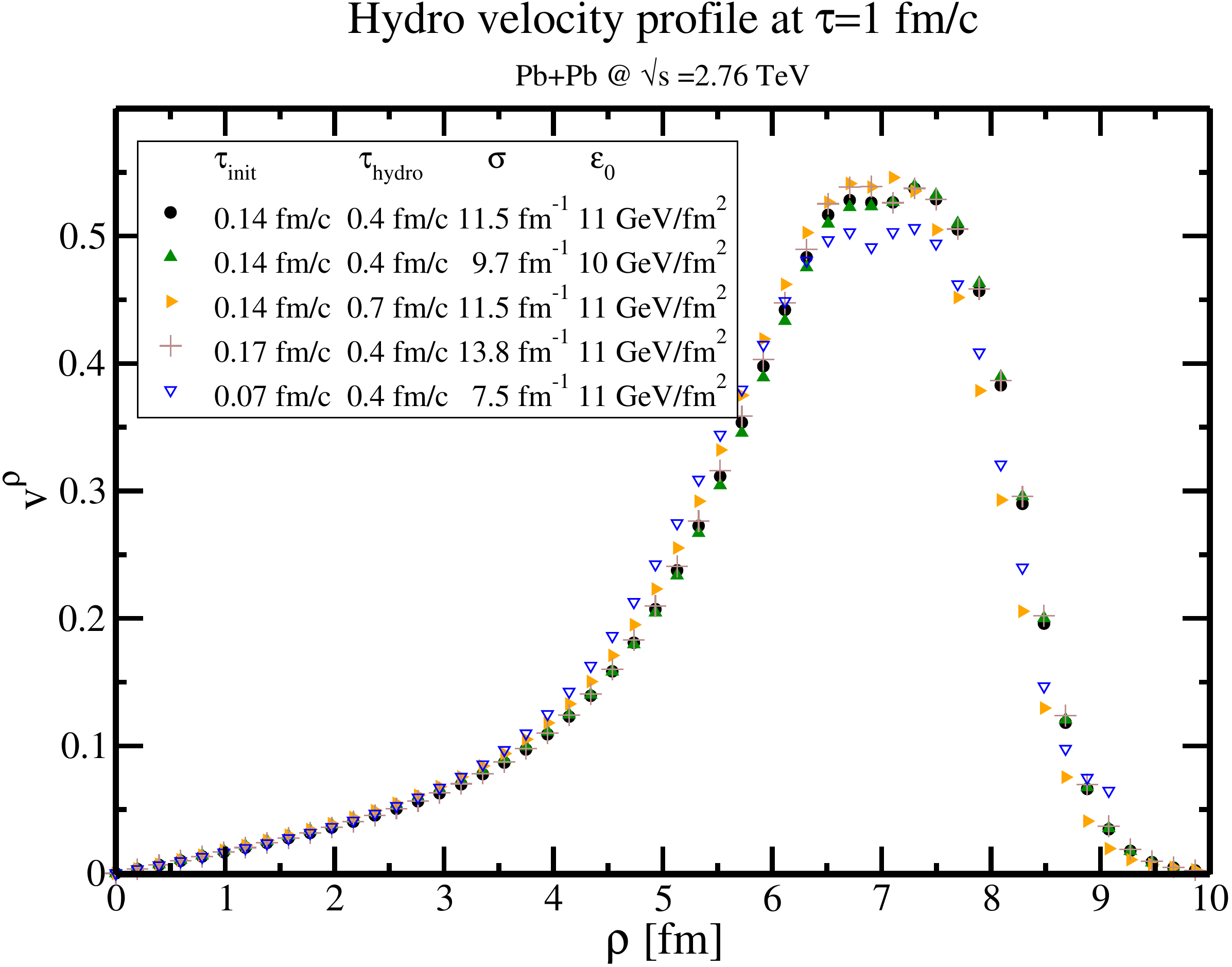} \caption{Radial velocity profile at $\tau=1$ fm/c for different AdS/CFT starting
times $\tau_{{\rm init}}$, different AdS/hydro switching times $\tau_{{\rm hydro}}>0.35$
and different values for the regulator $\sigma$ (in a.u.). One observes
that when normalized to the same final multiplicity, all these choices
lead to similar velocity profiles.\label{fig:vel1} }
\end{figure}

\begin{figure}[t]
\centering{}\includegraphics[width=7cm]{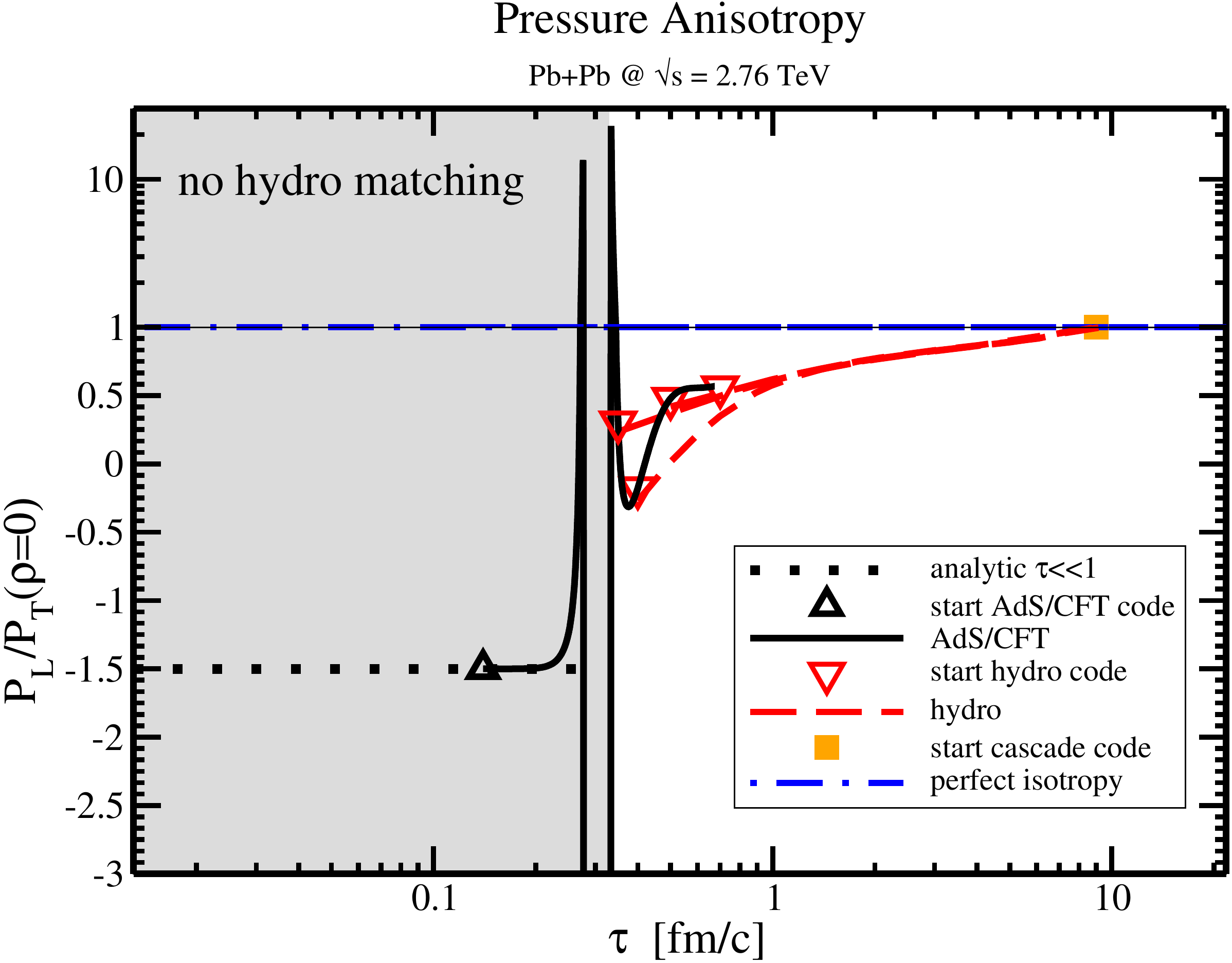} \caption{Time evolution of the pressure anisotropy $P_{L}/P_{T}$ at the center
of the fireball for single values of $\sigma,\tau_{{\rm init}}$ but
multiple AdS/hydro switching times $\tau_{{\rm hydro}}$. For $\tau\lesssim0.35$
fm/c, the pressure anisotropy is wildly varying, prohibiting a sensible
matching to hydrodynamics. At later times, matching to hydrodynamics
can be performed (indicated by triangles down) and leads to approximately
universal late-time evolution until freeze-out to the hadron cascade
(indicated by square). \label{fig:aniso1} }
\end{figure}

\begin{figure}[t]
\centering{}\includegraphics[width=8cm]{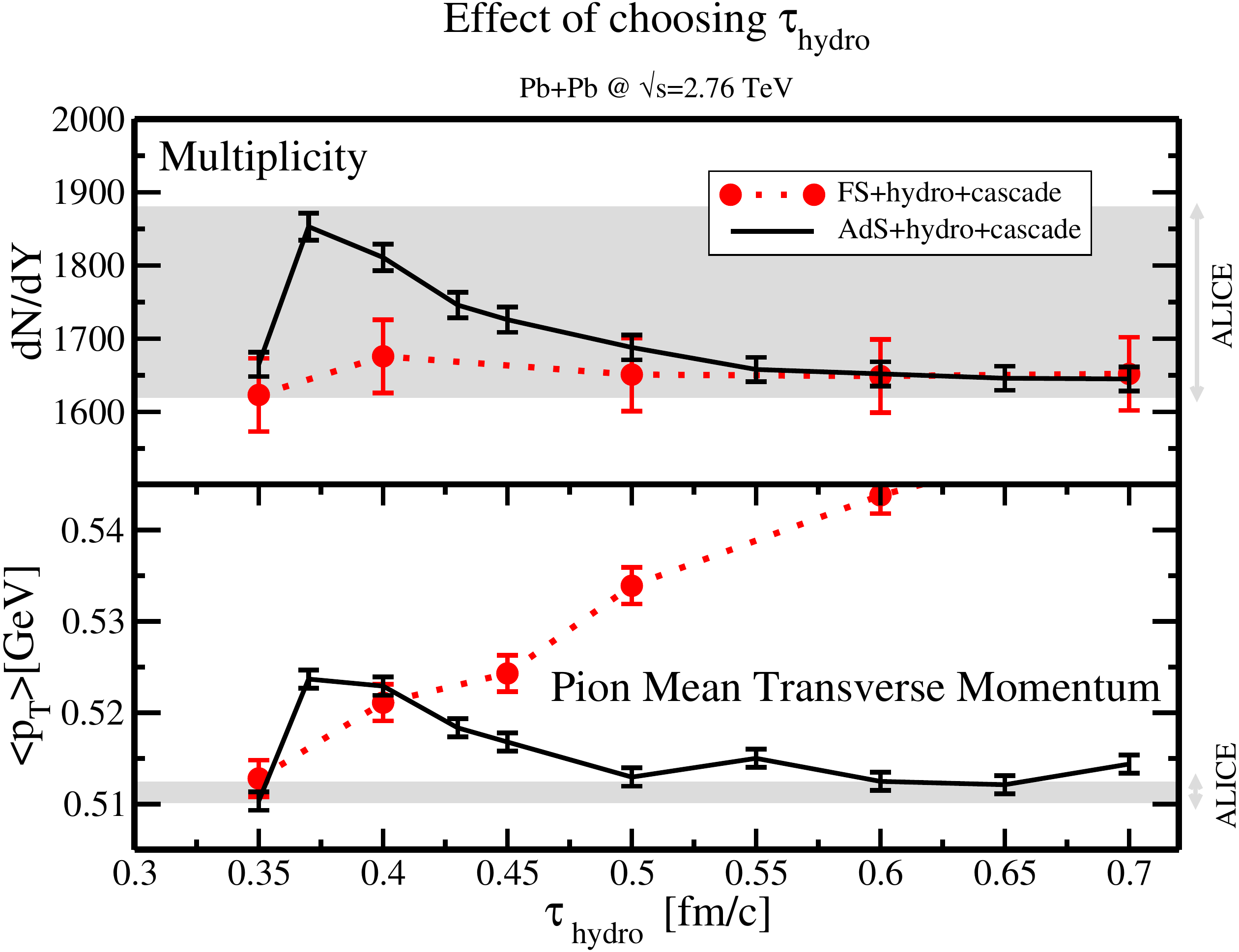} \caption{Final $dN/dY$ and pion $\langle p_{T}\rangle$ as a function of the
hydro switching time $\tau_{{\rm hydro}}$ for a single value of $\sigma,\tau_{{\rm init}}$
(AdS+hydro+cascade) compared to experimental data (ALICE \cite{Abelev:2012wca}).
AdS results seem to be independent of $\tau_{{\rm hydro}}$ provided
that $\tau_{{\rm hydro}}>0.5$ fm/c. By contrast, results for FS models
with $\tau_{{\rm init}}=0.05$ fm/c exhibit strong $\tau_{{\rm hydro}}$
dependence. Error bars correspond to accumulated numerical error.\label{fig:multi} }
\end{figure}

\begin{figure}[t]
\centering{}\includegraphics[width=8cm]{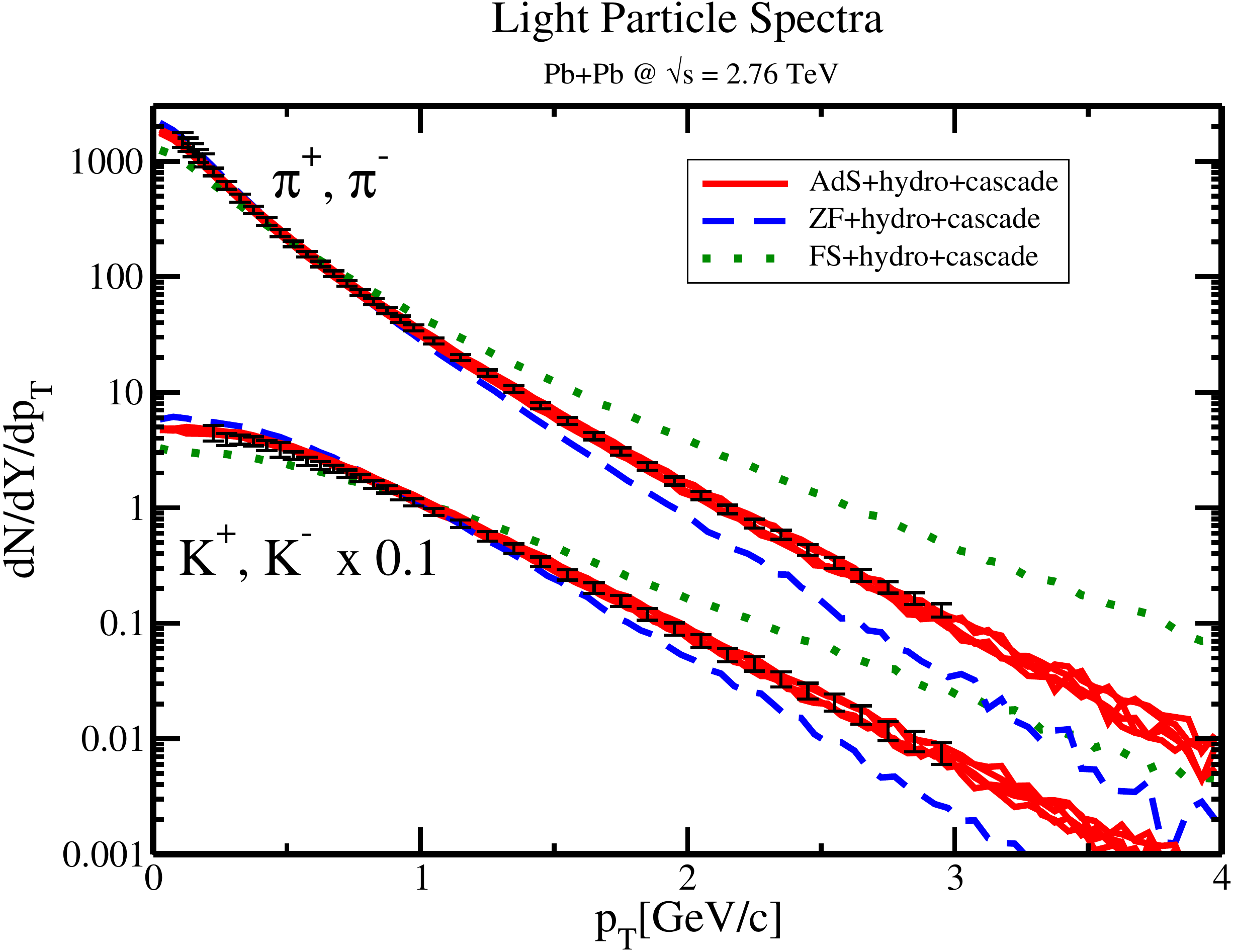} \caption{Pion and Kaon momentum spectra for 0-5\% most central $Pb+Pb$ collisions
at $\sqrt{s}=2.76$ TeV. Experimental measurements (ALICE \cite{Abelev:2012wca})
are compared to our AdS+hydro+cascade model (lines correspond to different
choices of $\epsilon_{0},\sigma$), the ZF model initialized from
eqn.~(\ref{Fermi}) at $\tau_{{\rm hydro}}=1$ fm/c, and the FS model
with $\tau_{{\rm hydro}}=0.5$ fm/c. \label{fig:spectra} }
\end{figure}

\noindent Matching our numerical relativity, viscous hydrodynamics
and hadron cascade simulations onto one another we obtain the time-evolution
of the energy density for $Pb-Pb$ collisions at $\sqrt{s}=2.76$
TeV (see Fig.~\ref{fig:Temp1}). The results depend on our choices
of $\epsilon_{0},\tau_{{\rm init}}$ and $\sigma$, which are all
parameters that in principle could be fixed by a more complete calculation.
Requiring that for constant $\tau_{{\rm init}}$ and $\sigma$ our
$dN/dY$ matches the experimental value fixes $\epsilon_{0}$. Different
combinations of $\tau_{{\rm init}}$ and $\sigma$ will have similar
late-time energy densities (cf. Fig.~\ref{fig:ed1}), but originate
from different early-time histories and the pre-equilibrium evolution
reported in Figs.~\ref{fig:ed1},\ref{fig:aniso1} should be considered
uncertain. However, we find that for fixed $dN/dY$ also the late
time radial flow velocity and final light hadron spectra are essentially
unaffected by our choice of $\tau_{{\rm init}}$ or $\sigma$ (see
Figs.~\ref{fig:ed1}--\ref{fig:spectra}).

This is evident when comparing the resulting hydrodynamic radial velocity
at $\tau=1$ fm/c shown in Fig.~\ref{fig:vel1}. Different values
for $\tau_{{\rm init}},\,\tau_{{\rm hydro}}$ and $\sigma$ collapse
onto an approximately universal velocity profile. Because the subsequent
evolution follows hydrodynamics, this is also true for the velocity
profile for all later times. We therefore expect our late time results
to be robust.

One is not completely free in specifying $\tau_{{\rm init}}$ or $\sigma$.
The coordinate singularity at $\tau=0$ prevents going to very early
times, while one naturally has to start the AdS/CFT code long before
the time hydrodynamics is expected to be applicable. In practice we
found $0.07\leq\tau_{{\rm init}}({\rm fm)\leq0.17}$ to be a good
range. For $\sigma$ one has to make sure the AdS spacetime is sufficiently
regular to allow for a stable evolution. In practice, we found $7.5\leq\sigma({\rm fm}^{-1})\leq14$
to work well.

The time evolution of the pressure anisotropy shown in Fig.~\ref{fig:aniso1}
indicates a strongly varying, and occasionally negative, longitudinal
pressure prohibiting any hydrodynamic description for $\tau<0.35$
fm/c. Besides the strongly varying anisotropy, we also typically encounter
a closed region in space-time where the system is so far from equilibrium
that a local rest frame does not seem to exist, which we plan to report
on in future work. In principle, one could choose any value of $\tau_{{\rm hydro}}>0.35$
fm/c; however, switching at very late times $\tau_{{\rm hydro}}\gg1$
fm/c is not recommended because of the prohibitive computational cost
of the numerical relativity code and the fact that at later times
the system has cooled down to temperatures where the QCD EoS is no
longer close to the conformal EoS in the AdS/CFT code. For $\tau>0.35$
fm/c we can attempt to match the pre-equilibrium phase onto viscous
hydrodynamics at $\tau=\tau_{{\rm hydro}}$, which surprisingly seems
to lead to roughly similar final results even when $P_{L}\simeq0$
(cf. Fig.~\ref{fig:aniso1}). A more refined result can be gained
by considering the dependence of the final light hadron spectra on
the choice $\tau_{{\rm hydro}}$ discussed below.

In order to compare our thermalising strongly coupled model we have
considered two other (extreme) possibilities for the initial stage
before $\tau_{hydro}$. The first has $P_{L}=0$, which gives zero
coupling boost-invariant free streaming (FS), whereas the second has
$P_{T}=0$, which hence has zero pre-equilibrium radial flow (ZF).
These models never lead to thermalisation, but operationally one can
switch to hydrodynamics at some time $\tau_{{\rm hydro}}$.

Fig. \ref{fig:multi} shows the dependence of the final multiplicity
and pion mean transverse momentum on the hydro switching time $\tau_{{\rm hydro}}$
for the AdS and FS models. For the final stage hadron cascade only
hydro information for $\tau>1$ fm/c is used. Fig. \ref{fig:multi}
indicates that final $dN/dY,\langle p_{T}\rangle$ in our AdS model
are constant, provided one switches to hydrodynamics after the far-from-equilibrium
regime has ended (at about $\tau\approx0.5$ fm/c). This suggests
that our model reaches hydrodynamics dynamically and hence is insensitive
to the choice of $\tau_{{\rm hydro}}$. 

In contrast, for the FS and ZF models $\langle p_{T}\rangle$ depends
on $\tau_{hydro}$, which hence only reproduces the data for a specific
value, not following from a theoretical calculation. So while measured
particle spectra do not rule out these models, the AdS model has the
conceptual advantage of naturally leading to hydrodynamics, thereby
making the model more constrained.

In Fig.~\ref{fig:spectra} we show the results for the final pion
and kaon transverse momentum spectra in comparison to data for central
$Pb+Pb$ collisions at $\sqrt{s}=2.76$ TeV from the ALICE experiment
\cite{Abelev:2012wca}. The integral over the momentum spectra corresponds
to the total multiplicity which we fixed by hand. However, Fig.~\ref{fig:spectra}
shows that our AdS+hydro+cascade model matches the shape of the experimental
data almost perfectly up to the highest transverse energies measured,
independent of our choices for $\tau_{{\rm init}},\,\tau_{{\rm hydro}}$
and $\sigma$.

In this section we have presented a fully dynamical multi-physics
simulation of central nuclear collision at LHC energies. This simulation
includes a simulation of the equilibration of the bulk of the system
using the AdS/CFT correspondence. When normalized to the same multiplicity,
our framework is approximately insensitive to the AdS initialization
time $\tau_{{\rm init}}$, the choice of bulk parameter $\sigma$
and the AdS/hydro switching time $\tau_{{\rm hydro}}$, provided the
switching occurs later than $\sim0.5$ fm/c. This is in constrast
to non-thermalising models such as FS+hydro+cascade where results
depend on choices for $\tau_{{\rm init}},\,\tau_{{\rm hydro}}$.

Because of the dynamical treatment of the pre-equilibrium stage and
the insensitivity to our free parameters, our model is more constrained
than a standard hydro+cascade model. In particular, we find that the
transverse pressure is consistently higher than the longitudinal pressure,
during most of the evolution (Fig.~\ref{fig:aniso1}). Very encouragingly,
the model turns out to have light particle spectra in excellent agreement
with experimental data for $Pb+Pb$ collisions at $\sqrt{s_{NN}}=2.76$
TeV.

We regard this work as the first step towards a truly realistic simulation
of high energy nuclear collisions. Many aspects of our work can and
should be improved in future work. For instance, we plan to do away
with the bulk parameter $\sigma$ by simulating the full shock-wave
collision process (cf.~chapter \ref{chap:Colliding-planar-shock})
and simulate event-by-event non-central collisions by employing a
linearised scheme. The latter would allow comparison of both $\langle p_{T}\rangle$
and angular correlations, thereby giving an even more meaningful comparison
of our model with experimental nucleus-nucleus or proton-nucleus data
\cite{Abelev:2013haa,Abelev:2008ab,Sickles:2013yna,Kolb:2003dz}.

\chapter{Conclusion and discussion}

In the preceding chapters three main lessons were learnt
\begin{enumerate}
\item A homogeneous strongly coupled conformal plasma starting in general
far-from-equilibrium states will thermalise very quickly, meaning
that hydrodynamics is applicable within a time $\sim1/T$, with $T$
the local temperature. Furthermore, linearising the initial state
around the final state can provide a much simpler and relatively accurate
description.
\item The longitudinal dynamics in a collision of gravitational shock waves
provides rich physics; if the shocks are wide relative to the energy
content, they will thermalise during the collision, come to a stop,
and explode hydrodynamically. If they are thin, however, they do not
have time to thermalise, but they pass right through each other, leaving
behind a plasma later on with a characteristic rapidity profile (section
\ref{sec:Rapidity-profile:-Bjorken}). In the latter case this profile
is insensitive to the microstructure of the shock wave: the shock
waves act coherently. In particular, this implies a universal shape
of the rapidity profile at high energies, which has experimental implications
outlined in this thesis.
\item Assuming boost-invariance in the longitudinal direction, it was possible
to realistically model a central heavy-ion collision in AdS/CFT. It
was shown explicitly that AdS/CFT continuously links far-from-equilibrium
dynamics with hydrodynamics, which is generally hard in other approaches.
The resulting particle spectra match LHC data well, perhaps unreasonably
well.
\end{enumerate}
Alongside these lessons special attention was paid to clarify the
set-up used for these problems in numerical relativity. They turn
out to be much simpler than typical problems in numerical relativity,
such as the merger of black hole. After the relatively straightforward
addition of electromagnetic and scalar fields this method is expected
to be useful in far more general settings within AdS/CFT, such as
in the context of holographic superconductors or other holographic
condensed matter studies \cite{Chesler:2013qla,Hartnoll:2008vx,Horowitz:2013jaa}.

\section{A comparison with experiments?\label{sec:A-comparison-with}}

In this thesis we have made an effort to compare our results with
experimental data, in particular high-lighting a major difference
in the rapidity distribution of proton-nucleus collisions computed
using perturbative QCD or our strongly coupled method (subsection
\ref{sub:Consequences-for-HIC}). Nevertheless it is clear that all
computations have been done in an oversimplified setting. We tried
to approximate real-world asymptotically free QCD with a conformal
theory, $\mathcal{N}=4$ SYM, notably at infinitely strong coupling
and with an infinite number of colours. Only gravity was considered,
whereas other forces may be relevant when modelling heavy-ion collisions.
Note, however, that gravity is the only force which gets stronger
with increasing energy classically. It is therefore expected that
at high energy heavy-ion collisions gravity is the dominant force.
Furthermore, lattice QCD computations have shown that QCD behaves
very similarly when increasing the number of colours, thereby validating
the assumption of a large number of colours \cite{Panero:2009tv}.

There is one other good argument to believe a relatively simple model
could capture many features of real collisions, which is the fortunate
separation of three scales: $1/T\ll r_{nucleon}\ll r_{nucleus}$,
with $T\sim1/(0.1\text{fm})$ the local temperature at thermalisation,
$r_{nucleon}\approx0.9$ fm the radius of a nucleon and $r_{nucleus}\approx6.7$
fm the radius of a nucleus. This separation can be very helpful in
a realistic dual of QCD; there one may imagine to model each nucleon%
\footnote{Here we model a nucleus as being simply the combination of many nucleons,
which can in principle be modelled holographically as bound states.
By neglecting the weak force such a configuration would not be stable
at rest, though it could be stable moving at the speed of light. Realistic
nuclei move a tiny bit slower and would therefore not be stable, but
we do not think this is important at the time scales involved in realistic
collisions.%
} as a source at a AdS radial position of about $1$ fm, where we used
the scale/radius duality. Alternatively we could place the sources
at about $1/T_{QCD}\approx1.2$ fm, with $T_{QCD}$ the QCD deconfinement
temperature. After some nucleons collide a horizon would form at a
depth of about $1/T\sim0.1$ fm, thereby completely hiding the original
sources behind the horizon, making their details of little importance.
On the other hand, the nucleus itself is much bigger (14 fm) than
the depth of the sources. This allows the horizon to fall off outside
the collision region sufficiently fast such that nucleons not directly
involved in the collision (spectators) can move on almost unperturbed.
Furthermore, the relatively big size of the nucleus makes it likely
that transverse dynamics can be treated independently of longitudinal
dynamics during the initial stage.

The above discussion does not say much about the difference between
an infinite coupling constant and the real-world QCD coupling constant,
which goes to zero at asymptotically high energies. At energies at
RHIC and LHC the QCD coupling is presumably not yet very weak, as
otherwise the success of hydrodynamic modelling of heavy ion and more
recently proton-lead collisions would be hard to explain. Nevertheless,
there is little reason to expect an infinite coupling approximation
to be valid. In fact, we believe that especially the total multiplicities
found in subsection \ref{sub:Multiplicities-and-a} suggest that at
infinite coupling there is more stopping than what is found in LHC
measurements. It is only natural that finite coupling corrections
would reduce the stopping, thereby getting closer to experimental
data. Presumably, this will reveal itself as a widening of the rapidity
profile (section \ref{sec:Rapidity-profile:-Bjorken}), which we found
to be universal at high energy and infinite coupling.

\section{Future directions}

Properly taking into account finite coupling effects within AdS/CFT
would most likely be the most promising avenue to make a more precise
link with experiments. Some efforts have been made by modifying the
UV geometry or including finite coupling effects in simple models
of thermalisation \cite{Gubser:2009sx,Kiritsis:2011yn,Steineder:2012si,Stricker:2013lma}.
It may also be necessary to critically compare the field content of
theories in AdS with QCD. It is generally believed that gluons dominate
most of the dynamics in both theories, but other fields definitely
play a role. In particular, it may be that quarks in QCD require a
different description than the one presented in this thesis.

On the other hand one may take a more phenomenological approach, which
has been the main motivation for chapter \ref{chap:Thermalisation-with-radial}.
There, we fitted the normalisation of the initial energy density in
order to match the total multiplicity. This fitting parameter indeed
turned out to be significantly different from the energy density we
would have found using a more complete calculation such as for instance
the colliding shock waves of chapter \ref{chap:Colliding-planar-shock}.
Optimistically one could therefore say that fitting the normalisation
of the energy density takes into account weak-coupling effects, and
in this specific case also violations of boost-invariance. We found
that this approach has appealing features, being most importantly
the dynamical transition to hydrodynamics and the almost perfect fit
of LHC measurements.

So in future we will most likely combine several of these ideas. It
would be good to try and do a full first principle finite coupling
calculation, but at the same time a lot can be achieved by using more
phenomenological inspired techniques. In particular, it would be interesting
to relax the rotational and boost symmetry imposed in chapter \ref{chap:Thermalisation-with-radial},
perhaps using a linearised approximation. It will also be worthwhile
to combine this study of thermalisation with other observables, most
notably the quenching of jets, or the production of photons.

Lastly, we would like to reiterate our introductory statement that
experimental heavy-ion data are highly constraining, and in future
this will lead to a much more complete understanding of the quark-gluon
plasma and QCD in general. We hope that our strongly coupled AdS/CFT
approach will be part of that understanding.

\chapter*{Appendix\markboth{\large APPENDIX}{\large APPENDIX}}

Here we write out the explicit form of the geodesic equations used
in subsection \ref{subsec:FG-to-EF}

\begin{equation}
\begin{array}{c}
\frac{\left(a'(\lambda)+1\right)^{2}+b'(\lambda)c'(\lambda)}{a(\lambda)+\lambda}=\frac{\lambda a''(\lambda)+2a'(\lambda)-h\lambda(a(\lambda)+\lambda)^{3}b'(\lambda)^{2}+2}{\lambda}\\
\frac{2\left(a(\lambda)-\lambda a'(\lambda)\right)b'(\lambda)}{\lambda(a(\lambda)+\lambda)}+b''(\lambda)=0\\
8h(a(\lambda)+\lambda)^{3}\left(a'(\lambda)+1\right)b'(\lambda)+\frac{2\left(\lambda a'(\lambda)-a(\lambda)\right)c'(\lambda)}{\lambda(a(\lambda)+\lambda)}+(a(\lambda)+\lambda)^{4}b'(\lambda)^{2}h'=c''(\lambda),
\end{array}
\end{equation}
where $h\equiv h[t+z+b(\lambda,\, t)]$ and we suppressed the time
dependence of $a$, $b$ and $c$, present through their boundary
conditions.

The Einstein equations for the shock wave collisions in chapter \ref{chap:Colliding-planar-shock}
can be reduced to \cite{Chesler:2010bi}:{\footnotesize 
\begin{equation}
\hspace{-0.5cm}\begin{array}{ccc}
S'' & = & -\frac{1}{2}SB'^{2}\\
S^{2}F'' & = & S\left(6\tilde{S}B'+4\tilde{S}'+3F'S'\right)+S^{2}\left(3\tilde{B}B'+2\tilde{B}'\right)-4\tilde{S}S'\\
12S^{2}\dot{S}' & = & e^{2B}\left(S\left(4\tilde{B}F'-7\tilde{B}^{2}-4\tilde{\tilde{B}}+2\tilde{F}'+F'^{2}\right)+2\tilde{S}\left(F'-8\tilde{B}\right)-8\tilde{\tilde{S}}+8\tilde{S}^{2}\right)+24S\left(S^{2}-\dot{S}S'\right)\\
6S^{3}\dot{B}' & = & e^{2B}\left(S\left(-\tilde{B}F'+\tilde{B}^{2}+\tilde{\tilde{B}}-2\tilde{F}'-F'^{2}\right)+\tilde{S}\left(\tilde{B}+4F'\right)+2\tilde{\tilde{S}}-4\tilde{S}^{2}\right)-9S^{2}\left(\dot{S}B'+\dot{B}S'\right)\\
2S^{3}A'' & = & e^{2B}\left(S\left(7\tilde{B}^{2}+4\tilde{\tilde{B}}-F'^{2}\right)+16\tilde{B}\tilde{S}+8\tilde{\tilde{S}}-32\tilde{S}^{2}\right)-2S^{3}\left(3\dot{B}B'+4\right)+24\dot{S}SS'\\
6S^{2}\dot{F}' & = & 3\left(8\dot{S}\tilde{S}+2S\left(S'\left(\tilde{A}+2\dot{F}\right)-6\dot{B}\tilde{S}-4\tilde{\dot{S}}-3\dot{S}F'\right)-S^{2}\left(2B'\left(\tilde{A}+2\dot{F}\right)+2\tilde{A}'+6\dot{B}\tilde{B}+4\tilde{\dot{B}}+A'F'\right)\right)\\
6S^{2}\ddot{S} & = & e^{2B}\left(S\left(2\tilde{B}\left(\tilde{A}+2\dot{F}\right)+\tilde{\tilde{A}}+2\tilde{\dot{F}}\right)+\tilde{S}\left(\tilde{A}+2\dot{F}\right)\right)+3S^{2}\left(\dot{S}A'-\dot{B}^{2}S\right),
\end{array}\label{eq:EEshock}
\end{equation}
}where $h'=\partial_{r}h$, $\dot{h}=\partial_{t}h+\frac{1}{2}Ah'$
and $\tilde{h}=\partial_{y}h-Fh'$. Note that these operators do not
commute and so that one has to be careful that $\tilde{h}'=(\tilde{h})'$.

The Einstein equations for the radial expanding plasma in chapter
\ref{chap:Thermalisation-with-radial} can be reduced to:{\tiny 
\begin{equation}
\hspace{-1cm}\begin{array}{ccc}
3S'' & = & -\frac{1}{6}S\left(B'C'+\left(B'\right)^{2}+\left(C'\right)^{2}\right)\\
2S^{2}F'' & = & S^{2}\left(\tilde{B}\left(2B'+C'\right)+\tilde{C}\left(B'+2C'\right)-2\tilde{B}'\right)+S\left(-6\tilde{S}B'+8\tilde{S}'+6F'S'\right)-8\tilde{S}S'\\
12S^{3}\dot{S}' & = & e^{-B}\left(-S^{2}\left(\tilde{B}\left(\tilde{C}+2F'\right)-2\left(\tilde{\tilde{B}}+\tilde{F}'-12e^{B}\dot{S}S'\right)+2\tilde{B}^{2}+\tilde{C}^{2}-\left(F'\right)^{2}\right)+2S\left(\tilde{S}\left(4\tilde{B}+F'\right)-4\tilde{\tilde{S}}\right)+4\tilde{S}^{2}+24e^{B}S^{4}\right)\\
6S^{3}\dot{C}' & = & e^{-B}\left(S\left(2\tilde{B}\left(2\tilde{C}+F'\right)+2\tilde{B}^{2}-2\tilde{\tilde{B}}+3\tilde{C}F'+\tilde{C}^{2}-3\tilde{\tilde{C}}-2\tilde{F}'-F'^{2}\right)+\tilde{S}\left(4F'-2\tilde{B}-3\tilde{C}\right)+2\tilde{\tilde{S}}-4\tilde{S}^{2}\right)-9S^{2}\left(\dot{S}C'+\dot{C}S'\right)\\
6S^{4}\dot{B}' & = & e^{-B}\left(-S^{2}\left(\tilde{B}\left(2\tilde{C}+F'\right)+\tilde{B}^{2}-\tilde{\tilde{B}}+2\tilde{C}^{2}-4\tilde{F}'-2F'^{2}\right)+S\left(\tilde{S}\left(\tilde{B}-8F'\right)-4\tilde{\tilde{S}}\right)+8\tilde{S}^{2}-9e^{B}S^{3}\left(\dot{S}B'+\dot{B}S'\right)\right)\\
2S^{4}A'' & = & e^{-B}\left(S^{2}\left(\tilde{B}\tilde{C}+2\tilde{B}^{2}-2\tilde{\tilde{B}}+\tilde{C}^{2}+24e^{B}\dot{S}S'-\left(F'\right)^{2}\right)+8S\left(\tilde{\tilde{S}}-\tilde{B}\tilde{S}\right)-4\tilde{S}^{2}-e^{B}S^{4}\left(\dot{B}+\dot{C}\left(B'+2C'\right)+8\right)\right)\\
2S^{2}\dot{F}' & = & 8\dot{S}\tilde{S}-S^{2}\left(2\tilde{A}'-B'\left(\tilde{A}+2\dot{F}\right)+\tilde{B}\left(2\dot{B}+\dot{C}\right)+\tilde{C}\left(\dot{B}+2\dot{C}\right)-2\tilde{\dot{B}}+A'F'\right)+2S\left(S'\left(\tilde{A}+2\dot{F}\right)+3\dot{B}\tilde{S}-4\tilde{\dot{S}}-3\dot{S}F'\right)\\
6S^{2}\ddot{S} & = & e^{-B}\left(S\left(\tilde{B}\left(-\left(\tilde{A}+2\dot{F}\right)\right)+\tilde{\tilde{A}}+2\tilde{\dot{F}}\right)+\tilde{S}\left(\tilde{A}+2\dot{F}\right)+3e^{B}\dot{S}S^{2}A'-e^{B}S^{3}\left(\dot{B}\dot{C}+\dot{B}^{2}+\dot{C}^{2}\right)\right),
\end{array}
\end{equation}
}where now $\tilde{h}=\partial_{\rho}h-Fh'$. Note that the tilded
derivative $\tilde{h}$ does not actually simplify computations much
as one still has to compute the tilded derivative explicitly. Here
it just serves to present the equations more compactly. In the actual
code all functions are regularised near the boundary, i.e. we redefine

\noindent {\small 
\begin{eqnarray}
A & = & r^{2}+\frac{Af}{r^{2}},\nonumber \\
B & = & -\frac{2}{3}\log(\tau\rho)+\frac{3r\tau(1-2r\tau)-2}{9r^{3}\tau^{3}}+\frac{Bf}{r^{4}},\nonumber \\
C & = & -\frac{2}{3}\log(\tau/\rho^{2})+\frac{3r\tau(1-2r\tau)-2}{9r^{3}\tau^{3}}+\frac{Cf}{r^{4}},\nonumber \\
S & = & \rho^{1/3}\frac{3r\tau(9r\tau(3r\tau(3r\tau+1)-1)+5)-10}{243r^{3}\tau^{11/3}}+\frac{Sf}{r^{4}},\nonumber \\
F & = & \frac{Ff}{r^{2}},\label{eq:near-boundary-radial-1}
\end{eqnarray}
where all computations are then done with $Af$, $Bf$, $Cf$, $Sf$,
$Ff$ and analogously $\dot{Bf}$ , $\dot{Cf}$ and $\dot{Sf}$, all
depending on $r,\,\tau$ and $\rho$. The gauge freedom $\xi(\tau,\,\rho)$
is still suppressed, but can easily be reinstated by letting $r\rightarrow r+\xi(\tau,\,\rho)$
and redefining $A$ and $F$ appropriately. Although these redefinitions
make the equations much longer, it has the advantage that one can
easily and accurately extract the boundary stress tensor, without
the need to compute derivatives.}{\small \par}

\chapter*{Nederlandse samenvatting}

\addcontentsline{toc}{chapter}{Nederlandse samenvatting}
\markboth{NEDERLANDSE SAMENVATTING}{NEDERLANDSE SAMENVATTING}

De meeste mensen zullen denken dat de kleinste elementaire deeltjes
weinig te maken hebben met de grootste en zwaarste objecten in ons
heelal. Toch lijken ontdekkingen uit de snaartheorie hier wel op te
wijzen. Niet letterlijk misschien, maar zwarte gaten kunnen een erg
goed model vormen voor het gedrag van het quark-gluon plasma, zoals
dat bij de LHC in Gen�ve geproduceerd wordt. 

Extremer kan het in de natuurkunde bijna niet worden; bij botsingen
van loodkernen in de LHC onstaat voor ongeveer $10^{-23}$ seconde
een plasma van quarks en gluonen met een temperatuur van $10^{12}$K
en versnellingen van wel $10^{31}g$. Voeg hieraan toe dat dit proces
misschien wel het best te beschrijven is met de vorming van een zwart
gat in een 5 dimensionaal universum, waarbij het zwarte gat ongeveer
de helft van dit universum inneemt, en het is duidelijk dat het hier
om extreme natuurkunde gaat. In dit proefschrift wordt de relatie
tussen loodkernen en zwarte gaten uitgelegd en wordt gekeken naar
hoe we dit in praktijk kunnen gebruiken.

\section*{Holografie }

Een oud idee uit 1974 van Gerard \textquoteright{}t Hooft \cite{'tHooft:1973jz}
is dat quarks en gluonen, zoals beschreven door de Kwantumchromodynamica
(QCD), equivalent kunnen zijn aan een theorie van snaren, waarbij
de snaren tussen de quarks spannen. Toen al was duidelijk dat een
dergelijke snaartheorie mogelijk kon worden versimpeld, maar de precieze
uitwerking hiervan was erg ingewikkeld. 

De doorbraak werd uiteindelijk in 1997 door Juan Maldacena gevonden
\cite{Maldacena:1997zz}, die een precieze snaartheorie vond die overeen
kwam met een precieze kwantumtheorie. Deze kwantumtheorie heeft dan
wel supersymmetrie, maar lijkt toch vrij veel op normale niet-supersymmetrische
QCD. Achteraf gezien was er in 1974 allereerst veel meer kennis over
snaartheorie nodig, maar een ander verrassend aspect was dat de snaartheorie
��n extra ruimtelijke dimensie heeft: de snaartheorie is 4+1 dimensionaal,
in tegenstelling tot de 3+1 dimensionale kwantumtheorie. Hierdoor
heeft het de naam holografie gekregen.

Berekeningen in snaartheorie zijn ontzettend ingewikkeld, maar in
de situatie dat de snaartjes heel klein zijn (puntdeeltjes) reduceert
de snaartheorie tot normale \hyphenation{natuur-kunde}, met zwaartekracht
en de andere krachten. In deze limiet krijgen we een \textquoteleft{}normaal\textquoteright{}
universum met onder andere zwarte gaten terug, maar dan wel met ��n
dimensie meer. In dit simpele geval is equivalente QCD juist (bijna)
onmogelijk op te lossen, zodat een zwart gat echt een versimpeling
voor het quark-gluon plasma is!

\section*{Botsingen van loodkernen }

E�n maand per jaar botsen er in de Large Hadron Collider (LHC) loodkernen
op elkaar. De rest van het jaar worden protonen gebotst die het meest
bruikbaar zijn voor het vinden van het Higgsboson. Loodkernen zijn
echter veel zwaarder dan protonen en aangezien de snelheid gelijk
is, geeft deze botsing veel meer energie. 

\renewcommand{\figurename}{Figuur}

\begin{figure}
\begin{centering}
\includegraphics[width=12cm]{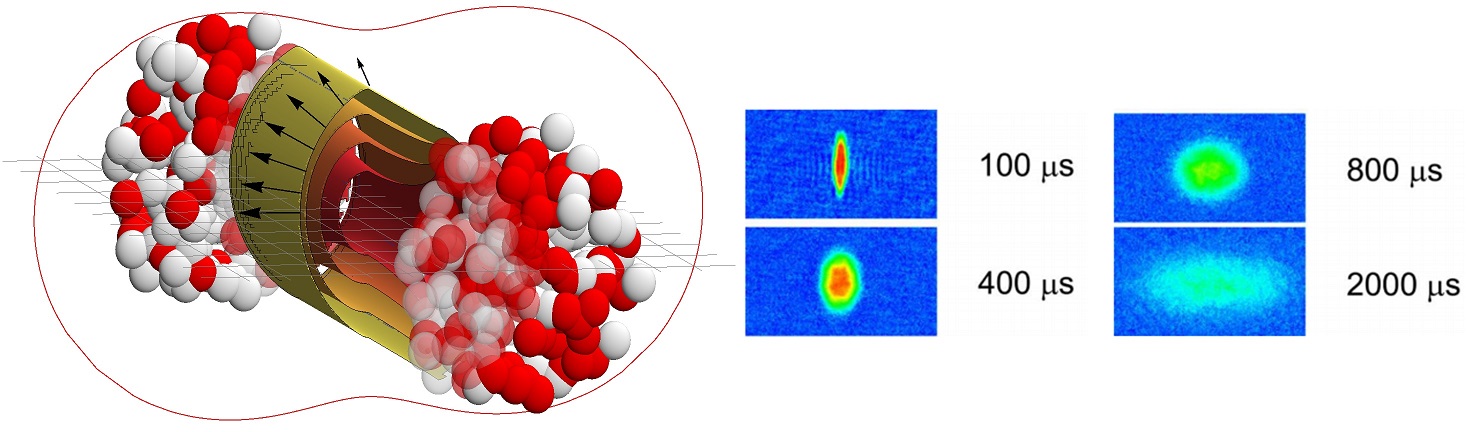}
\par\end{centering}

\caption{Een botsing van twee loodkernen. In de regio waar de kernen overlappen
onstaat een heet quark-gluon plasma, dat ongeveer ellipsvormig is.
In de detector worden meer deeltjes gevonden in de richting van de
korte as, wat ge�llustreerd wordt door de rode lijn. De expansie van
het quark-gluon plasma zal nooit gefotografeerd worden, maar het lijkt
veel op het rechter plaatje. Dit zijn moment opnames van bepaalde
expanderende zeer koude atomen\cite{O'Hara:2002zz}. Het leuke hiervan
is dat deze atomen net als het plasma door een moeilijke kwantumtheorie
worden beschreven; het idee is dan ook dat ook hier zwarte gaten een
model vormen voor deze vloeistof met zeer lage viscositeit. In zekere
zin zijn de koudste en warmste vloeistoffen op aarde dus erg vergelijkbaar!\label{fig:elliptic-flow}}
\end{figure}

Met een paar honderd botsende energetische protonen en neutronen zou
je kunnen verwachten dat alle deeltjes een paar keer botsen en in
een betrekkelijk willekeurige richting in de detector belanden. Dit
is echter niet wat in de LHC gevonden wordt; de deeltjes bewegen voornamelijk
in de richting van de korte kant, zoals in figuur \ref{fig:elliptic-flow}
ge�llusteerd. Deze zogenoemde \textquoteleft{}elliptic flow\textquoteright{}
toont aan dat er veel interacties zijn, die de deeltjes in de x-as
duwen. 

Hoewel er nu veel experimentele data zijn, resteren nog veel open
vragen over het quark-gluon plasma. Allereerst is het experimenteel
een gigantische taak om naar het quark-gluon plasma te \textquoteleft{}kijken\textquoteright{}.
Het plasma zelf bestaat daar namelijk veel te kort voor en experimenteel
zijn dus alleen de wegvliegende deeltjes lang na de botsing te detecteren.
Hier zit echter een schat aan informatie in, zoals bijvoorbeeld de
vorm van de distributie in figuur \ref{fig:elliptic-flow} afgebeeld.
Maar ook het type deeltje, de snelheid, de verdeling in de z-richting
en andere data kunnen allemaal nauwkeurig gemeten worden.

\section*{Een theoretische beschrijving }

De uitdaging is uiteraard om al deze data theoretisch te voorspellen.
Dit lukt heel aardig, maar er is een aantal cruciale aannames en waardes
die moeilijk theoretisch te onderbouwen zijn. Het typische model is
nu dat de individuele quarks en gluonen zich heel erg snel als een
vloeistof gaan gedragen; daarna expandeert het plasma volgens relativistische
hydrodynamica, met de kleinste viscositeit ooit gemeten (zie ook figuur
\ref{fig:elliptic-flow}). Het plasma wordt daarom ook wel de meest
perfecte vloeistof genoemd. Op een gegeven moment is de energiedichtheid
zo laag dat zo\textquoteright{}n 30.000 deeltjes ontstaan, die dan
nog weer veel later in de detector worden gemeten. 

Het is met name erg moeilijk te beantwoorden waarom en hoe de deeltjes
zo snel een vloeistof vormen. Ook de lage viscositeit is niet uit
te rekenen binnen QCD. Holografie, echter, geeft op een heel natuurlijke
wijze een lage viscositeit \cite{Policastro:2001yc}, aangezien in
zekere zin ook de horizon van een zwart gat zich gedraagt als een
perfecte vloeistof. Het hieronder gepresenteerde onderzoek probeert
een (holografisch) beeld te vormen van de allereerste evolutie, nog
voor de deeltjes een echte vloeistof vormen.

\section*{De formatie van zwarte gaten }

Zoals eerder al geanticipeerd, is het erg moeilijk om met QCD precieze
berekeningen te maken, in het bijzonder in de chaotische botsing van
twee atoomkernen. Dit komt doordat QCD de sterke kracht beschrijft,
die zo sterk is dat hij zich moeilijk laat benaderen. 

Het mooie (en soms lastige) van holografie is dat als de kwantummechanische
deeltjestheorie moeilijk is, de snaartheorie juist makkelijk is (en
vice versa). In dit geval versimpelt snaartheorie naar klassieke zwaartekracht
met eventueel andere krachten. Deze zwaartekracht, beschreven door
Einsteins algemene relativiteitstheorie, zou dus goed een model kunnen
vormen voor een moeilijke kwantumtheorie! 

Het uitgangspunt in deze berekening is dat het quark-gluon plasma
op een bepaalde manier goed beschreven kan worden door eigenschappen
van de horizon van een zwart gat. De botsing zelf is dan equivalent
aan de formatie van een zwart gat. Op een gegeven moment zal de horizon
van dit zwarte gat goed worden beschreven door hydrodynamica en de
grote vraag is hoe het plasma er op dat moment uitziet. Het gaat hierbij
dan met name om wanneer hydrodynamica werkt en wat het snelheidsprofiel
van de vloeistof op dat moment is. 

Berekeningen in algemene relativiteitstheorie zijn een vakgebied op
zichzelf, en het heeft daarom ook een tijd geduurd om de formatie
van een zwart gat te beschrijven. In dit proefschrift is dit op een
aantal manieren geprobeerd. Hoofdstuk \ref{chap:General-relativity-in}
bestudeert het betrekkelijk triviale geval van een volledig homogeen
zwart gat. Toch was dit al interessant, want hier kwam uit dat dit
zwarte gat altijd erg snel door hydrodynamica beschreven wordt.

Hoofdstuk \ref{chap:Colliding-planar-shock} beschrijft de botsing
van schokgolven van zwaartekracht. In aanvulling op eerder werk \cite{Chesler:2010bi}
vonden we grote verschillen tussen schokgolven met veel en schokgolven
met weinig energie. Met name dit eerste regime van hoge energie botsingen
leidt tot interessante vergelijkingen met botsingen in LHC. Het belangrijkste
resultaat is mogelijk een universeel snelheidsprofiel in de richting
van de botsing (sectie \ref{sec:Rapidity-profile:-Bjorken}), hetgeen
fundamenteel anders is dan de twee bestaande profielen van Landau
\cite{Landau:1953gs} en Bjorken \cite{Bjorken:1982qr}.

Mogelijk het meest innovatief bestudeert hoofdstuk \ref{chap:Thermalisation-with-radial}
een model dat dynamica en expansie in het vlak transversaal op de
botsing heeft. Hoewel dit nog steeds rotatie-symmetrisch is, is het
hiermee mogelijk met de uitkomsten in bestaande hydrodynamische modellen
verder te rekenen. Dit leidde uiteindelijk tot meetbare deeltjesspectra,
die verrassend goed overeenkomen met LHC meetresultaten (figuur \ref{fig:spectra}).

\section*{Discussie }

Hoewel veel technische details achterwege moesten blijven, is hopelijk
toch duidelijk geworden waarom quarks en gluonen misschien wel het
best door zwarte gaten beschreven kunnen worden. E�n van de leuke
dingen hiervan is dat zwaartekracht in het algemeen \textquoteleft{}moeilijke\textquoteright{}
kwantumtheorie�n beschrijft, zoals het quark-gluon plasma, maar ook
bijvoorbeeld bepaalde koude atomen (figuur \ref{fig:elliptic-flow}).
Doordat beide systemen door zwarte gaten beschreven kunnen worden,
gedragen ze zich erg vergelijkbaar; iets wat erg onverwacht is voor
de heetste en koudste vloeistoffen op aarde. 

Holografie is essentieel gebleken om experimentele data van het quark-gluon
plasma te verklaren. Een beschrijving in termen van zwarte gaten geeft
hierbij een natuurlijke verklaring waarom het plasma zich zo snel
als vloeistof gedraagt en waarom de viscositeit zo laag is. Met holografie
is het nu ook mogelijk de allereerste momenten van de botsing te simuleren,
als het plasma zich nog niet als een vloeistof gedraagt. E�n van de
belangrijkste successen daarvan is een realistisch snelheidsprofiel,
zowel in de richting van de botsing, als loodrecht daarop (hoofdstuk
\ref{chap:Colliding-planar-shock} en \ref{chap:Thermalisation-with-radial}).
Voor het loodrechte profiel blijkt dit goed overeen te komen met experimentele
data, voor het longitudinale profiel is dit minder duidelijk (sectie
\ref{sec:A-comparison-with}).

Een heel ander recent succes is een model van zwarte gaten met supergeleiding
\cite{Hartnoll:2008vx}. In dit geval zijn het meer ingewikkelde zwarte
gaten, met scalaire en elektrische velden. Deze extra\textquoteright{}s
leveren een ingewikkelder fasediagram op, waar ook een supergeleidende
fase in blijkt te zitten. Hoewel het niet heel duidelijk is wat voor
precieze kwantumtheorie zulke zwarte gaten zou moeten beschrijven,
is er natuurlijk de hoop dat het iets te maken heeft met hoge temperatuur
supergeleiders. Deze zijn nog erg slecht begrepen, ook weer omdat
het een moeilijke kwantumtheorie betreft. 

Het is ook leuk om op te merken dat holografie helemaal niet ontdekt
is met oog op dit soort toepassingen vanuit de snaartheorie. Onderzoekers
waren veelal ge�nteresseerd in het vinden van een \textquoteleft{}theorie
van alles\textquoteright{}. Hoewel holografie hier zeker aan bijdraagt,
zal het misschien nog wel veel nuttiger blijken voor het begrijpen
van ingewikkelde kwantumsystemen. En wie weet, begrijpen we uiteindelijk
QCD, zwarte gaten of het mechanisme achter supergeleiding bij hoge
temperatuur.

\addcontentsline{toc}{chapter}{Acknowledgements}

\begingroup \renewcommand{\vspace}[2]{}% Gobble 2 arguments after \vspace

\chapter*{Acknowledgements}

\endgroup

First of all I would like to thank my advisors, Gleb Arutyunov, Thomas
Peitzmann and also Raimond Snellings. Even though it may be hard to
really bridge experimental physics with mathematical physics, our
discussions and your physical insights have always helped. But most
of all I got complete freedom and trust to do and go what and where
I thought best, which made these four years a very exciting project.

This work would never have gone anywhere without excellent collaborators.
Firstly this was Micha\l{}, giving me a good start in numerics and
quickly setting up the collaboration with David. The intensive weeks
discussing physics with David have definitely been a highlight of
this project, and of course the excellent non-physics times in Spain
were great. Also Jorge and Miquel joined, strengthening the team with
heavy ion knowledge and numerical skills. My time with Paul in Colorado
has also guaranteed fun discussions and productive work together with
Scott.

I have always enjoyed excellent education, in particular I would like
to thank Hans Jordens for his efforts in the physics olympiad during
my time in secondary school, Toine Arts and Henk Stoof for providing
so much freedom in my Bachelor's and Master's, and Gerard 't Hooft
during my Master's for showing me great intuition and research in
general.

This research was generously supported, most of all by a Utrecht University
Foundations of Science grant. I also thank Universitat de Barcelona
and Perimeter Institute for long time stays and the University of
Colorado and MIT for support and hospitality during shorter stays.
This thesis has benefited from interesting discussions with Koenraad
Schalm, Paul Chesler, Larry Yaffe, Roberto Emparan, Derek Teaney,
Krishna Rajagopal, Umut G\"{u}rsoy, Luis Lehner and Rob Myers.

Then of course it was great to have such nice colleagues. In Utrecht
there were Alessandro, Chiara, Dra\^{z}en, Nava, Jan, Philipp, Anne, Javier,
Alessandra, Jasper, Hedwig, Martijn, Chris, and Vivian. I greatly
enjoyed my time in Barcelona, mainly due to Marina, Adriana, Blai,
Markus, Ivan, Miquel, Javier again, and I thank Daniel also for comments
on this manuscript. And lastly everyone at Perimeter, Pablo, Dami\'{a}n,
Anton, Yangang, Heidar, Anthony, Dalit and Philipp again.

Outside of physics I want to thank my friends for all the nice times:
Alexander, Arno, Dieuwertje, Jan, Lysanne, Niels, Pieter, Renee, Sietse,
Thomas, and especially Ronnie and Yvette as paranymphs.

Finally, I am very grateful to have such a supportive and caring family,
Margo, Henk, Marlies and Haitske; thank you for all your love and
encouragement. And most importantly Ino, for having such a happy life
together.

\chapter*{Curriculum Vitae}

The author was born on February 3, 1987 in Wageningen, the Netherlands.
He completed secondary school in 2004 at Het Streek in Ede, after
which he moved on to study physics and mathematics at Utrecht University,
obtaining bachelor's degrees in both programmes in 2008 (\emph{cum
laude}). The next two years he continued there for a master's degree
(\emph{cum laude}) in the programme Theoretical Physics, with a master
thesis `Black holes as Information Scramblers' supervised by prof.
Gerard 't Hooft. 

From 2010 till 2014 the current PhD work was performed, under supervision
of prof. Gleb Arutyunov, prof. Thomas Peitzmann and prof. Raimond
Snellings, with the goal of bringing recent progress on heavy ion
collisions using AdS/CFT closer to experiments. After obtaining his
doctorate the author will continue his research at the Center of Theoretical
Physics at the Massachusetts Institute of Technology in Cambridge,
USA.

\bibliographystyle{bibstyle}
\bibliography{outlinefinal}

\end{document}